\documentclass{aastex631}

\usepackage{natbib}
\usepackage{amsmath}
\usepackage{graphicx}
\usepackage{multirow}
\usepackage{CJKutf8}

\newcommand{\pkg}[1]{\texttt{#1}}

\newcommand{\jlname}{\begin{CJK}{UTF8}{gbsn}(李嘉霖)\end{CJK}}

\newcommand{\farcsec}{\hbox{$.\!\!^{\prime\prime}$}}

\newcommand{\hr}{HR~4796A}
\newcommand{\sbf}{$SB_{disk} / F_{star}$ (arcsec$^{-2}$)}
\newcommand{\iband}{$i'$-band}
\newcommand{\zband}{$z'$-band}
\newcommand{\rband}{$r'$-band}
\newcommand{\gband}{$g'$-band}
\newcommand{\adi}{KLIP-ADI}
\newcommand{\rdi}{KLIP-RDI}

\begin{document}

\title{A Multiband Study of the HR 4796A Disk in the Optical Using MagAO-X}

\shorttitle{Multiband MagAO-X Imaging of HR 4796A}
\shortauthors{Kueny et al.}

\author[0000-0001-8531-038X]{Jay K. Kueny}
\affiliation{Steward Observatory, University of Arizona, Tucson, 933 N Cherry Ave, Tucson, AZ 85721, USA}
\affiliation{James C. Wyant College of Optical Sciences, University of Arizona, 1630 E. University Blvd., Tucson, AZ 85721, USA}
\affiliation{National Science Foundation Graduate Research Fellow}

\author[0000-0001-6654-7859]{Alycia J. Weinberger}
\affiliation{Earth and Planets Laboratory, Carnegie Institution for Science, 5241 Broad Branch Road NW, Washington, DC 20015-1305}

\author[0000-0001-7233-4171]{Zhe-Yu Daniel Lin}
\affiliation{Earth and Planets Laboratory, Carnegie Institution for Science, 5241 Broad Branch Road NW, Washington, DC 20015-1305}

\author[0000-0003-1905-9443]{Joseph D. Long}
\affiliation{Center for Computational Astrophysics, Flatiron Institute, 162 5th Ave, New York, NY}

\author[0000-0002-2346-3441]{Jared R. Males}
\affiliation{Steward Observatory, University of Arizona, Tucson, 933 N Cherry Ave, Tucson, AZ 85721, USA}

\author[0000-0002-4934-3042]{Joshua Liberman}
\affiliation{James C. Wyant College of Optical Sciences, University of Arizona, 1630 E. University Blvd., Tucson, AZ 85721, USA}
\affiliation{Steward Observatory, University of Arizona, Tucson, 933 N Cherry Ave, Tucson, AZ 85721, USA}

\author[0000-0002-8110-7226]{Jialin Li \protect\jlname}
\affiliation{Steward Observatory, University of Arizona, Tucson, 933 N Cherry Ave, Tucson, AZ 85721, USA}
\affil{National Science Foundation Graduate Research Fellow}

\author[0000-0001-5130-9153]{Sebastiaan Haffert}
\affiliation{Leiden Observatory, Leiden University, PO Box 9513, 2300 RA Leiden, The Netherlands}
\affiliation{Steward Observatory, University of Arizona, Tucson, 933 N Cherry Ave, Tucson, AZ 85721, USA}

\author[0000-0002-2167-8246]{Laird M. Close}
\affiliation{Steward Observatory, University of Arizona, Tucson, 933 N Cherry Ave, Tucson, AZ 85721, USA}

\author[0000-0003-0843-5140]{Eden McEwen}
\affiliation{Steward Observatory, University of Arizona, Tucson, 933 N Cherry Ave, Tucson, AZ 85721, USA}
\affiliation{James C. Wyant College of Optical Sciences, University of Arizona, 1630 E. University Blvd., Tucson, AZ 85721, USA}
\affil{National Science Foundation Graduate Research Fellow}

\author[0000-0003-3253-2952]{Maggie Y. Kautz}
\affiliation{Steward Observatory, University of Arizona, Tucson, 933 N Cherry Ave, Tucson, AZ 85721, USA}
\affiliation{James C. Wyant College of Optical Sciences, University of Arizona, 1630 E. University Blvd., Tucson, AZ 85721, USA}

\author[0000-0002-1097-9908]{Olivier Guyon}
\affiliation{Subaru Telescope, National Observatory of Japan, Hilo, HI}
\affiliation{Steward Observatory, University of Arizona, Tucson, 933 N Cherry Ave, Tucson, AZ 85721, USA}
\affiliation{James C. Wyant College of Optical Sciences, University of Arizona, 1630 E. University Blvd., Tucson, AZ 85721, USA}

\author[0000-0003-3904-7378]{Logan Pearce}
\affiliation{Department of Astronomy, University of Michigan, Ann Arbor, MI}

\author[0009-0005-5534-7495]{Parker T. Johnson}
\affiliation{Steward Observatory, University of Arizona, Tucson, 933 N Cherry Ave, Tucson, AZ 85721, USA}
\affil{National Science Foundation Graduate Research Fellow}

\author[0009-0002-9752-2114]{Katie Twitchell}
\affiliation{James C. Wyant College of Optical Sciences, University of Arizona, 1630 E. University Blvd., Tucson, AZ 85721, USA}
\affiliation{Steward Observatory, University of Arizona, Tucson, 933 N Cherry Ave, Tucson, AZ 85721, USA}

\author[0000-0002-5559-1544]{Alex Hedglen}
\affiliation{Northrop Grumman Corporation, 600 South Hicks Rd, Rolling Meadows, IL}

\author{Avalon Gower}
\affiliation{Draper Laboratory, 555 Technology Square, Cambridge, MA}

\author{Warren Foster}
\affiliation{Steward Observatory, University of Arizona, Tucson, 933 N Cherry Ave, Tucson, AZ 85721, USA}

\author{Jhen Lumbres}
\affiliation{Northrop Grumman in Pasadena, CA}

\author{Lauren Schatz}
\affiliation{Starfire Optical Range, Kirtland Air Force Base, Albuquerque, NM}

\author[0009-0006-4370-822X]{Elena Tonucci}
\affiliation{Leiden Observatory, Leiden University, PO Box 9513, 2300 RA Leiden, The Netherlands}

\begin{abstract}

We present total intensity images of the debris disk around HR 4796A from observations spanning 2023 to 2025 with the Magellan extreme adaptive optics instrument (MagAO-X). We detected the disk at high signal-to-noise ratios at $g'\ (527$ nm), $r'\ (615$ nm), $i'\ (762$ nm), and $z'\ (909 $ nm). Additionally, we present images collected using the ``star-hopping" technique that show the entirety of the disk, including the dramatic forward-scattering at the minor axis. We subjected our images to a battery of modeling techniques to constrain the geometry and photometry of the disk. Leveraging our clear detections of the disk's minor axis, we modeled the scattering phase function (SPF) using a basis of the Legendre polynomials. To mitigate self-subtraction artifacts in our angular differential imaging, we implemented a forward-modeling pipeline that generates a pixel-based freeform disk forward model leading to a deconvolved image of the disk. Our best-fit disk models reveal: (1) highly forward-scattering SPFs with a minimum at the $\sim65^{\circ}$ scattering angle, (2) a faint halo of dust just exterior to the spine of the disk that is not well-described by a broken power law density profile, (3) a red spectral slope for the dust, and finally (4) a compact, clump-like feature in the freeform disk models. Our empirically-measured SPFs suggest that the scattering is dominated by large, highly-absorptive grains. However, we emphasize the need for testing advanced irregular grain models using our SPFs to learn more about the physical and chemical properties of this complex system.

\end{abstract}


\section{Introduction} 
\label{sec:intro}

\subsection{Relevant Background}
\label{sec:background}
Studies of debris disks, which are massive analogs to our own zodiacal dust, main asteroid belt, and Kuiper belt, can illuminate the composition of exoplanets as they are generated by the building blocks of planets. Furthermore, the interaction between starlight and their dust encodes the physical properties of the dust grains (e.g., \citealt{mcguire_experimental_1995}; \citealt{pawellek_debris_2024}). Small grains have short lifetimes as they get ejected via radiation pressure \citep{strubbe_dust_2006}, but are continuously replenished through collisions of the larger planetesimals in the disk (so-called ``collisional cascade"). Debris disks encompass a late stage of circumstellar disk evolution and are typically characterized by low gas content, relatively low surface brightness, and low optical depth \citep{hughes_debris_2018} which allows for unobstructed access to the scattering properties of the dust, especially micron-sized dust grains which are very efficient scatterers of light at visible wavelengths. Morphological features in the dust may hint at perturbances brought on by exoplanets --- an idea exemplified by the warp in the disk around $\beta$ Pictoris that alerted observers to an unseen planet \citep{mouillet_planet_1997} which was eventually imaged directly by \cite{lagrange_giant_2010}.

Directly imaging debris disks in different wavelength regimes is critical for probing the full spectrum of grain sizes in a disk \citep{olofsson_halo_2022}. Observations at any wavelength are most sensitive to grains of size approximately equal to that wavelength. Scattered light observations are made at visible to near-infrared (NIR) wavelengths and emission from dust grains is measured at mid-infrared through mm wavelengths.

Another advantage brought by the direct imaging of debris disks in scattered light is the measurement of the disk's scattering phase function (SPF). Dust grains strongly forward-scatter, and the shape of the SPF depends on size and composition and is therefore valuable for retrieving the physical properties of the grains (\citealt{milli_near-infrared_2017}; \citealt{arnold_stumbling_2022}; \citealt{lin_glitterin_2025}). Moderately inclined disks make it possible to measure the broadest range of scattering angles. However, the projected separation of the disk's minor axis to its host star shrinks with increasing inclination, heightening the demand for excellent imaging performance to suppress noise. Several analyses from recent years have included modeling or measurement of a disk's SPF (e.g., \citealt{olofsson_azimuthal_2016}; \citealt{milli_near-infrared_2017}; \citealt{esposito_direct_2018}; \citealt{ren_exo-kuiper_2019}; \citealt{engler_hd_2020}; \citealt{desgrange_dust_2025}). These studies highlight the many challenges that limit our ability to infer the physical properties of the dust through direct images. One such limitation stems from the geometry of the disk in question, which needs to be well-constrained in order to accurately measure the SPF \citep{olofsson_challenge_2020}. Subtracting the stellar PSF from the images (and post-processing in general) also presents another major obstacle towards unbiased measurement of a disk's SPF.

\subsection{Overview of HR 4796A}

\begin{table}[!ht]
    \centering
    \small
    \caption{\hr{} and Reference Star Properties.}
    \begin{tabular}{lcc}
        \hline
         & \hr{}           & HR 4748 \\ \hline
        R.A. (J2000)    & 12 36 01.03     &   12 28 22.47   \\
        Decl. (J2000)   & -39 52 10.22    &   -39 02 28.19 \\
        Spectral Type   & A0V      &      B8/9V     \\
        $V$ (mag)       & $5.87 \pm 0.01$ &       $5.39 \pm 0.01$    \\
        $J$ (mag)       & $4.09 \pm 0.01$ &      $5.515 \pm 0.02$     \\ 
        Distance        & $72.248$ pc      & ---                      \\
        Age             & $8 \pm 2$ Myr     & ---                     \\\hline
    \end{tabular}
    \label{tab:properties}
\end{table}

\hr{} is a young A0 star of estimated age $8 \pm 2$ Myr \citep{stauffer_age_1995} located at a distance of 72.248 pc \citep[Table \ref{tab:properties};][]{gaia_collaboration_gaia_2023}. It also has a possibly bound M dwarf companion HR 4796B which is located at an apparent separation of $7\farcsec7$. \hr{} hosts a bright, highly inclined ring of dust that makes it ideally suited for benchmarking AO instrument performance (it was used to gauge first light performance for SPHERE and for GPI's polarimetry mode, for example; \citealt{perrin_polarimetry_2015}) and also for studying its grain properties because of the broad range of scattering angles that can be measured. That latter point coupled with the recent advances in coronagraphic instruments has subjected the \hr{} disk to numerous studies based on its measured SPF: \cite{schneider_stis_2009} using HST/STIS at visible wavelengths, \cite{milli_near-infrared_2017} using VLT/SPHERE in the NIR, \cite{chen_multiband_2020} using Gemini/GPI in the NIR, and \cite{arriaga_multiband_2020} using Gemini/GPI polarimetry in the NIR. Complementing these studies are prior works that instead inferred the composition of the dust by fitting grain models to the disk spectrum (\citealt{augereau_hr_1999}; \citealt{li_modeling_2003}; \citealt{rodigas_morphology_2014}; \citealt{milli_near-infrared_2017}). \cite{milli_near-infrared_2017} performed a comprehensive analysis of the disk's SPF using SPHERE NIR images but the results raised further questions. Particularly, the surprisingly high values in the $< 30^{\circ}$ scattering angle regime in their measured SPF hint that a significant population of very large grains on the order of tens of microns is needed to explain the high-degree of forward scattering. However, a population of mostly large grains cannot explain the infrared (IR) spectrum of the disk leading to the conclusion that more complicated grain models are needed to fit both the observed SPF \textit{and} the disk spectrum. One promising type of grain model is agglomerated dust particles \citep{zubko_effect_2015, arnold_effect_2019} since porous grains can replicate strongly forward-scattering grains with a smaller size parameter (i.e. the ratio between the grain size and the wavelength) compared to simpler grain models such as compact spheres. Because of the efforts and conclusions outlined above, \hr{} has naturally been established as a laboratory for testing advanced grain models. 

The debris disk around \hr{} has been resolved across a very broad range of wavelengths including: visible \citep{schneider_stis_2009}, near-IR (\citealt{schneider_nicmos_1999};\citealt{thalmann_images_2011}; \citealt{wahhaj_gemini_2014}; \cite{milli_new_2015}; \citealt{milli_near-infrared_2017}; \citealt{chen_multiband_2020}; \citealt{arriaga_multiband_2020}), mid-IR (\citealt{koerner_mid-infrared_1998}; \citealt{moerchen_asymmetric_2011}; \citealt{lagrange_insight_2012}; \citealt{lisse_infrared_2017}), submillimeter \citep{sheret_submillimetre_2004}, and millimeter \citep{greaves_dust_2000}. One of the many advantages of directly imaging the disk includes the ability to measure the color of the dust. Furthermore, the photometry conducted on this disk in the visible to near-IR regime has revealed a red spectral slope. As a result, there has been speculation on whether the dust is red due to its composition \citep{debes_complex_2008} or due to the scattering behavior of the grains \citep{kohler_complex_2008}. The former point is interesting because complex organic molecules like tholins might explain the red color. Molecules of this type have been detected on numerous bodies in our own outer solar system \citep[e.g.,][]{stern_initial_2019, emery_tale_2024} and have been theorized to be linked to the origin of life \citep[e.g.,][]{khare_organic_1984}. In contrast, grain modeling efforts by \cite{mulders_why_2013} and \cite{tazaki_effect_2019} have shown that a dust grain's shape alone can present as red coloring in detected scattered light by a debris disk.

In this paper, we overview our observations and data reduction in Section \ref{sec:obs-title}. We then describe our disk and SPF modeling procedures in Section \ref{sec:model_description}. We provide a high-level overview of our freeform modeling routine to complement the full description of this novel forward modeling technique in a companion paper (Kueny et al. 2025, submitted) in Section \ref{sec:freeform-describe-sec}. We detail all our disk and SPF modeling results in Section \ref{sec:results-modeling}. We discuss our geometric disk and SPF modeling results in Section \ref{sec:discussion} and Section \ref{sec:spf-discussion-sec} respectively. Finally, we conclude with a summary and suggestions for future works in the last section of the paper.

\section{Observations and Data Reduction} \label{sec:obs-title}

\subsection{Observations}
\label{sec:obs}

\begin{table*}[!ht]
\centering
\caption{MagAO-X Observation Log of \hr{} and PSF Reference}
\begin{tabular}{lcclccccccc}
\hline
Date        & Time (UT)    & Seeing$^a$ & Object     & Band  & Air Mass & $t_{\text{exp}}$ (s) & $N_{\text{exp}}$ & $N_{\text{coadds}}$ & $\theta (^{\circ})$ &  \\ \hline
2023 Mar 10 & 05:49/07:36 & 0.6 & HR 4796A & $i'$  & 1.0/1.1  & 1                    & 5047    &     169         & -21.5/65              &            \\
2023 Mar 10 & 05:52/07:36 & 0.6 & HR 4796A & $z'$  & 1.0/1.1  & 1                 & 4827     &     165        & -18.5/65.0               &            \\
2023 Mar 13 & 04:46/07:15 & 0.6 & HR 4796A & $r'$  & 1.1/1.0  & 0.23                 & 35355   &   234               & -58.5/61.5             &             \\
2023 Mar 13 & 04:42/06:47 & 0.6 & HR 4796A & $g'$  & 1.1/1.0  & 0.23                 & 27319   &    205               & -60.0/47.0              &           \\
2024 Mar 29 & 05:22/05:31 & 0.5 & HR 4748 & $i'$  & 1.2/1.2  & 1                  & 1540      &    83            &  0.0/87.3              &         \\
2024 Mar 29 & 05:58/06:03 & 0.5 & HR 4748 & $z'$  & 1.1/1.1  & 1                  & 1548       &    83            & 31.4/75.6               &         \\
2024 Mar 29 & 06:44/06:52 & 0.5 & HR 4796A & $i'$  & 1.1/1.1  & 1                    & 2583    &    143              &  0.8/69.6              &           \\
2024 Mar 29 & 06:56/07:19 & 0.5 & HR 4796A & $z'$  & 1.1/1.1  & 1                    & 2683   &    141               & 8.0/69.6               &           \\
2025 Apr 08 & 03:29/05:47 & 0.6 & HR 4796A & $g'$  & 1.0/1.0  & 0.5                  & 15299    &    183              & -42.0/49.0               &         \\
\hline
            &              &            &       &          &                      &                     &                  &                   &                  
\end{tabular}

\footnotesize
\raggedright
\textbf{Notes.} Time: start/end observation; Air Mass: start/end air mass; $t_{\text{exp}}$: single frame exposure time; $N_{\text{exps}}$: number of raw science images; $N_{\text{coadds}}$: number of image cutouts used for PSF subtraction;  $\theta$: start/end parallactic angle. The total integration time for a given dataset is found by multiplying $t_{\text{exp}}$ and $N_{\text{exp}}$. \\
\vspace{0.5mm}
$^a$ median values in arcseconds.
\label{tab:obslog}
\end{table*}

We observed \hr{} using the 6.5 m Magellan-Clay telescope at Las Campanas Observatory (LCO) during three separate epochs\footnote{here we use ``epoch" to refer to all \hr{} observations made in a single semester}: 2023A, 2024A, and 2025A. For all epochs, we entered stationary pupil mode with MagAO-X to facilitate Angular Differential Imaging (ADI, \citealt{soummer_detection_2012}) and Reference Differential Imaging (RDI; \citealt{ruane_reference_2019}) using the star HR 4748 as a point-spread function (PSF) reference; see Tables \ref{tab:obslog} and \ref{tab:properties}. Additionally, we made use of the small Lyot coronagraph (focal plane mask diameter of $3 \lambda/D$) for all observations. The instrument has a plate scale of 0\farcsec0059 \citep{long_astrometric_2025} and a field-of-view of $6\farcsec0 \times 6\farcsec0$. After every coronagraphic science observation, as part of collecting our calibration data, we removed the focal plane mask and collected about 5 minutes of unsaturated images of the star for photometric calibration and convolution with our disk models for each passband.

MagAO-X has a a 97-actuator ``woofer" deformable mirror (DM), a 2k-actuator ``tweeter" DM, and a new 1k-actuator DM within the coronagraph that was installed after the observing run in 2023A. The internal mechanisms underneath the face sheet of the 2k- and 1k-actuator DMs cause them to have a prominent periodic structure on their reflective surfaces which add intensity patterns reminiscent of a diffraction grating in the image plane (see \citealt{males_magao-x_2022} for an example MagAO-X PSF with these artifacts). Furthermore, we can generate four additional artificial speckles (sparkles hereafter; \citealt{mcewen_-sky_2024}) by setting a sinusoidal pattern on the tweeter DM that are useful for astrometry and photometry as explained in the following paragraphs. The speckles and sparkles directly measure the Strehl ratio of the PSF. For all our observations, we generated sparkles at a distance of $15 \lambda/D$ from the star.

We conducted ADI observations during 3 epochs in 2023A and 2025A. For all but the observations in 2025A, we made use of MagAO-X's dual-imaging mode to observe the star through 2 filters simultaneously. To reject frames with poor image quality (due to short spikes in seeing), we analyzed the distribution of aperture photometry of the sparkles and speckles to exclude frames whose speckle photometry fell significantly below the ensemble average. This aperture photometry was performed in elliptical apertures.

For the 2023A epoch, our exclusions criteria resulted in excluding $10\%$ and $7.5\%$ of the frames for $z'$ and $i'$ datasets respectively. Despite good seeing, the turbulence was fast and caused a substantial wind-driven halo \citep[WDH;][]{cantalloube_wind-driven_2020} in the images on the nights of UT 2023 March 12. The \gband{} data collected on UT 2023 Mar 12 suffered heavily from the WDH, which was unfortunately aligned almost perfectly along the minor axis of the disk, and from a problem with the beam-splitter throughput that led to low signal-to-noise. We decided to exclude  our \gband{} data from this night from our analysis as modeling results were misleading. We rejected images in the same fashion as described in the previous paragraph and rejected 4\% of the frames for \rband.

For the 2024A epoch, the conditions were sufficiently stable for RDI observations using methods similar to those presented in \cite{wahhaj_search_2021}. We used $2 \times 2$ binning at both science cameras before readout to increase our sensitivity to diffuse extended features in the disk by reducing the impact of detector readout noise. We selected frames as before and excluded 3\% of the lowest quality $z'$-band images and 4\% of $i'$-band.

For the 2025A epoch we observed the star only through a $g' (527 $nm) filter, without the beamsplitter. We imaged the star using a single science camera with exposure times of 0.5 seconds. We excluded lower-quality images using the above-mentioned procedure leading to an exclusion rate of 8\%. Additionally, these \gband{} data also suffered heavily from the WDH so we went through each cutout (by eye) and excluded those that showed an exceptionally strong WDH. These additional exclusions encompassed the last $\sim 40$ minutes of data leading to a total usable integration time of $\sim 90$ minutes for this dataset. All of the relevant observational parameters for each of our observations are shown in Table \ref{tab:obslog}.

\subsection{Data reduction and PSF subtraction}
\label{sec:data_reduction}

\begin{figure}[!ht]
    \centering
    \includegraphics[width=0.5\linewidth]{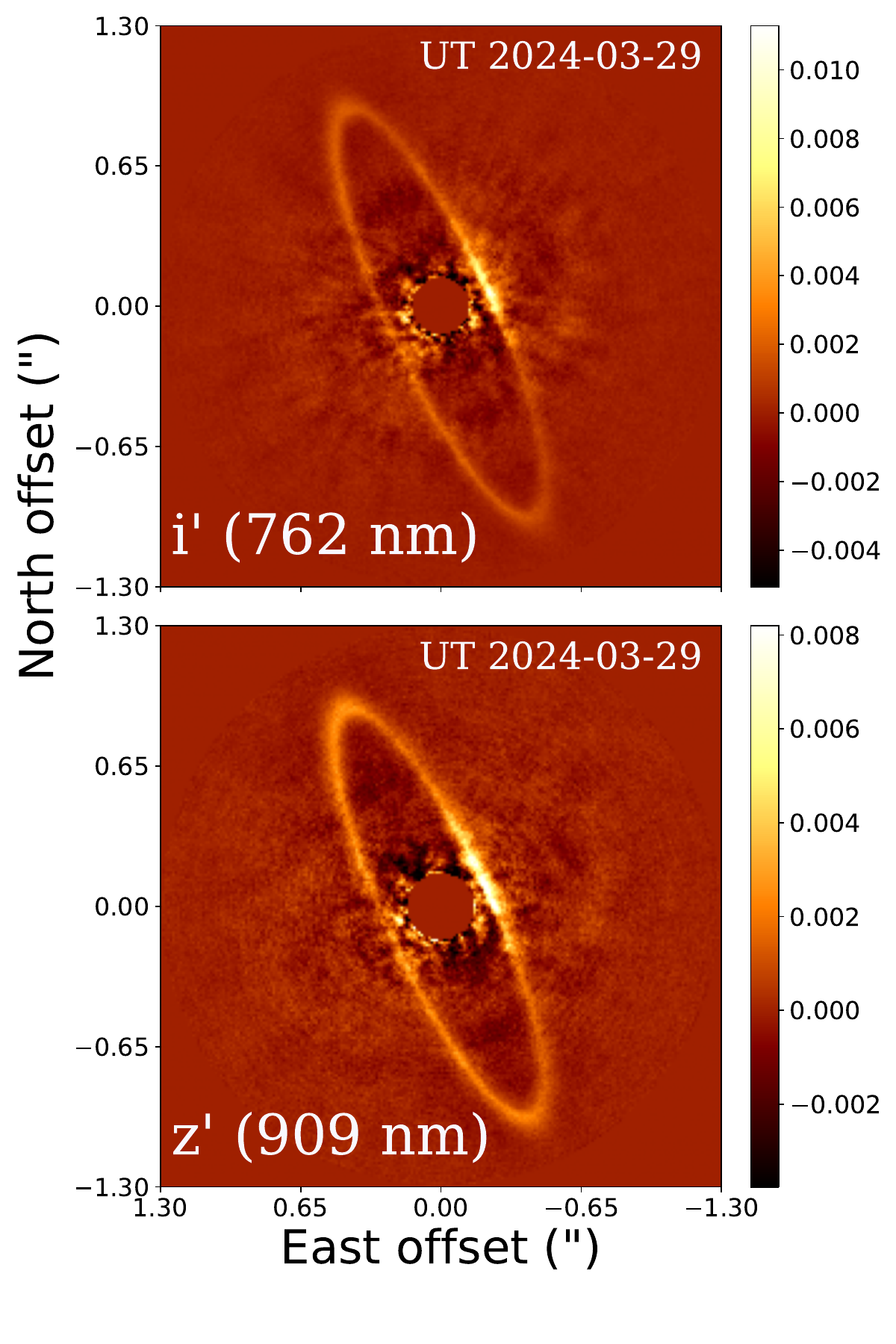}
    \caption{Final KLIP-RDI reductions using MagAO-X data at $i'$ and $z'$ showing the disk in a north-up, east-left orientation. The colorbar shows \sbf{} units. The central, high speckle noise region and field exterior to $1\farcsec34$ have been blocked with a software mask. The star is located at (0,0).}
    \label{fig:finalklip_rdi}
\end{figure}

\begin{figure*}[!ht]
    \centering
    \includegraphics[width=0.8\linewidth]{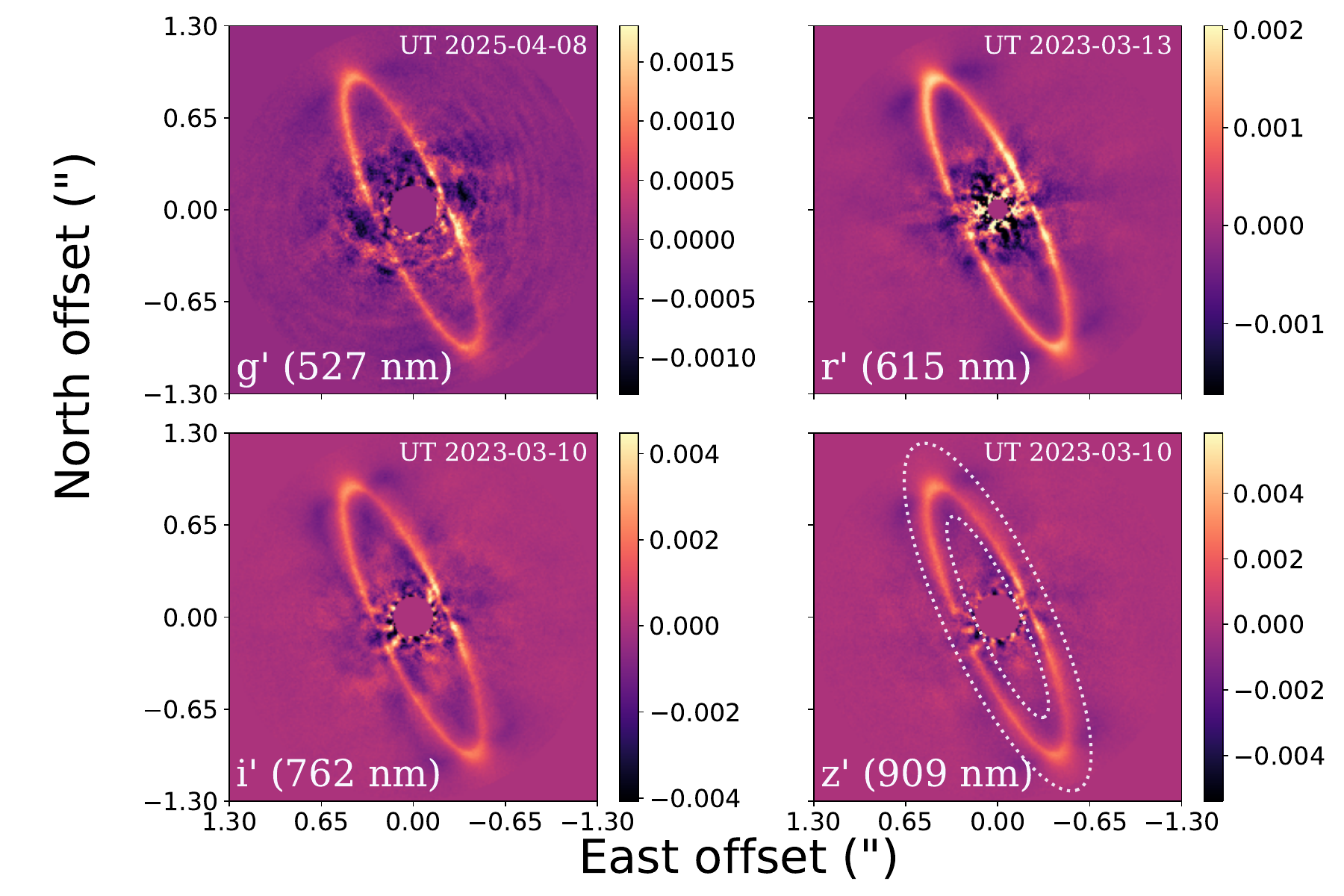}
    \caption{Final KLIP-ADI reductions using MagAO-X data showing the disk in a north-up, east-left orientation. We show the disk images at $g'r'i'z'$ in a left-to-right progression. The colorbar shows \sbf{} units. The two white dotted lines trace the elliptical region where the likelihood was maximized by the MCMC for all our disk images.}
    \label{fig:finalklip_adi}
\end{figure*}

All image processing steps described in this section apply to all image data included in this analysis. After excluding lower quality images as described in the above section, we prepared the raw images for PSF subtraction by subtracting a camera dark calibration frame and then normalizing the image pixel values by the exposure time. We note that we made use of smaller image cutouts to reduce data set size and computation times, taking care to retain all pixels that contain disk flux in our cutouts. For image registration and defining image cutouts, we computed the approximate location of the star behind the coronagraph using the pixel coordinates of the centroids of the sparkles and DM speckles in a median stacked image made using the entire dataset. We used this coordinate to extract $224 \times 224$ ($2\farcsec69$ square) image cutouts with the star centered in the image cutout with about one pixel precision. 

To prepare for PSF subtraction, we needed to precisely register all our images using a common reference with a phase cross-correlation routine \citep{guizar-sicairos_efficient_2008}. To make this registration reference, we median-stacked the entire set of (post-exclusions) cutouts. We assume that this stacked reference image encompasses enough integration time to sufficiently median out any small-scale jitter in the PSF. To further increase the effectiveness of this registration reference, we performed two pre-processing steps: (1) compute and subtract the median radial profile and apply an unsharp mask to remove the background and enhance the DM speckles, and (2) apply a bespoke DM speckle mask to zero everything in the image besides these sharpened DM speckles using elliptical apertures. These two steps were also repeated for each of our science image cutouts. We then cross-correlated these isolated, sharpened DM speckle patterns to compute the needed x- and y-shift to center the PSF. Finally, we applied these computed x- and y-shifts for every raw science image cutout to center them on the star. We performed a sanity check using our unsaturated PSF calibration images to confirm that averaging the x- and y-coordinates of the DM speckle centroids retrieves the coordinates of the core.

To reduce the computational burden during PSF subtraction and modeling procedures (see Section \ref{sec:disk_modeling}), we spatially downsampled our ADI images with $2\times2$ pixel binning and coadded (i.e., binning in-time) using parallactic angle bins. We coadded using angle bins of $0.5^{\circ}$ and $0.25^{\circ}$ for our scattered-light (SCL) parametric disk modeling and freeform disk modeling respectively. Using different metrics in coadding allows us to leverage the computational efficiency of \textsc{Jax} (see Section \ref{sec:freeform-describe-sec}); using a large dataset for our SCL disk modeling would be computationally infeasible. For our RDI images, we again downsampled using $2\times2$ pixel binning but coadded by integration time since sky rotation is not as critical as in the ADI case \citep{wahhaj_search_2021}. We used 20 and 10 second bins for our SCL disk and freeform disk modeling, respectively, for the same aforementioned reasons. Importantly, we chose these criteria for coadding to mitigate smearing of the disk due to the changing parallactic angle across each dataset. Because we were able to coadd less aggressively for our freeform modeling, our coadd criteria ensures the disk does not smear more than the width of half a pixel (5.9 mas) at the disk's outmost points (1060 mas from the star), fully satisfying the Nyquist-Shannon sampling theorem. For datasets used for our parametric disk modeling, the disk does not smear more than the width of a pixel (11.8 mas). This particular choice is not optimal, but we made this compromise since the minor axis of the disk is the focus of our modeling analysis.

\begin{figure*}[!ht]
    \centering
    \includegraphics[width=\linewidth]{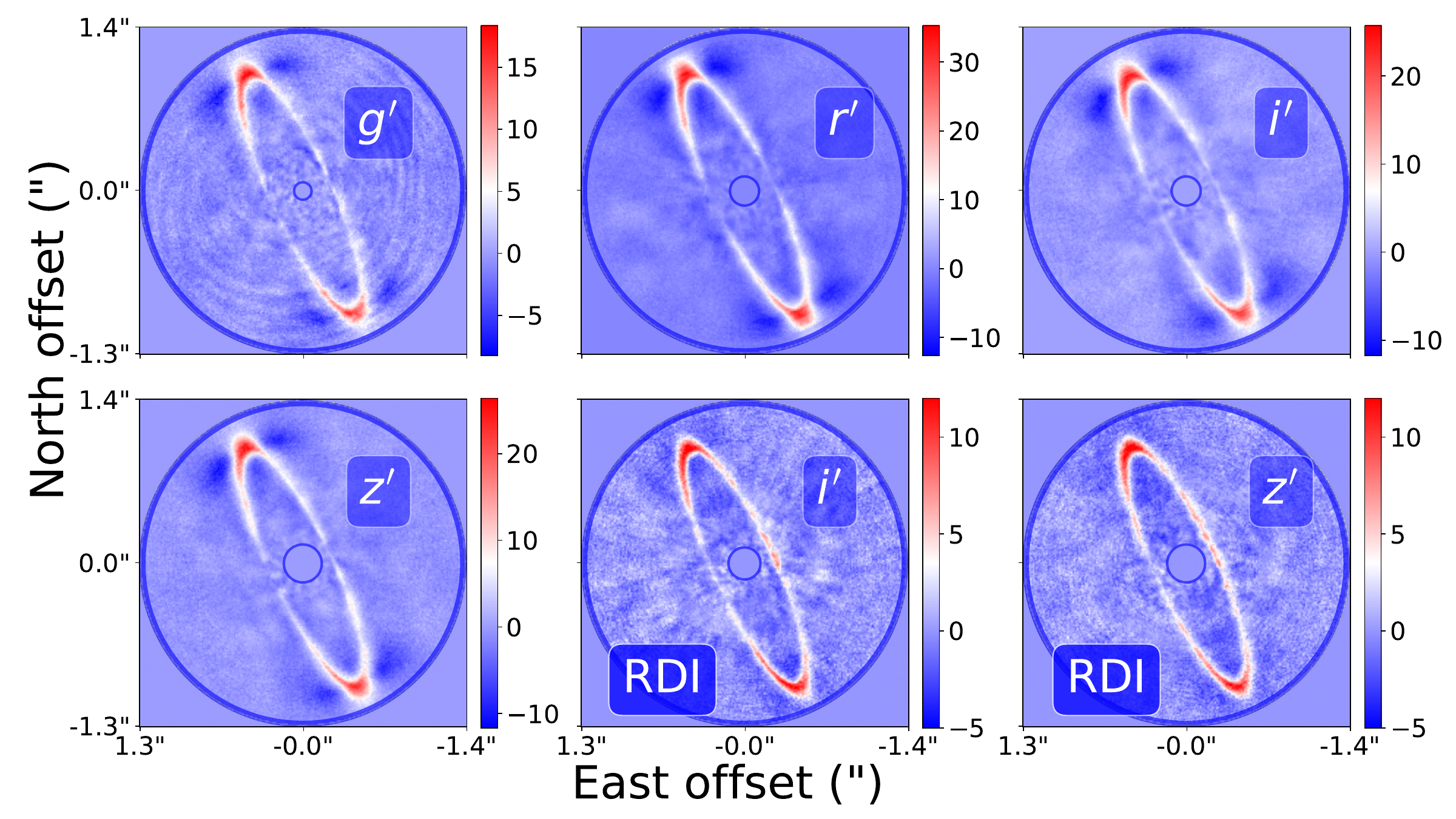}
    \caption{Signal-to-noise per pixel maps of our KLIP-reduced disk images at $g'$, $r'$, $i'$, and $z'$. The star and field outside of $1\farcsec34$ have been blocked with a software mask. The colorbar represents $S/N$.}
    \label{fig:snr_adi}
\end{figure*}

\begin{figure*}[!ht]
    \centering
    \includegraphics[width=\linewidth]{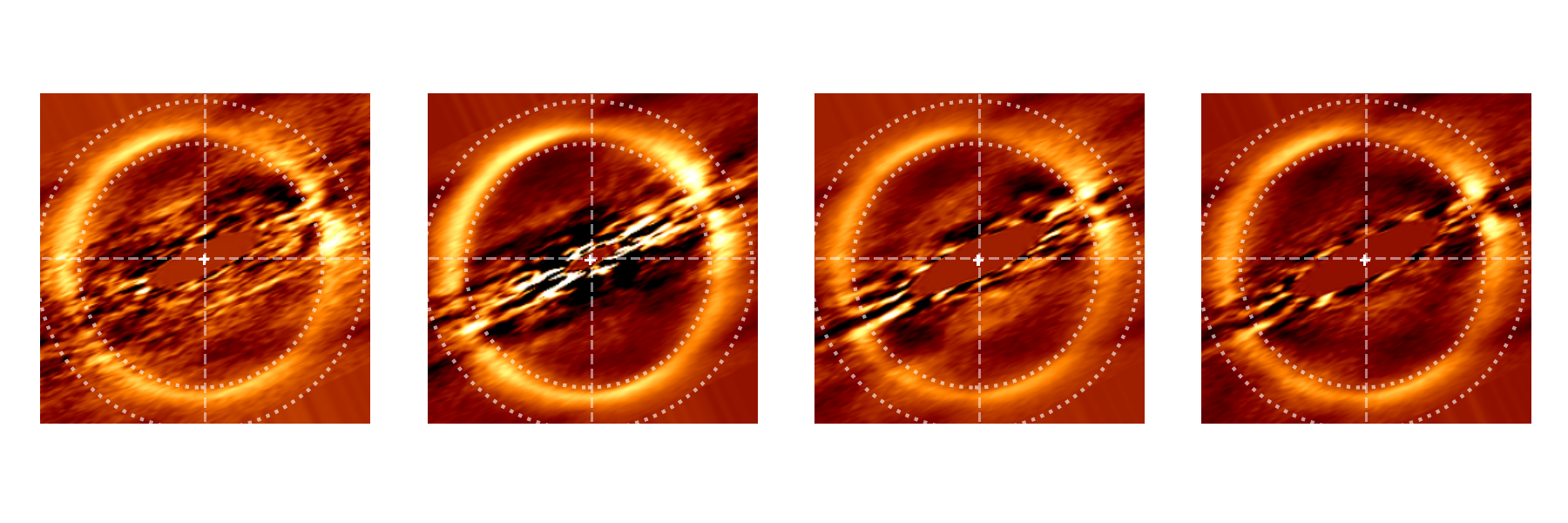} \\
    \includegraphics[width=\linewidth]{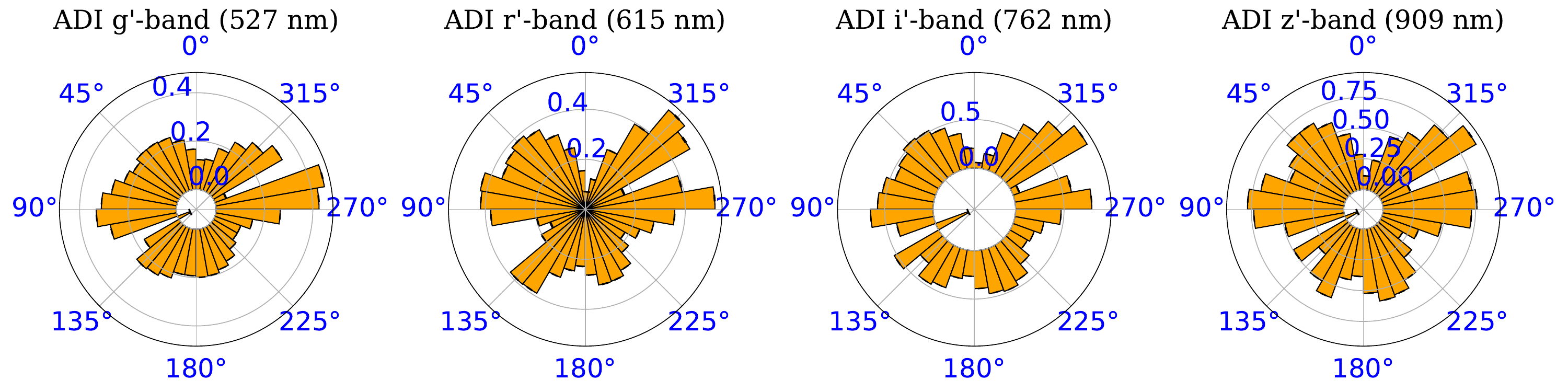} 
    \caption{\textit{Top row:} Deprojected \adi{} images of the disk at $g'r'i'z'$ as labeled. We retained the original disk position angle in our deprojected images for a better comparison to the images. The region where we quantify the azimuthal flux ranges between 74 to 91 au from the star. The dashed white lines intersect at the location of the star while the white cross marks the measured center of the disk. \textit{Bottom row:} Radar charts portraying the sum of pixel values of the deprojected KLIP-ADI disk images in Figure \ref{fig:finalklip_adi} across 36 annular wedges showing the azimuthal flux distribution. The radial plot markers denote sums in \sbf{} units and axially-inward bars denote negative sums.}
    \label{fig:wedge_phot_adi}
\end{figure*}

\begin{figure}[!ht]
    \centering
    \includegraphics[width=0.75\linewidth]{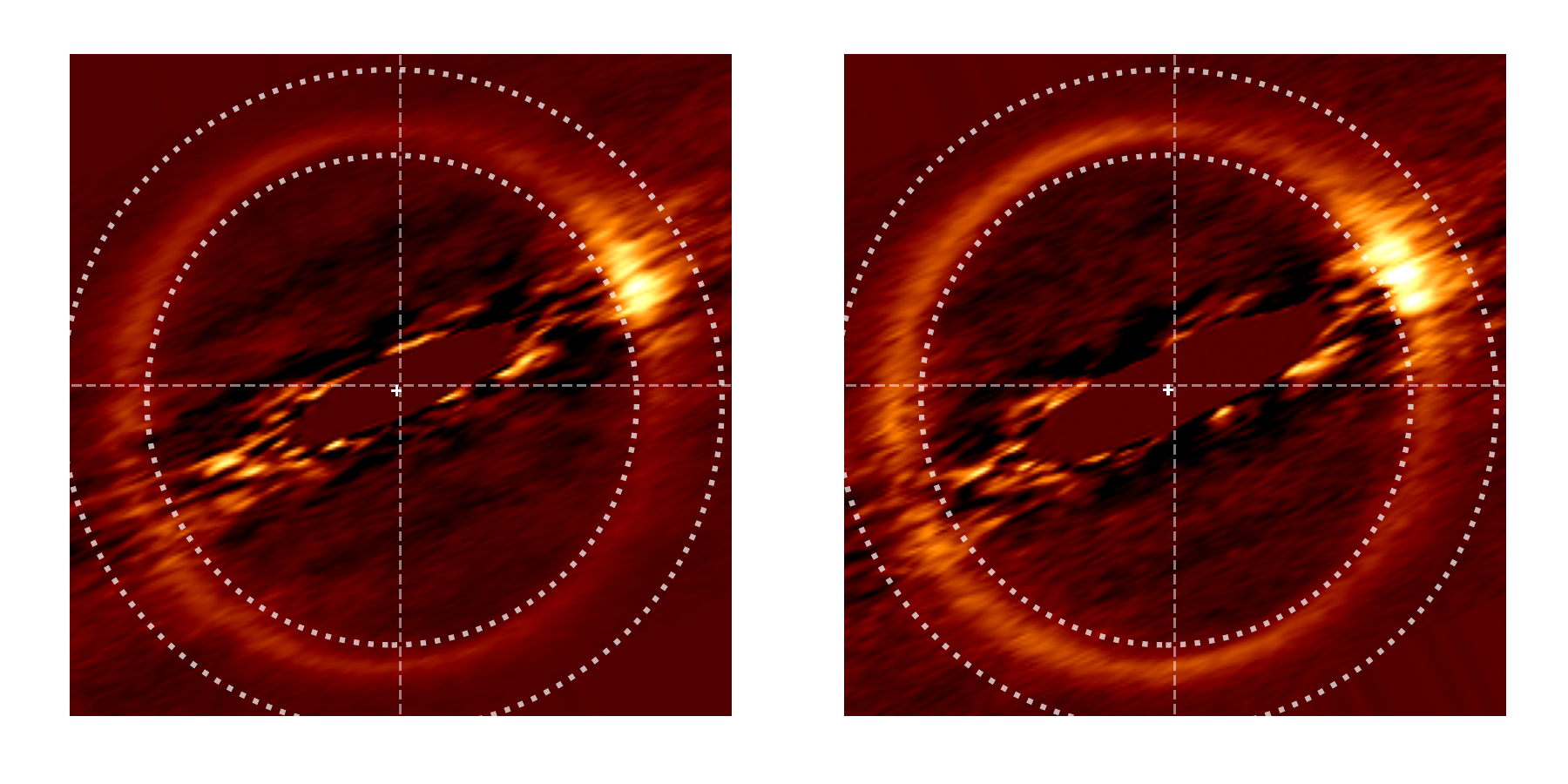} \\
    \includegraphics[width=0.75\linewidth]{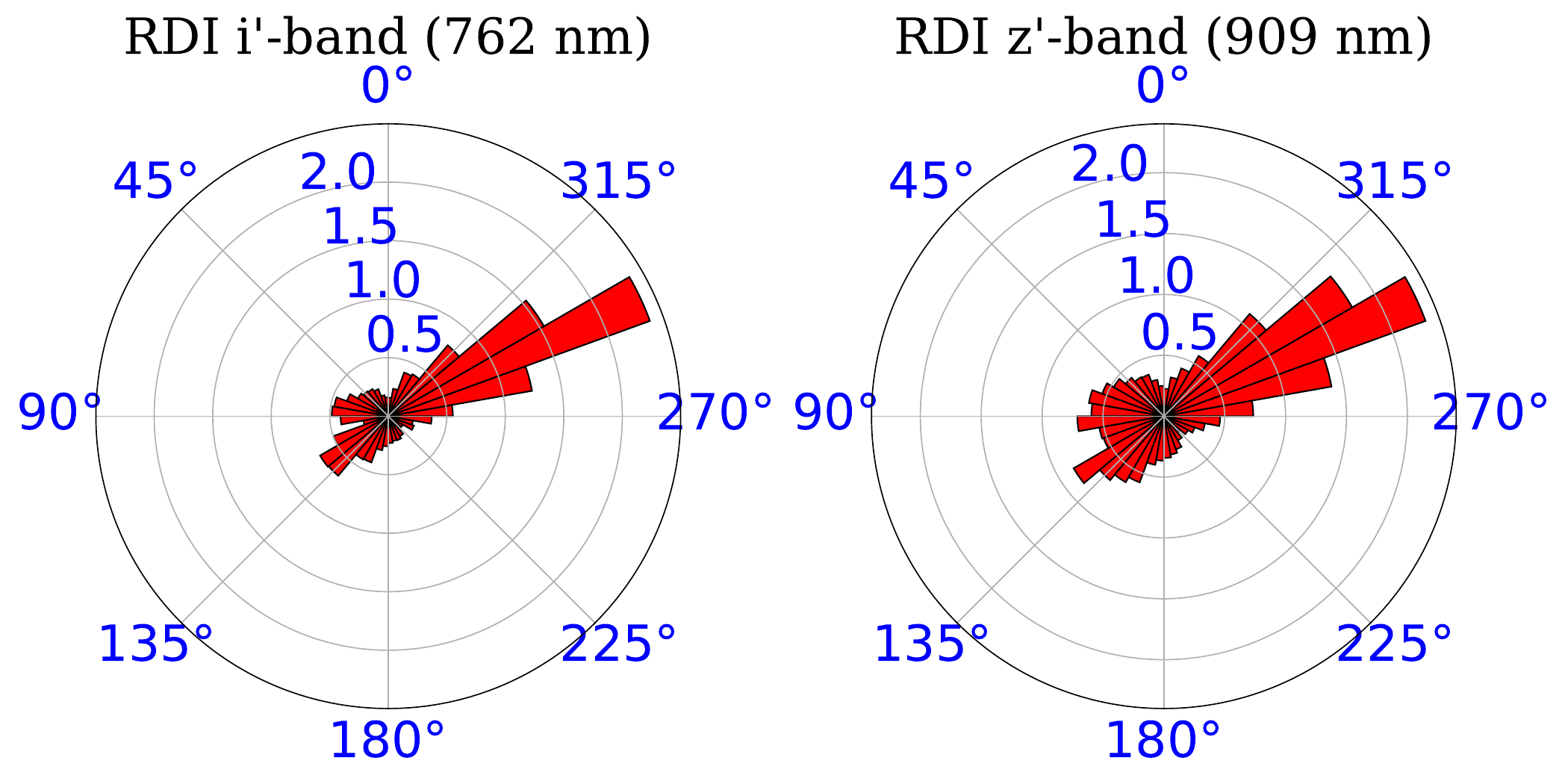} 
    \caption{The same as Figure \ref{fig:wedge_phot_adi} but for our \rdi{} images at \iband{} (left) and \zband{} (right).}
    \label{fig:wedge_phot_rdi}
\end{figure}

We made use of the Karhunen-Lo$\grave{\text{e}}$ve Image Projection (KLIP; \citealt{soummer_detection_2012}) algorithm to handle subtraction of the PSF, which is an application of principal component analysis assuming discrete-to-discrete operations and Gaussian statistics. KLIP can be coupled to the classical PSF subtraction techniques ADI and RDI for greater efficiency in removal of the speckle halo (\adi{} and \rdi{} hereafter, respectively). For the 3 epochs across 2023A and 2025A, we used \adi{} to subtract the PSF. For the night of UT 2024 March 28, the observing conditions were favorable and resulted in a very stable PSF, so we subtracted the PSF using \rdi{}. For implementing KLIP, we utilized the \texttt{pyklip} Python library \citep{wang_pyklip_2015} which offers a convenient way to tune the KLIP hyperparameters (i.e., the number of modes used, the parallactic angle movement constraint, and the size of the IWA). We also made use of \texttt{pyklip}'s built-in high-pass filtering which uses a Gaussian window in Fourier space to suppress low spatial frequencies. We used the star HR 4748 for our library of reference PSF images which we processed in the same fashion as our target PSF images. Our final \rdi{} and \adi{} images are showcased in Figure \ref{fig:finalklip_rdi} and Figure \ref{fig:finalklip_adi} respectively.

\begin{table}
\centering
\caption{Chosen KLIP parameters based on signal-to-noise Optimization.}
\label{tab:klip_params}
\begin{tabular}{llcccc}
\hline
     & Date Obs. (UT) & $N_{\text{modes}}$ & IWA (") & $\theta_{\text{min}} (^{\circ})$ & HP  \\ \hline
$g'$ & 2025-04-08     & 26/218             & 0.144   & 16                             & --- \\
$r'$ & 2023-03-12     & 12/234             & 0.060   & 7                              & --- \\
$i'$ & 2023-03-09  & 22/171             & 0.144   & 10                             & --- \\
$z'$ & 2023-03-09  & 26/164             & 0.156   & 11                             & --- \\
$i'$ & 2024-03-28  & 26/83              & 0.132   & ---                            & 20  \\
$z'$ & 2024-03-28  & 35/83              & 0.156   & ---                            & 22  \\ \hline
\end{tabular} \\
\footnotesize

\vspace{1mm}

\raggedright \textbf{Notes.} $N_{\text{modes}}$: number of KLIP modes; IWA: inner-working angle; $\theta_{\text{min}} (^{\circ})$: minimum rotation; HP; High-pass filter kernel size.
\end{table}

We optimized the signal-to-noise of the disk in the reduced images by performing a grid search over the \texttt{pyklip} hyperparameters, as described in \cite{kueny_probing_2024}. We optimized the number of KLIP modes, the size of the central software mask (IWA), the built-in \texttt{pyklip} minimum rotation parameter, and the size of the high-pass filter kernel. \texttt{pyklip} performs high-pass filtering by using a Gaussian window in Fourier space to suppress the lowest spatial frequencies according to the spread of the core in the kernel, $\sigma_{\xi} = \text{d} / (\text{HP}*2\sqrt{2\ln{2}})$, where d is the size of the image cutout (224 pixels for all our images) and HP is an integer; larger values of HP correspond to less aggressive high-pass filtering (see Table \ref{tab:klip_params}). Additionally, we tested if the reduced images benefitted from subtracting a median radial profile from each PSF image cutout and the only dataset that did not benefit from radial profile-subtracted PSF images was the \zband{} \rdi{} image.

To estimate the noise maps used for optimization and for our forward model fitting, we followed \cite{lawson_scexaocharis_2020} and computed the standard deviation at every radial location using concentric measurement annuli of 1 pixel width in each of our KLIP-reduced images. To improve our estimation, we excluded the disk from our noise calculation with a software mask. In short, we evaluated the signal-to-noise of the disk using the IWAs in Table \ref{tab:klip_params} and an outer-working angle of $0\farcsec72$ so as to: (1) focus the algorithm's efforts on removing the WDH which is confined to the AO control region only and (2) best preserve the disk's minor axis from algorithmic signal loss \citep{milli_impact_2012} especially in the case of the \adi{} images. See  our final optimized \rdi{} and \adi{} images in Figure \ref{fig:finalklip_rdi} and Figure \ref{fig:finalklip_adi}, respectively. To better illustrate algorithmic signal losses among our datasets, we include the azimuthal flux distribution in our final deprojected \adi{} and \rdi{}images in Figure \ref{fig:wedge_phot_adi} and Figure \ref{fig:wedge_phot_rdi}, respectively. The ansae of the disk are located in the seeing-limited region while the minor axis is within the AO control radius. If we optimized the KLIP parameters to achieve the highest signal-to-noise of the whole disk, the result was very bright ansae and oversubtraction at the minor axis. Table \ref{tab:klip_params} shows the optimal PSF subtraction parameters for each of our datasets. We present the signal-to-noise per pixel maps of all our KLIP-reduced images in Figure \ref{fig:snr_adi}.

\section{Disk modeling}
\label{sec:disk_modeling}

\subsection{Scattered light disk model}
\label{sec:model_description}

PSF subtraction using KLIP heavily distorts the signal from the disk (see Figure \ref{fig:wedge_phot_adi} and Figure \ref{fig:wedge_phot_rdi}). To best recover the true features of the disk, we performed disk forward modeling using our KLIP-reduced images. Our SCL disk modeling procedures are generally based on the workflow presented in \cite{kueny_probing_2024}. Briefly, we built a pipeline incorporating the \textsc{DiskFM} module \citep{mazoyer_diskfm_2020} within the \texttt{pyklip} package which forward models disk objects based on the theory presented in \cite{pueyo_detection_2016}. 

For generating our SCL disk models, we used the model description presented in \cite{ren_exo-kuiper_2019} which is a modified version of the model used by \cite{millar-blanchaer__2015}. In short, we fit the dust density profile $\rho(r)$ with a radius of maximal dust density $R_c$ and two radial power laws ($\alpha_{in}$ and $\alpha_{out}$ to describe the dust density profile that is interior and exterior to $R_c$:

\begin{equation}
    \rho(r) \propto \left[\left(\frac{r}{R_c}\right)^{-2\alpha_{\text{in}}} + \left(\frac{r}{R_c} \right)^{-2\alpha_{\text{out}}} \right]^{-\frac{1}{2}},
\end{equation}

\noindent where $r$ denotes the radius from the star in au. The dust distribution in the vertical direction $Z(r,z)$ (perpendicular to the disk midplane) is modeled with a Gaussian profile,

\begin{equation}
    Z(r,z) \propto \exp{\left[-\left(\frac{z}{h_0r^{\beta}} \right)^2 \right]},
\end{equation}

\noindent where $z$ is the vertical distance, $h_0$ is the aspect ratio, and the exponent $\beta$ denotes the flaring parameter. For this analysis, we assumed no flaring is occurring, so $\beta = 1$ for all models we tested.

The disk model is generated using a stellocentric coordinate system $(\hat{x}, \hat{y}, \hat{z})$ where the free parameter $\theta_{PA}$ applies a coordinate transformation to adjust the position angle of the model with respect to the sky:

\begin{equation}
    \begin{split}
        x_{PA} = \hat{y}\sin\theta_{PA} - \hat{x}\cos{\theta_{PA}}, \\ y_{PA} = \hat{y}\cos\theta_{PA} + \hat{x}\sin\theta_{PA},
    \end{split} 
\end{equation}

\noindent where positive values of $\theta_{PA}$ translate to a rotation east of north. Another free parameter $\phi_{inc}$ applies a tilt to the coordinate system with respect to the observer (i.e., the inclination). Finally, free parameters $dx$ and $dy$ apply offsets to the disk model's center coordinate with respect to the star and we show the final form of the model coordinate system:

\begin{equation}
    \begin{split}
    x = x_{PA}\sin{\phi_{inc}} - dx \\ y = y_{PA} - dy, \\ z = x_{PA}\cos{\phi_{inc}} - \hat{z}\sin{\phi_{inc}}.
    \end{split}
\end{equation}

We generated the intensity $I$ for each pixel $(x', y')$ in the disk model using a brightness integral along the line of sight $z'$ (primed coordinates are with respect to the detector frame; not to be confused with \zband). In order to integrate the intensity along the line of sight numerically, we take R2 as the outer disk radius where I(x’,y’) is 0 which is much larger than the vertical extent of the disk:

\begin{equation}
    I(x',y') = I_0 + \int_{z' = -R_2}^{R_2} dz' \frac{N_0}{r^2}  \rho(r) Z(r,z) P(\theta),
\end{equation}

\noindent where $P(\theta)$ is the SPF, $N_0$ the flux normalization factor, and $I_0$ is a constant offset. 

The scattering angle $\theta(x', y', z')$ is a function of the position of the image. We note that we fixed the constant offset to $I_0 = 0$ because preliminary modeling trials showed that the KLIP algorithm was sufficient for removing the background. We also fixed $R_2$ to be 110 au based on the modeling results in \cite{chen_multiband_2020}.

\subsection{DiskFM + MCMC for parameter estimation}
\label{sec:mcmc_description}

To place strict limits on the morphology that best fits our reduced disk images, we used the Markov-Chain Monte Carlo (MCMC) sampler \texttt{emcee} package \citep{foreman-mackey_emcee_2013}. The range of values probed for all fitted parameters are shown in Table \ref{tab:diskfm-results} with the exception of the model flux scaling parameter $N_0$ and coefficient values for our basis of Legendre polynomials.  We assumed uniform prior probability distributions for all fitted parameters. Our SCL disk models are forward modeled using DiskFM \citep{mazoyer_diskfm_2020} to estimate the signal distortion and throughput loss caused by KLIP. An overview of the full forward modeling procedure is as follows:

\begin{enumerate}
    \item Generate a disk model using the method outlined in \ref{sec:model_description} and \ref{sec:spf_description}.
    \item Convolve the disk model with the empirically-measured, unsaturated instrument PSF (see Sec. \ref{sec:obs}) to simulate a model image.
    \begin{enumerate}
        \item If median radial profile subtraction was used as part of the data reduction, compute and subtract the median radial profile from this model image.
        \item If high-pass filtering was used as part of the data reduction, high-pass filter this model image.
    \end{enumerate}
    \item Use the \texttt{pyKLIP} DiskFM pipeline to produce a forward model (FM) using the same KLIP parameters as the KLIP-reduced image.
    \item Gauge the goodness of fit of these results using the standard $\chi^2$ metric:
    \begin{equation}
    \label{eq:chisquare}
        \chi^2 = \sum_S \frac{(\text{Data} - \text{FM})^2}{\text{Uncertainty}^2},
    \end{equation}
\end{enumerate}

\noindent where $\mathcal{S}$ is the search region where the likelihood $\mathcal{L} = e^{-\chi^2/2}$ is maximized. We generate the disk model and evaluate the likelihood for the MCMC in an elliptical region that generously captures the entire disk; see Figure \ref{fig:finalklip_adi} for visualization of this zone in the $z' (909)$ nm image marked with the dashed white ellipses. How we compute the estimate for the uncertainty is described in Section \ref{sec:data_reduction}.

\subsection{Fitting a custom SPF}
\label{sec:spf_description}

One of the core objectives of our analysis is studying the SPF of the disk. Since our \rdi{} images clearly display the strong forward-scattering at the disk's minor axis, we used these images to fit a custom SPF constructed from the first few Legendre polynomials. This method is one way to empirically measure the SPF since we fit all of the disk geometrical parameters in tandem during the forward modeling procedure. Importantly, we emphasize that these measured SPFs can be leveraged in future studies to rigorously test grain size distribution, composition, and other grain properties without the need for expensive disk forward modeling efforts. 

One drawback of the ADI observing strategy is the phenomenon of disk self-subtraction \citep{milli_impact_2012}. Due to the geometry of the \hr{} disk, the minor axis is almost entirely self-subtracted and the forward-scattering peak is not well-realized (compare Figure \ref{fig:finalklip_adi} and Figure \ref{fig:finalklip_rdi}). Indeed, preliminary modeling trials using our custom Legendre polynomial-based SPF did not produce good results because including enough Legendre basis components to make a well-fitting SPF to our data resulted in overfitting the self-subtraction artifacts at the minor axis of the disk. We note that while other data reduction strategies may have potential to reduce the self-subtraction associated with ADI (e.g., a masked ADI approach as in \citealt{milli_near-infrared_2017} or library of archival PSF images as demonstrated in \citealt{xie_reference-star_2022}), we leave exploration of these additional strategies for future work. As such, we elected to use the two-parameter Henyey-Greenstein \citep[HG;][]{henyey_diffuse_1941} SPF for fitting SCL models to our \adi{} images. The HG SPF is a convenient mathematical function that approximates the forward- and back-scattering behavior seen in cosmic dust with only three free parameters: $g_1$ and $g_2$ are the forward- and back-scattering asymmetry parameters respectively and $\alpha_1$ is a weighting factor. $g_1, g_2,$ and $\alpha_1$ are constrained to values between zero and one.

For both the HG and custom SPF cases, we normalized the SPF at the ansae (i.e., the $90^{\circ}$ scattering angle) before generating a given disk model.

To best estimate the azimuthal brightness distribution of the disk in our \rdi{} images using a custom SPF, we implemented a careful workflow and regularization to best fit our data while minimizing over- or under-fitting. As an initial guess for the MCMC, we used the best-fit HG SPFs from the $i'$-band and $z'$-band KLIP-ADI images because we observed decent results with those SPFs when modeling our KLIP-ADI images (see Section \ref{sec:results-modeling}). Additionally, we fixed the zeroth Legendre polynomial (the constant offset) to the mean of the best-fit HG solution because we found that this free parameter exhibited unbounded growth due to the SPF normalization at 90 degrees. We further regularized the MCMC with a penalty factor $p(\boldsymbol{a})$ applied to the vector of coefficients $\boldsymbol{a}$,

\begin{equation}
    p(\boldsymbol{a}) = -\lambda \sum_{k=1}^K a_k^2
\end{equation}

\noindent where $K$ is the number of Legendre polynomials used in the basis, $a_k$ is the coefficient of polynomial $k$, and $\lambda$ is a small scale factor. During our modeling, we found that a value of $\lambda = 0.1$ was sufficient for preventing strong correlations in the posterior distributions of the coefficients while still permitting sufficient walker movement in the MCMC. We then combined $p(\boldsymbol{a})$ with Equation (\ref{eq:chisquare}) to obtain the complete likelihood function:

\begin{equation}
\ln \mathcal{L}(\boldsymbol{\theta | D, a}) = 
- \frac{1}{2} \sum_{ij} \left( \frac{D_{ij} - M_{ij}(\boldsymbol{\Theta, a})}{\sigma_{ij}} \right)^2
+ p(\boldsymbol{a})
\end{equation}

\noindent where $\boldsymbol{\Theta}$ is a vector of our free geometrical disk parameters, $\boldsymbol{a}$ is a vector of the Legendre polynomial coefficients used in the SPF basis, $D_{i,j}$ is the KLIP-reduced image data at pixel $(i,j)$, $M_{ij}$ is the forward model at that pixel, and $\sigma_{ij}$ is the noise at that pixel. The first term is computed for all pixels in the search region $(i,j) \in \mathcal{S}$.
We chose to use the first 16 and 18 of the Legendre polynomials for fitting disk models to our \rdi{} \iband{} and \zband{} images, respectively. We used more Legendre polynomials for the \zband{} \rdi{} image because the signal-to-noise at the minor axis is higher compared to \iband{}. As a check, we confirmed that using these bases to reconstruct the HG SPF solution from the KLIP-ADI $i'$- and $z'$-band images resulted in a RMS error of 0.01 for both passbands, so these bases should contain a sufficient number of degrees of freedom to best fit the disk (we assumed the HG SPF is already close to the optimum fit, see Section \ref{sec:results-spfs}).

\subsection{Photometric Calibration}
\label{sec:phot-calib}
We used our unsaturated PSF images taken at the end of each science observation for photometric calibration. To maintain a consistent PSF, we left the  Lyot stop in the beam and only changed the EMgain and exposure time. We reduced the images by subtracting a camera dark calibration frame and then dividing by the exposure time to convert the pixel values to counts per second. We then averaged all processed unsaturated PSF images. We computed the sum of the counts in the largest circular aperture possible such that the aperture boundary did not go off the edge of the chip (this differed between datasets as the star was not perfectly aligned to the chip center in all cases). We inspected curves of growth (i.e., summed counts as a function of aperture radius) for each stacked unsaturated PSF image and confirmed that the curve assumed an asymptote indicating that the background was properly removed. To convert our final reduced images and disk models from counts/s to Jy/arcsec$^2$, we used the ratio of the expected flux density of the star (in Jy units) to the counts/s detected in our unsaturated PSF image of \hr{}.

To calculate the expected flux density from the star in each of our filters, we downloaded the spectral energy distribution of Alpha Lyrae from CALSPEC and normalized it to catalog photometry of \hr{} in Johnson $B$ (GSC-II; \citealt{lasker_second-generation_2008}), Gaia $G$, Gaia $Gbp$, Gaia $Grp$ (Gaia DR3; \citealt{gaia_collaboration_gaia_2023}), and $J$ (2MASS; \citealt{skrutskie_two_2006}). We made use of publicly-available bandpasses for $g', r', i',$ and $z'$ available on the MagAO-X instrument handbook\footnote{\url{https://magao-x.org/docs/handbook/observers/filters.html}}. These bandpasses already have a Las Campanas Observatory atmosphere model\footnote{Calculated by J. Males using the BTRAM code.} applied to them with an assumed distance from zenith and precipitable water vapor; these defaults were reasonable assumptions for our observations. We then multiplied the shifted spectrum by the filter transmission profile and integrated this result to get the expected flux density.

\section{Fitting a freeform disk model}
\label{sec:freeform-describe-sec}

Our \adi{} images suffer from substantial self-subtraction artifacts at the minor axis. During preliminary model fitting to these ADI data we found it very difficult to fit a custom Legendre-based SPF because the model would start to fit the self-subtraction artifacts before producing residuals that were qualitatively acceptable. In an effort to measure SPFs from our ADI images, we constructed the freeform disk forward modeling pipeline \texttt{ffortissimo} (Kueny et al. 2026, accepted for publication in the Astronomical Journal). This new pipeline is based on \textsc{Jax} \citep{bradbury_jax_2018}, which is a Python library that enables accelerated array computing via the use of graphics processing units (GPUs) and gradient descent optimization through automatic differentiation. We were able to optimize freeform models using datasets twice as large as the datasets used in the procedures described in the previous section. Namely, we binned our images in time using parallactic angle bins of $0.25^{\circ}$ instead of $0.5^{\circ}$ (see Section \ref{sec:data_reduction}) which lead to fully Nyquist sampled final KLIP images with higher signal-to-noise. This was only possible due to gains in computational efficiency using our JAX-based pipeline; we found that using datasets this large with the nominal DiskFM pipeline is infeasible even when parallelized on a large computing cluster. The freeform modeled datasets consisted of 332 images for \zband{}, 341 images for \iband{}, 465 images for \rband{}, and 435 images for \gband{}. A detailed description of this freeform modeling pipeline is in Kueny et al. 2025 (submitted).

\subsection{Freeform model description}
\label{sec:freeform-describe-subsec}

Our aforementioned fitting of SCL disk models (and traditional disk model fitting in general) assumes a fixed parametric form for the morphology, often leading to poor fits to high-contrast images. To overcome this as well as improve upon our model residuals, we treat individual pixels in a region of interest (see the region bounded by the dotted white line in Figure \ref{fig:finalklip_adi}) as optimizable parameters to fit the disk intensity distribution in our reduced images. With these new methods, the number of free parameters then goes up by many orders of magnitude. However, by encompassing the DiskFM pipeline with a \textsc{Jax} framework, we leverage GPU-accelerated operations and automatic differentiation to gain the efficiency needed to converge to a model within 2 to 3 hours for \rband{} and \gband{} (our largest datasets) depending on regularization parameters and a tunable learning rate.

The workflow for a single iteration is essentially identical to the DiskFM pipeline description given in Section \ref{sec:disk_modeling} in that a disk model is (1) generated, (2) convolved with the instrument PSF, (3) forward-modeled, and then (4) compared to the KLIP-reduced image within a loss function. As part of the startup procedure for our modeling pipeline, the``first guess" disk model is initialized as an image of uniform noise (i.e., $U\sim(0,1)$). During the main training loop, the optimizer iterates until the loss values reach an asymptote. The main differences between this modeling pipeline and that of the pipeline described in the previous section are the regularization methods needed to keep the freeform disk model from overfitting the data.

\begin{figure*}[!ht]
    \centering
    \includegraphics[width=\linewidth]{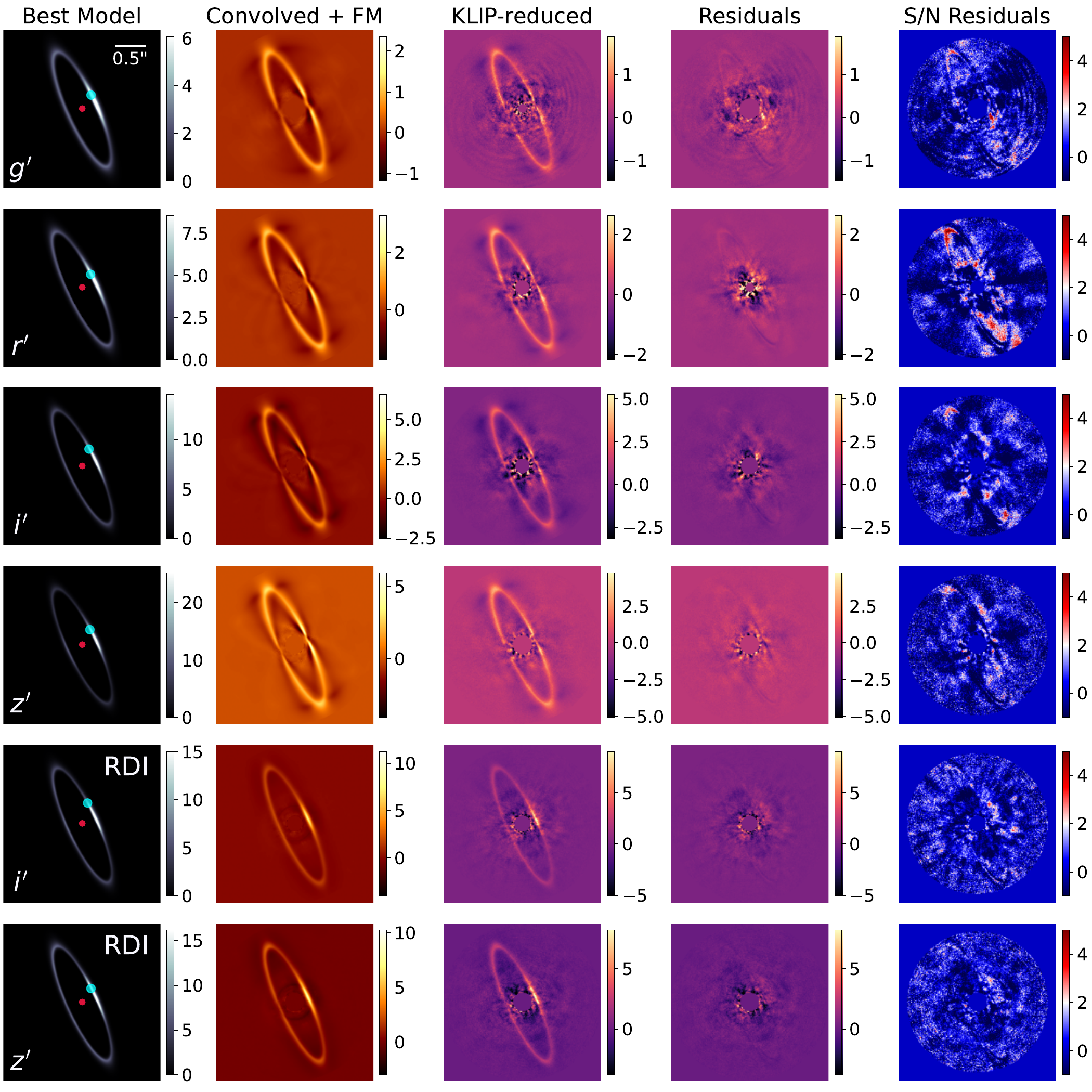}
    \caption{Forward modeling results for all our images at $g'$, $r'$, $i'$, and $z'$. The columns show, in left to right progression, our best fitting disk model, the disk model convolved with the instrument PSF and then forward modeled, our KLIP-reduced image, and the residuals between the best-fit forward model and the KLIP-reduced image for each of the passbands. The colorbars in the first 4 columns show surface brightness $SB_{disk}/F_{star}$ (arcsec$^{-2})$ and the colorbars in the last column represent $S/N$. The red and cyan circles denote the location of the star and argument of pericenter, respecfully. Larger versions of the KLIP-reduced images are found in Figure \ref{fig:finalklip_adi} and Figure \ref{fig:finalklip_rdi}.}
    \label{fig:diskfm_all}
\end{figure*}

\begin{table*}[!ht]
\centering
\caption{DiskFM Modeling Results for \hr{}}
\hspace*{-2cm}\begin{tabular}{lccccccccc}
\hline \hline
\multirow{2}{*}{Parameter$^\text{a}$} &  & \multirow{2}{*}{Range$^\text{b}$} &  & \multicolumn{6}{c}{Best Fit Values$^\text{c}$}                                                                    \\ \cline{5-10} 
                    &  &                 &  & $g'$ (ADI)                            & $r'$ (ADI)                         & $i'$ (ADI)                        &  $z'$ (ADI)                         &  $i'$ (RDI) & $z'$ (RDI)  \\ \hline
$R_c$ (au) &  & (70, 120)        &       & $76.57 \pm 0.08$                         & $77.06 \pm 0.07$                   & $76.83 \pm 0.10$                  & $76.95 \pm 0.32$                    & $76.69 \pm 0.07$ & $76.70 \pm 0.08$ \\
$\alpha_{in}$       &  & (1, 100)        &  & $34.96 \substack{+1.57 \\ -1.48}$     & $47.97 \substack{+2.48 \\ -2.23}$  & $45.86 \substack{+2.96 \\ -2.60}$ & $45.31 \substack{+10.44 \\ -6.91}$  & $46.00 \pm 2.10$ & $46.62 \pm 2.22$  \\
$-\alpha_{out}$      &  & (-1, -60)       &  & $12.37 \pm 0.16$                     & $12.70 \pm 0.11$                   & $15.00 \pm 0.20$                 & $14.23 \pm 0.54$                   & $14.20 \pm 0.20$ & $15.21 \pm 0.24$ \\
$h_0$ (\%)          &  & (0.01, 10)      &  & $1.69 \pm 0.09$                       & $1.50 \pm 0.05$                    & $1.05 \pm 0.06$                   & $1.00 \substack{+0.16 \\ -0.18}$ &  ---  & ---               \\
$\phi_{inc} (^{\circ})$      &  & (70, 85)        &  & $76.41 \pm 0.03$                      & $76.75 \pm 0.02$                 & $76.62 \pm 0.02$                  & $76.62 \pm 0.05$                  & $76.49 \pm 0.02$ & $76.38 \pm 0.02$   \\
$\theta_{PA} (^{\circ})$     &  & (20, 30)        &  & $26.64 \pm 0.16$                          & $26.91 \pm 0.12$               & $26.78 \pm 0.12$                  & $26.68 \pm 0.13$                    & $26.32 \pm 0.15$ & $26.15 \pm 0.15$  \\
$dx$ (au)           &  & (-15, 15)       &  & $-1.56 \pm 0.07$                      & $-2.14 \pm 0.05$                   & $-1.89 \pm 0.07$                  & $-1.98 \pm 0.20$                    & $-1.61 \pm 0.06$ & $-2.66 \pm 0.02$  \\
$dy$ (au)           &  & (-10, 10)       &  & $0.99 \pm 0.03$                       & $1.31 \pm 0.02 $                   & $1.75 \pm 0.04$                   & $1.62 \pm 0.10$                     & $2.02 \pm 0.03$ & $1.70 \pm 0.03$   \\
$g1 (\%)$                &  & (0.1, 99.9)  &  & $79.64 \pm 1.04$                    & $81.70 \pm 0.89$                   & ($90.45 \pm 1.72$)                & ($95.65 \substack{+3.73 \\ -5.31}$)     & ---             & --- \\
$g2 (\%)$                &  & (0.1, 99.9)  &  & $-20.26 \pm 0.49$                    & $-20.15 \pm 0.32$                 & ($-14.52 \pm 0.35$)               & ($-14.27 \pm 0.96)$                     & ---             & ---  \\
$\alpha1 (\%)$           &  & (0.1, 99.9)  &  & $28.72 \pm 0.79$                    & $29.60 \pm \substack{+0.77 \\ -0.69}$                   & ($36.51 \pm 4.3$)                 & ($52.34 \substack{+36.27 \\ -18.73}$)   & --- & ---\\ \hline
$F$ (mJy)$^\text{d}$&  &                 &  & $25.84 \pm 0.13$                      & $18.62 \pm 0.18$                   & $17.66 \pm 0.15$                  & $16.07 \pm 0.12$        & $18.38 \pm 0.18$   & $18.69 \pm 0.14$ \\ \hline
\end{tabular}

\footnotesize
\raggedright
\vspace{1mm}
\textbf{Notes.} $i'$ (ADI) and $z'$ (ADI) include results from fitting disk models with our custom \iband{} and \zband{} SPFs, but we include the HG SPF results from separate modeling runs in parentheses for comparison to the other HG solutions. We applied the rotation angle offset from Appendix \ref{sec:app_astrometry} to $\theta_{PA}$ for $g'$ (ADI) and from \citet{long_astrometric_2025} for all other datasets. \\
\vspace{1mm}
$^{\text{a}}$ We included parameter $N_0$ in our modeling to handle the flux normalization, but it is not shown in this table. \\
$^{\text{b}}$ We assumed uniform prior distributions for all parameters. \\
$^\text{c}$ 16th, 50th, and 84th percentiles. \\
$^\text{d}$ We include the total flux of the modeled disk in mJy units which we computed by summing the best-fit model; see text. \\
\label{tab:diskfm-results}

\end{table*}

\begin{figure*}[!ht]
    \centering
    \includegraphics[width=\linewidth]{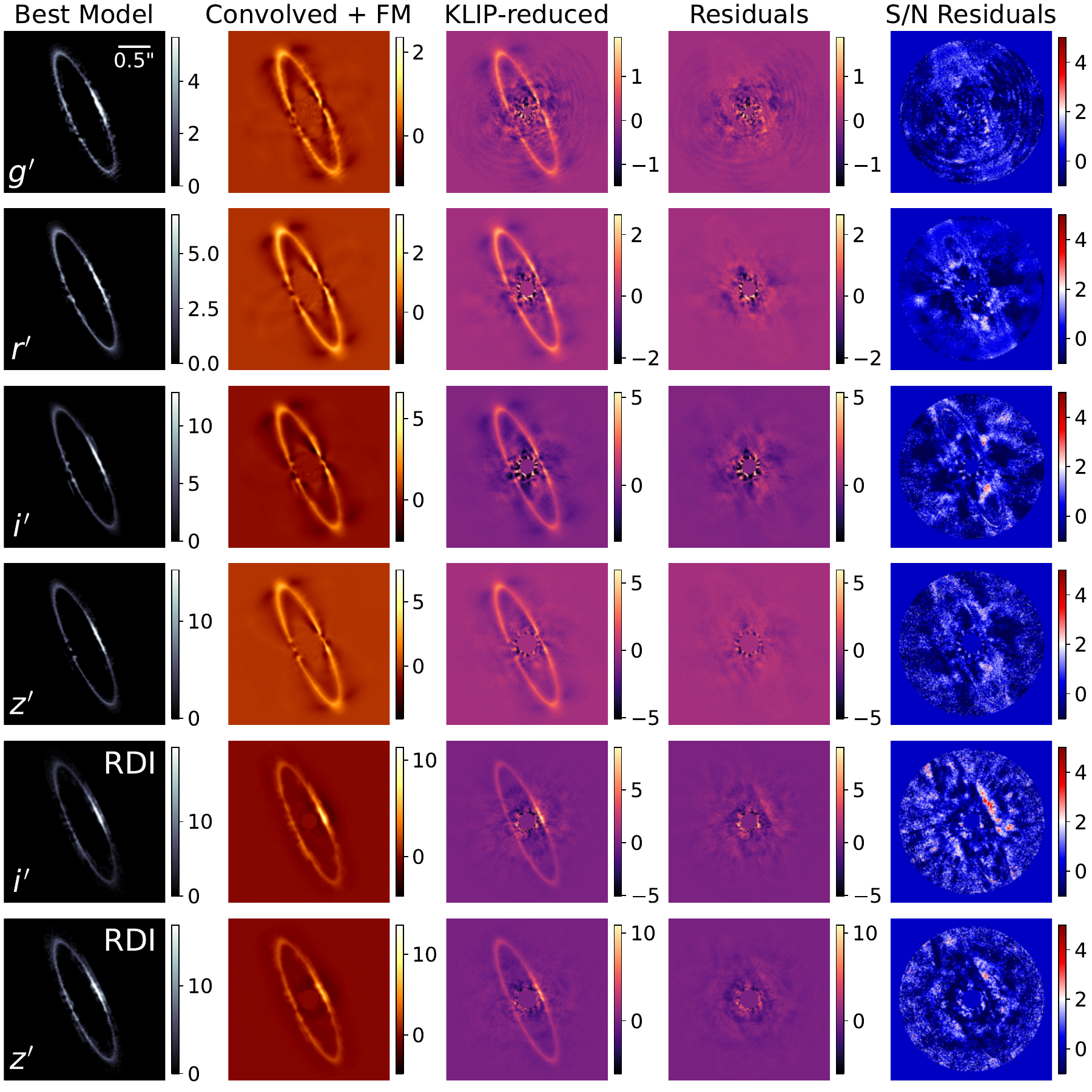}
    \caption{Similar to Figure \ref{fig:diskfm_all}, but featuring our freeform disk modeling results.}
    \label{fig:freeform_models_all}
\end{figure*}

\begin{table}
\centering
\caption{Selected DiskFM Results from \cite{chen_multiband_2020}.}
\begin{tabular}{ccc}
\hline
  Parameter      & $J$ (GPI)              & $H2$ (SPHERE) \\ \hline
$R1$ (au)        & $74.74 \pm 0.37$  & $74.31 \pm 0.10$     \\
$R2$ (au)        & $98.69 \pm 1.67$  &  $98.99 \pm 0.20$ \\
-$\alpha_{out}$   & $12.13 \pm 0.96$  & $14.08 \pm 0.30$          \\
$inc (^{\circ})$ & $76.93 \pm 0.13$  & $76.82 \pm 0.07$     \\
$PA (^{\circ})$  & $26.83 \pm 0.11$  & $26.94 \pm 0.04$         \\
$dx$ (au)        & $-0.78 \pm 0.43$  & $-3.16 \pm 0.14$         \\
$dy$ (au)        & $1.48 \pm 0.25$   & $1.49 \pm 0.07$    \\
$g1 (\%)$        & $79.81 \pm 3.67$  & $89.93 \pm 3.01 $ \\
$g2 (\%)$        & $-20.80 \pm 2.62$ & $-17.44 \pm 0.71$     \\
$\alpha1 (\%)$   & $33.30 \pm 3.07$  & $31.84 \pm 7.31$     \\ \hline
\end{tabular}
\label{tab:modeling_comparison}
\end{table}

\subsection{Regularizing the freeform model fit}
\label{sec:freeform-regularization}

Since every pixel in the region of interest is a tunable parameter, regularization is critical to curb overfitting. Our preliminary trials using unregularized freeform models revealed near-zero residual values in the entire region of interest, indicating that the model was also fitting the background and noise artifacts, leading to a completely unphysical model disk that only appeared disk-like after convolution. As such, we detail our efforts to regularize the freeform model fit below.

A detailed walkthrough of how we regularize the freeform disk models is given in Kueny et al. (2025, submitted). We provide a high-level overview of the regularization process here: First, we impose a positivity constraint on the model pixel values since we are fitting the distribution of disk intensity in the KLIP image. We regularize high-spatial frequency artifacts in the model by using the spatial frequency spectrum of a reference disk model to use as a threshold for penalizing excessively high spatial frequencies. In short, for a given iteration, we compute the magnitude of the Fourier transform of the current freeform model and compute the mean of the power in excess to the aforementioned reference spectrum and add that value to the loss.

\subsection{Inferring physical parameters from the freeform models}
\label{sec:freeform-fit-physical}

Our freeform models do not directly provide information on the physical parameters of the disk or their uncertainty. To quantify the morphology of the optimized freeform models, we fit the same SCL disk models to our \adi{} images in an MCMC architecture as described in the previous section. We used the noise map measured from the KLIP-reduced image as described in Section \ref{sec:data_reduction} to estimate the uncertainty in each fitted disk parameter.

Our optimized freeform models can fit complex features and asymmetries (Kueny et al. 2025, submitted) so we chose to extract the SPFs from the north- and south-halves of the freeform models separately. Using the best-fit inclination values learned from the MCMC DiskFM analysis, we performed an automated ellipse-fitting routine to the trace of the brightest ridge in azimuth of the disk. Along this trace we positioned elliptical apertures with major and minor radii of $0\farcsec1$ and $0\farcsec1\cos{\phi_{inc}}$ (following \citealt{milli_near-infrared_2017}), respectively, and extracted the pixel sums across the north and south disk halves. For each disk half, we used 18 apertures to satisfy the Nyquist sampling criterion given the arc length calculated from the fitted ellipse results. To estimate the error on a given point on these SPF curves, we used the uncertainty maps as before.

\section{Results}
\label{sec:results}

\begin{figure}
    \centering
    \includegraphics[width=0.45\linewidth]{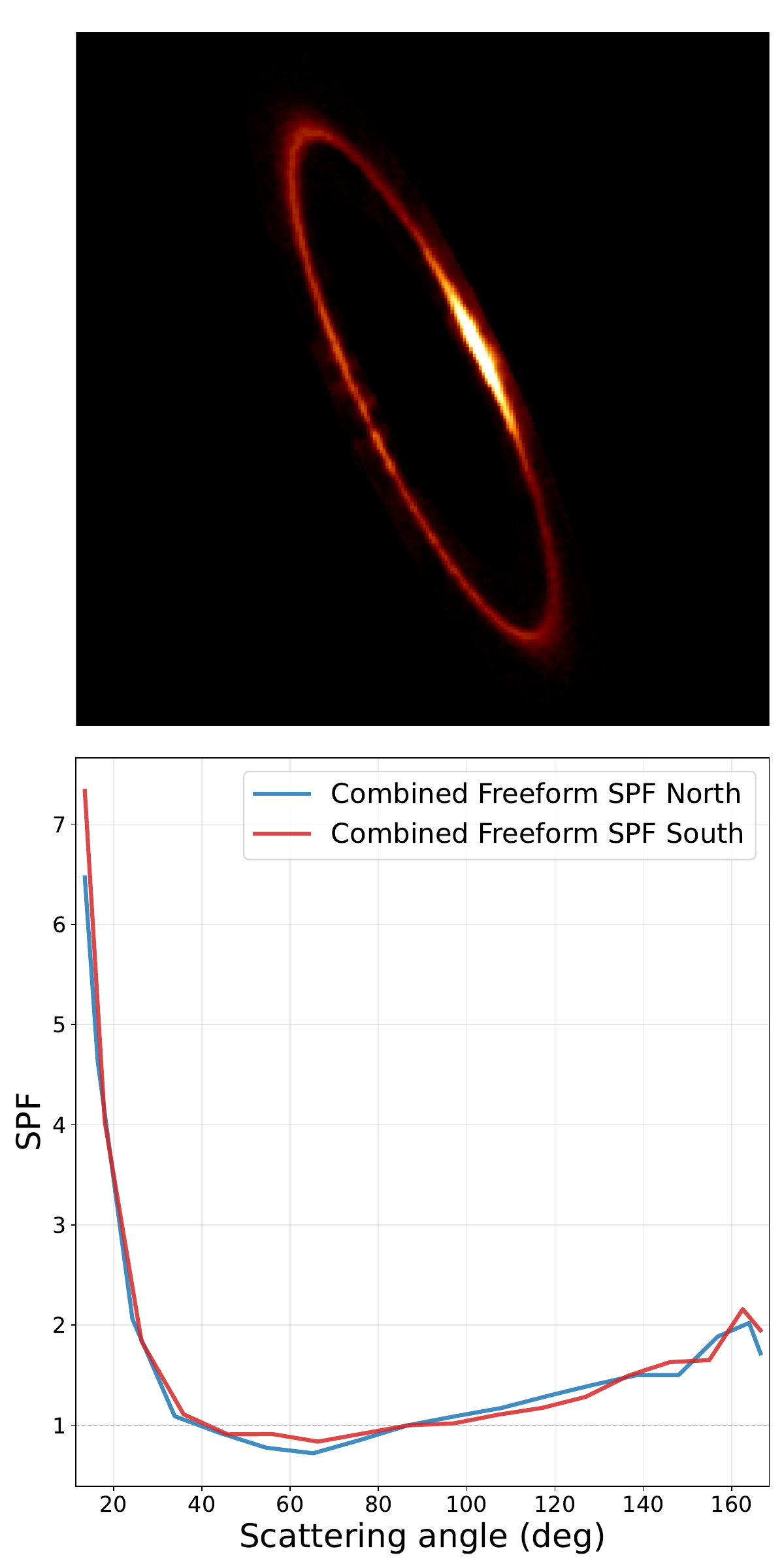}
    
    \caption{Top: averaged freeform disk model shown in a linear stretch representing a deconvolved image of the disk in a wide $\sim 500$ nm to 900 nm bandpass. Bottom: The  SPF extracted separately from the north- and south-half of the averaged optimized freeform disk model shown with blue and red lines, respectively. The curves have been normalized such that the north-half SPF at the 90 degree scattering angle is 1.}
    \label{fig:freeform_averaged}
\end{figure}

\begin{figure*}
    \centering
    \includegraphics[width=0.32\linewidth]{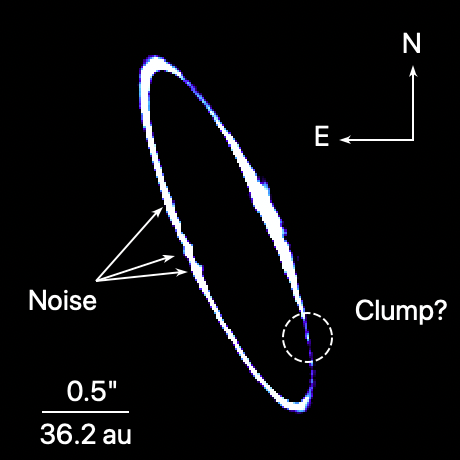}
    \includegraphics[width=0.32\linewidth]{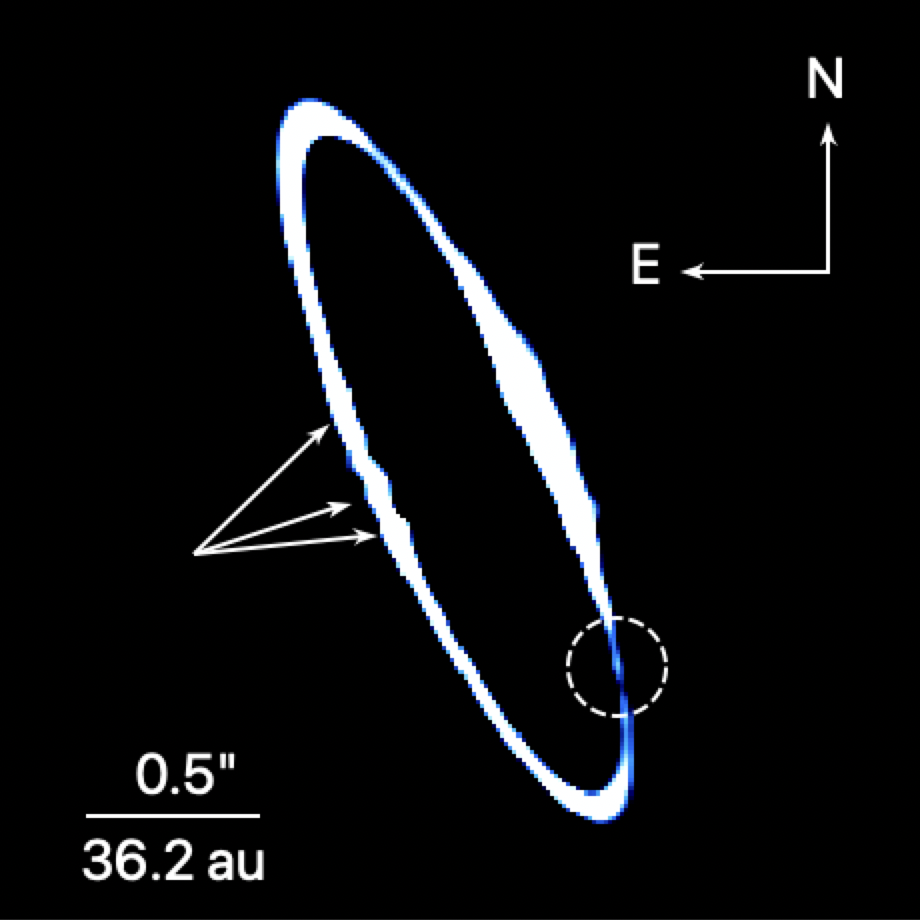} \\
    \includegraphics[width=0.2\linewidth]{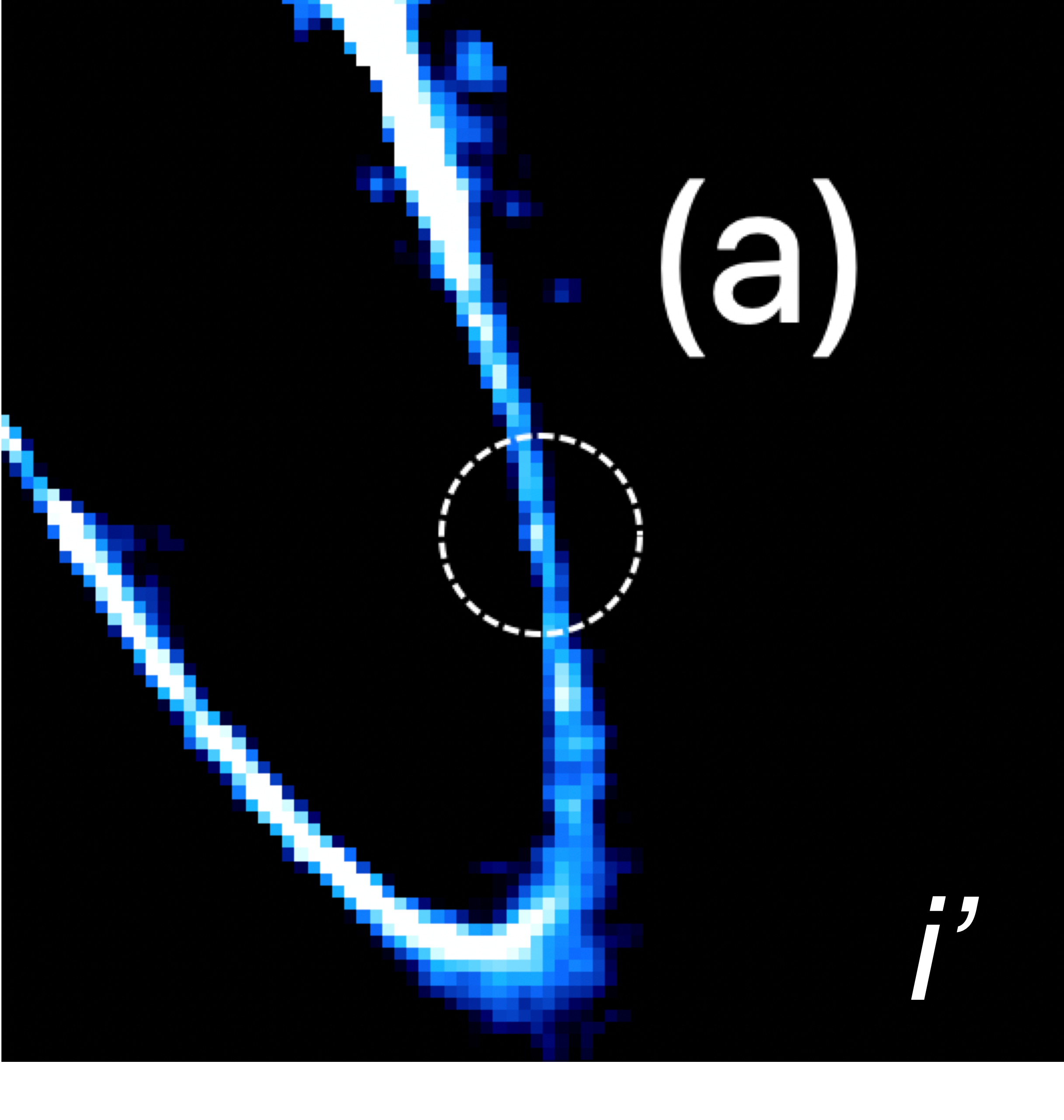}
    \includegraphics[width=0.2\linewidth]{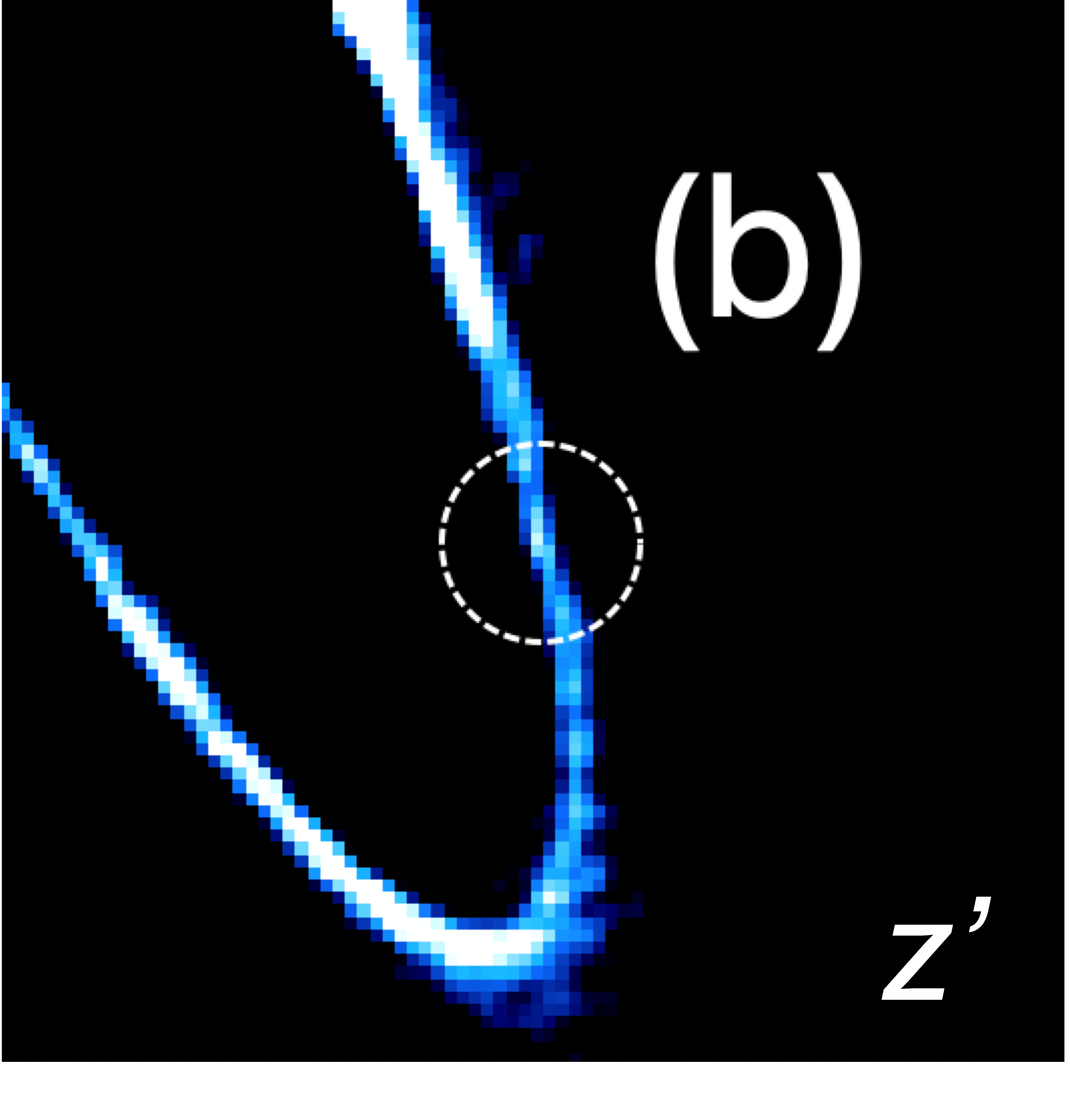}
    \includegraphics[width=0.2\linewidth]{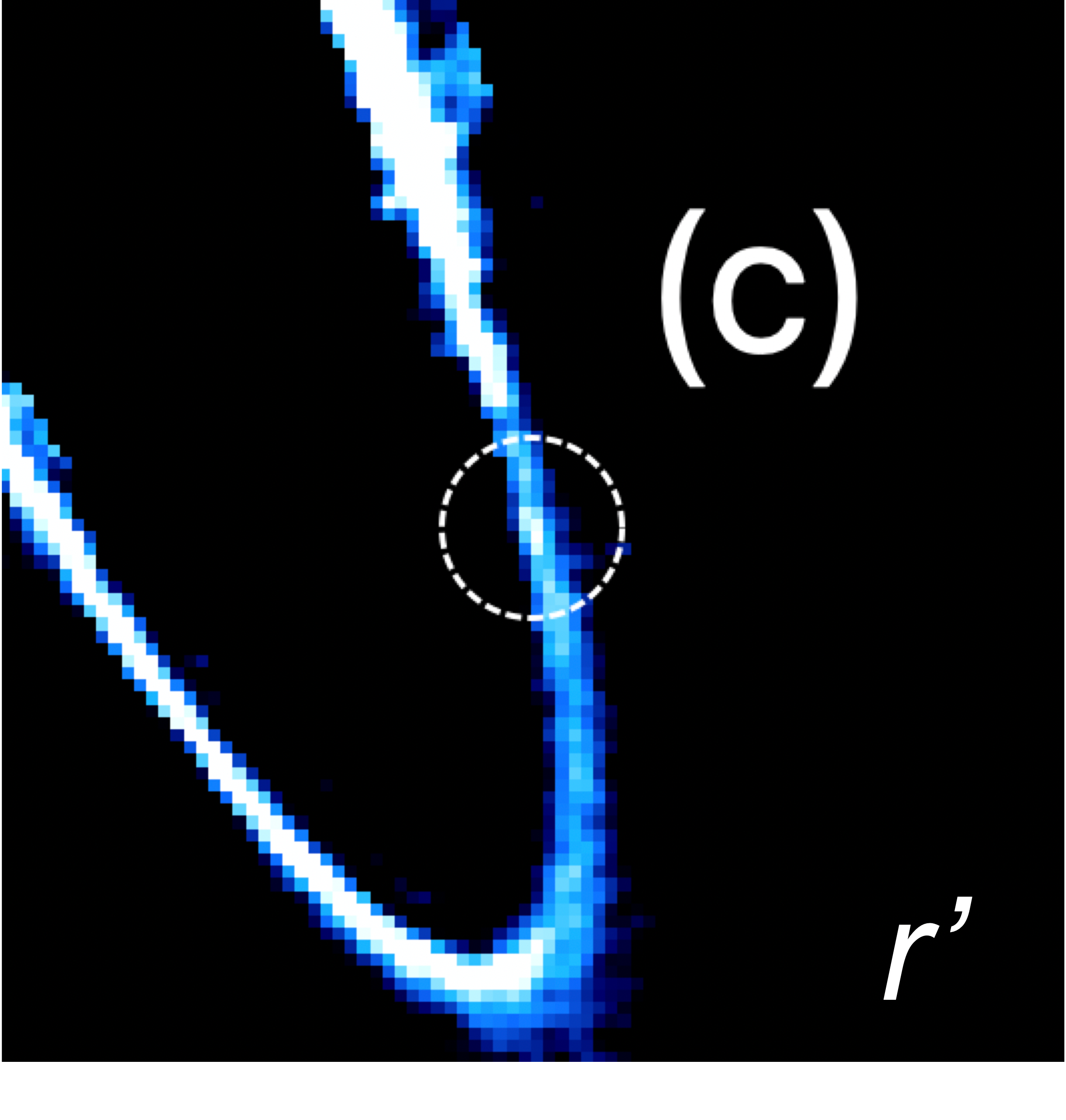} \\
    \includegraphics[width=0.2\linewidth]{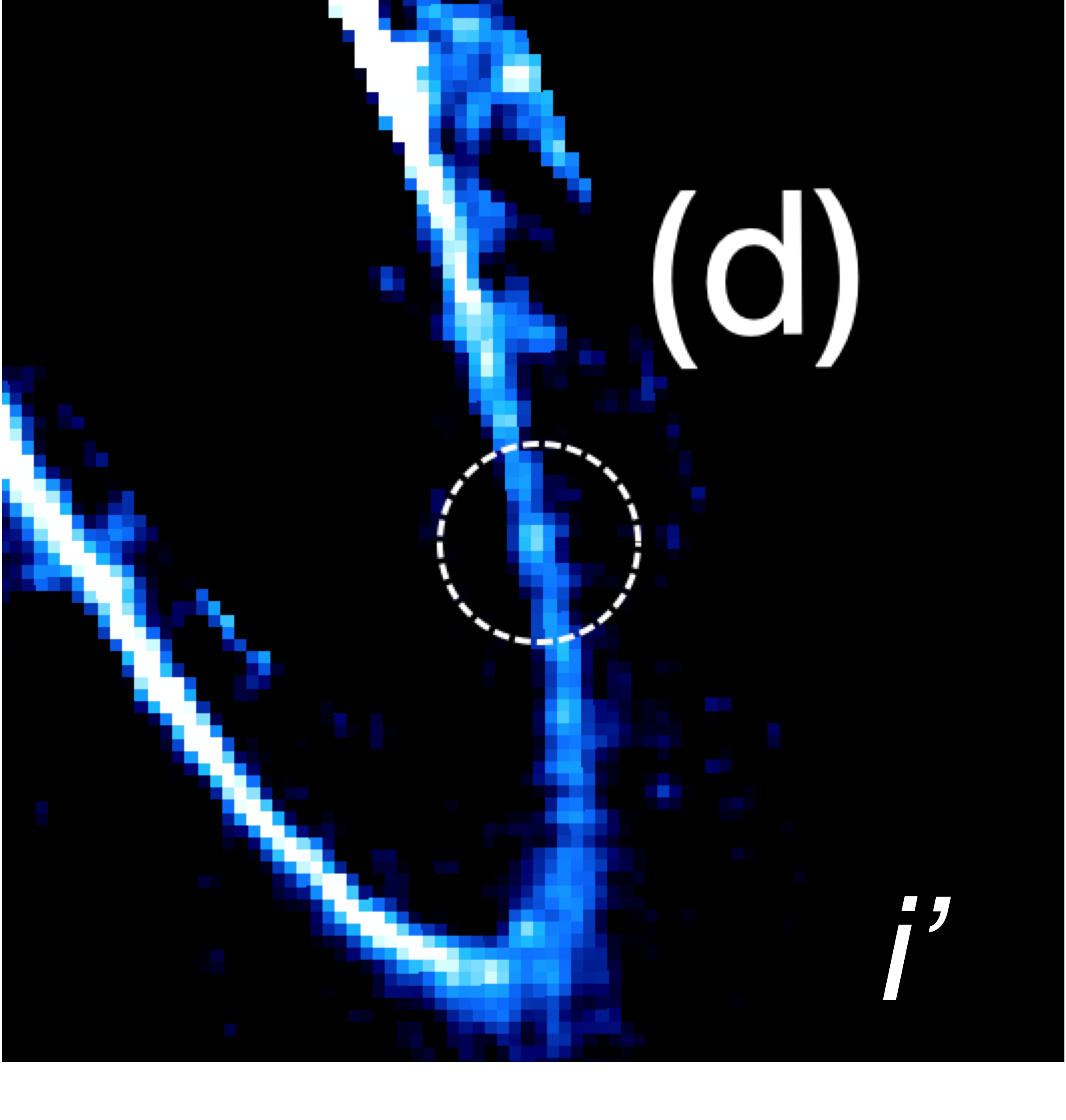}
    \includegraphics[width=0.2\linewidth]{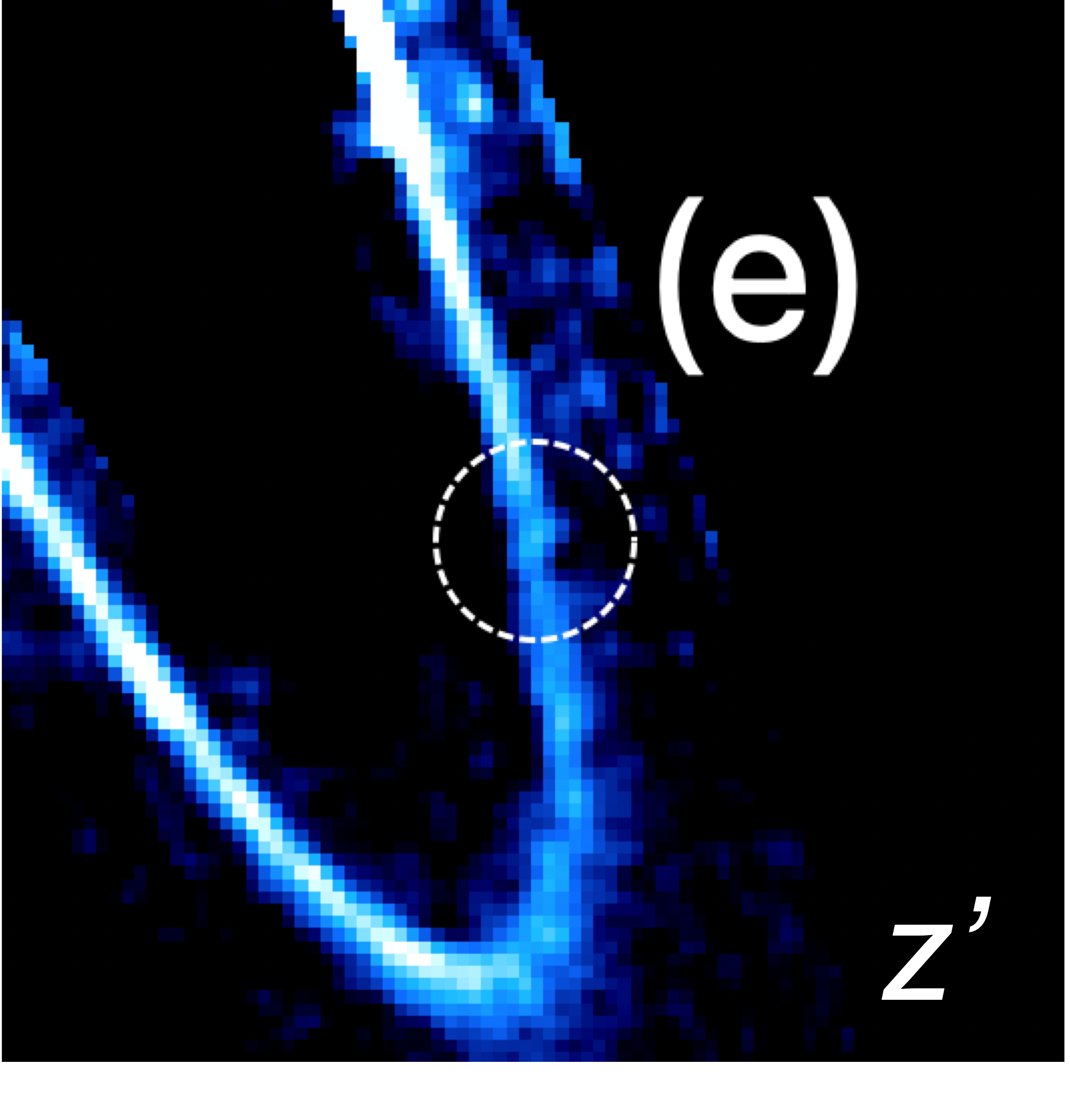}
    \includegraphics[width=0.2\linewidth]{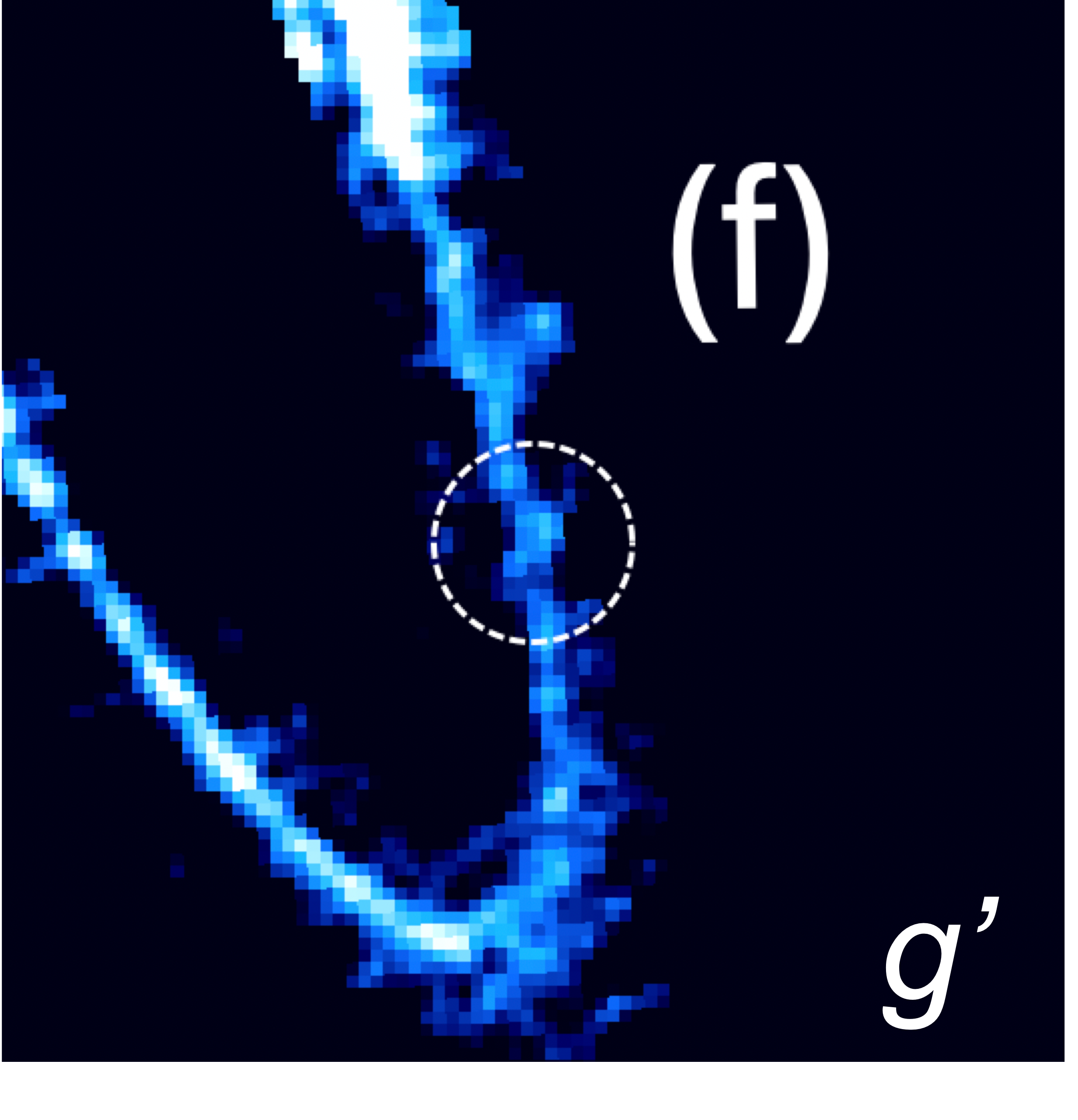}
    
    \caption{The averaged freeform model made by combining all our data can theoretically reveal structure that spans fewer pixels than a resolution element (i.e., $1$ $\lambda/D$). We made this model by averaging optimized freeform models using all $g'r'i'z'$ data from 2023A, 2024A, and 2025A. Top row: the same averaged freeform model image in two flavors: (left) with a hard colorbar stretch and (right) Gaussian smoothed and with a hard colorbar stretch. The six smaller cutouts show optimized freeform models fit to datasets from \textbf{(a)} UT 2023 Mar. 10 (\iband) \textbf{(b)} UT 2023 Mar 10 (\zband) \textbf{(c)} UT 2023 Mar. 13 (\rband) \textbf{(d)} UT 2024 Mar 29 (\iband) \textbf{(e)} UT 2024 Mar 29 (\zband) \textbf{(f)} UT 2025 Apr. 9 (\gband). The suspected dust feature is highlighted with the dashed circle while speckles imprinted around the minor axis are identified with white arrows. Importantly, the suspected feature is at a sufficient distance from the star such that it is not likely to be a speckle.}
    \label{fig:deconvolution}
\end{figure*}

\subsection{MCMC Modeling results}
\label{sec:results-modeling}

To assure reliable posterior distributions for our free parameters for our \rdi{} images, we iterated to exceed 50 times the autocorrelation time of the chains, which amounted to testing 1.9 million and 2 million disk models for \iband{} and \zband{} bands respectively. Our \adi{} forward modeling takes more time per iteration because \textsc{DiskFM} has to calculate the self-subtraction impact on the model due to having the disk in the reference images that make the KL basis. Because of this increase in computation demand, we ended our MCMC runs after testing at least 1 million disk models even if the chains did not fully reach 50 times the autocorrelation time. For all our fitted geometrical parameters, our sampling results ended with posterior distributions that were approximately Gaussian and showed no significant correlations. For fitting the Legendre polynomials for our custom \rdi{} SPF modeling, the posterior distributions exhibited some correlations mostly between the coefficients associated with the first few of the Legendre polynomials. We present further details pertaining to our \rdi{} MCMC modeling in Appendix \ref{sec:app_mcmc}. For each passband, Figure \ref{fig:diskfm_all} shows our best-fit disk models, their forward models, the final KLIP-reduced images, and the signal-to-noise of the residuals in columns one through four respectively. For each best-fit model (first column), we mark the location of the star and argument of pericenter as the red and cyan dot, respectively.

The best-fit parameter values from our forward modeling are summarized in Table \ref{tab:diskfm-results}. We also include GPI $J$-band and SPHERE $H2$ results from \cite{chen_multiband_2020} in Table \ref{tab:modeling_comparison} (who also made use of \textit{DiskFM} for their modeling) for comparison. It is worth mentioning that the \texttt{DiskFM} pipeline underestimates error bars in the fitted parameters \citep{mazoyer_diskfm_2020}. The exact underlying mechanisms that lead to this effect are not known. One source of this uncertainty could be from our estimated 2D noise maps that assume azimuthally-symmetric noise. Another potential source of uncertainty is in our assumed Gaussian statistics when computing our noise map. In actuality, speckle noise follows a modified Rician distribution \citep{fitzgerald_speckle_2006,soummer_speckle_2007}. One could artificially inflate the estimated noise map with a scale factor to study the effects on the uncertainties in the posterior distributions, but we leave that for a future analysis. In any case, our best-fit values for the $R_c, PA,$ and $inc$ parameters are generally consistent between datasets. The geometrical parameters responsible for describing the shape of the radial dust density profile are not as consistent and we discuss the implications of a complex dust distribution in the disk affecting these parameters in Section \ref{sec:discuss-radprof} below.

\subsection{Freeform modeling results}
\label{sec:results-freeform}

\begin{table*}[!ht]
\centering
\caption{Freeform Disk Forward Modeling Results for \hr{}}
\hspace*{-3cm}\begin{tabular}{lccccccc}
\hline \hline
\multirow{2}{*}{Parameter$^\text{a}$} &  & \multirow{2}{*}{Range$^\text{b}$} &  & \multicolumn{4}{c}{Best Fit Values$^\text{c}$}                                                                    \\ \cline{5-8} 
                    &  &                 &  & $g'$ (ADI)                    & $r'$ (ADI)                  & $i'$ (ADI)                              &  $z'$ (ADI)       \\ \hline
$R_c$ (au) &  & (70, 120)        &       & $77.29 \pm 0.02$           & $77.68 \pm 0.02$      & $77.64 \pm 0.04$                  & $77.87 \pm 0.17$                    \\
$\alpha_{in}$       &  & (1, 100)        &  & $46.70 \pm 0.63$        & $45.89 \pm 0.57$      & $42.82 \substack{+0.83 \\ -0.79}$ & $48.43 \substack{+4.94 \\ -4.24}$   \\
$-\alpha_{out}$      &  & (-1, -60)       &  & $21.36 \pm 0.11$       & $21.79 \pm 0.10$      & $23.62 \pm 0.20$                  & $21.96 \substack{+0.89 \\ -0.82}$   \\
$h_0$ (\%)          &  & (0.01, 10)      &  & $1.31 \pm 0.11$         & $1.03 \pm 0.01$       & $1.04 \pm 0.01$                   & $1.10 \pm 0.04$                     \\
$\phi_{inc} (^{\circ})$      &  & (70, 85)        &  & $76.33 \pm 0.01$      & $76.53 \pm 0.01$      & $76.43 \pm 0.02$                  & $76.57 \pm 0.03$             \\
$\theta_{PA} (^{\circ})$     &  & (20, 30)        &  & $26.67 \pm 0.15$        & $26.91 \pm 0.12$      & $26.68 \pm 0.12$                  & $26.62 \pm 0.11$           \\
$dx$ (au)           &  & (-15, 15)       &  & $-2.54 \pm 0.07$        & $-2.56 \pm 0.01$      & $-2.55 \pm 0.02$                  & $-2.61 \pm 0.07$                    \\
$dy$ (au)           &  & (-10, 10)       &  & $1.11 \pm 0.03$         & $1.44 \pm 0.01 $      & $1.48 \pm 0.02$                   & $1.65 \pm 0.07$                     \\ \hline
$F$ (mJy)$^\text{d}$&  &                 &  & $22.77 \pm 0.12$        & $14.82 \pm 0.11$      & $15.28 \pm 0.13$                  & $13.66 \pm 0.13$                    \\ \hline
\end{tabular}

\footnotesize
\raggedright
\vspace{1mm}
\vspace{1mm}
$^{\text{a}}$ We included parameter $N_0$ in our modeling to handle the flux normalization, but it is not shown in this table. \\
$^{\text{b}}$ We assumed uniform prior distributions for all parameters. \\
$^\text{c}$ 16th, 50th, and 84th percentiles. \\
$^\text{d}$ We include the total flux of the modeled disk in mJy units which we computed by summing the best-fit model; see text. \\
\label{tab:freeform-results}

\end{table*}

We illustrate the results from our freeform forward modeling in Figure \ref{fig:freeform_models_all}. We report the results from fitting our physical SCL disk models generated with a Legendre polynomial-based SPF to our freeform disk models in Table \ref{tab:freeform-results} (see Section \ref{sec:freeform-fit-physical} for the methods and motivation for carrying out this procedure). The majority of the disk parameters retrieved with this method are consistent with the results from the MCMC model fit of the physical disk models with the exception of the outward radial power law for the dust density $\alpha_{out}$; the physical model fits to the freeform models reveal a preference towards a much steeper decay and thus a thinner ring in the radial direction. 

To increase the signal-to-noise of fine spatial features, we averaged our optimized freeform models using two different configurations: an average using 4/6 models (excluding the $i'z'$ \adi{} models corresponding to the night of UT 2023 Mar 09) and an average including all 6 of our freeform models. The reason for these choices is to give equal weighting among all 4 passbands $g'r'i'z'$ and the minor axis of the disk at $i'z'$ has a higher signal-to-noise in the \rdi{} images (from the night of UT 2024 Mar 28). We show the image made from averaging 4/6 of our models in the top panel in Figure \ref{fig:freeform_averaged}. We note that this image perhaps represents an effective deconvolved image of the disk across a wide bandpass spanning $g'r'i'z'$ (i.e., about 500 to 900 nm). We also used this averaged model to extract the SPF using the procedure described in Section \ref{sec:freeform-fit-physical}, which is shown in the bottom plot by the blue and red lines respectively. Despite the markedly reduced amount of noise at the minor axis, we note that the sharp features at the highest scattering angles in this averaged SPF (i.e., $> 160^{\circ}$) are due to lingering noise artifacts at the back-most part of the disk.

The averaged model image made using all 6 of our freeform results reveals a clump-like feature in the southwest portion of the disk, which is shown with a dashed circle and hard colorbar stretch in the topmost panel in Figure \ref{fig:deconvolution}. We show the individual freeform models used in our averaged model as the 6 smaller images in the bottom half of Figure \ref{fig:deconvolution} which correspond to the $g'r'i'z'$ \hr{} datasets shown in Table \ref{tab:obslog}. We mark the locations of other clump-like features close to the minor axis that are likely to be surviving speckles with white arrows, however, the north-most arrow marks the blob that aligns very closely with one of the blobs seen in ALMA data in \cite{kennedy_alma_2018}'s Figure 3.

\subsection{Dust distribution in the disk}
\label{sec:results_dust}

\begin{figure}[!ht]
    \centering
    \includegraphics[width=0.65\linewidth]{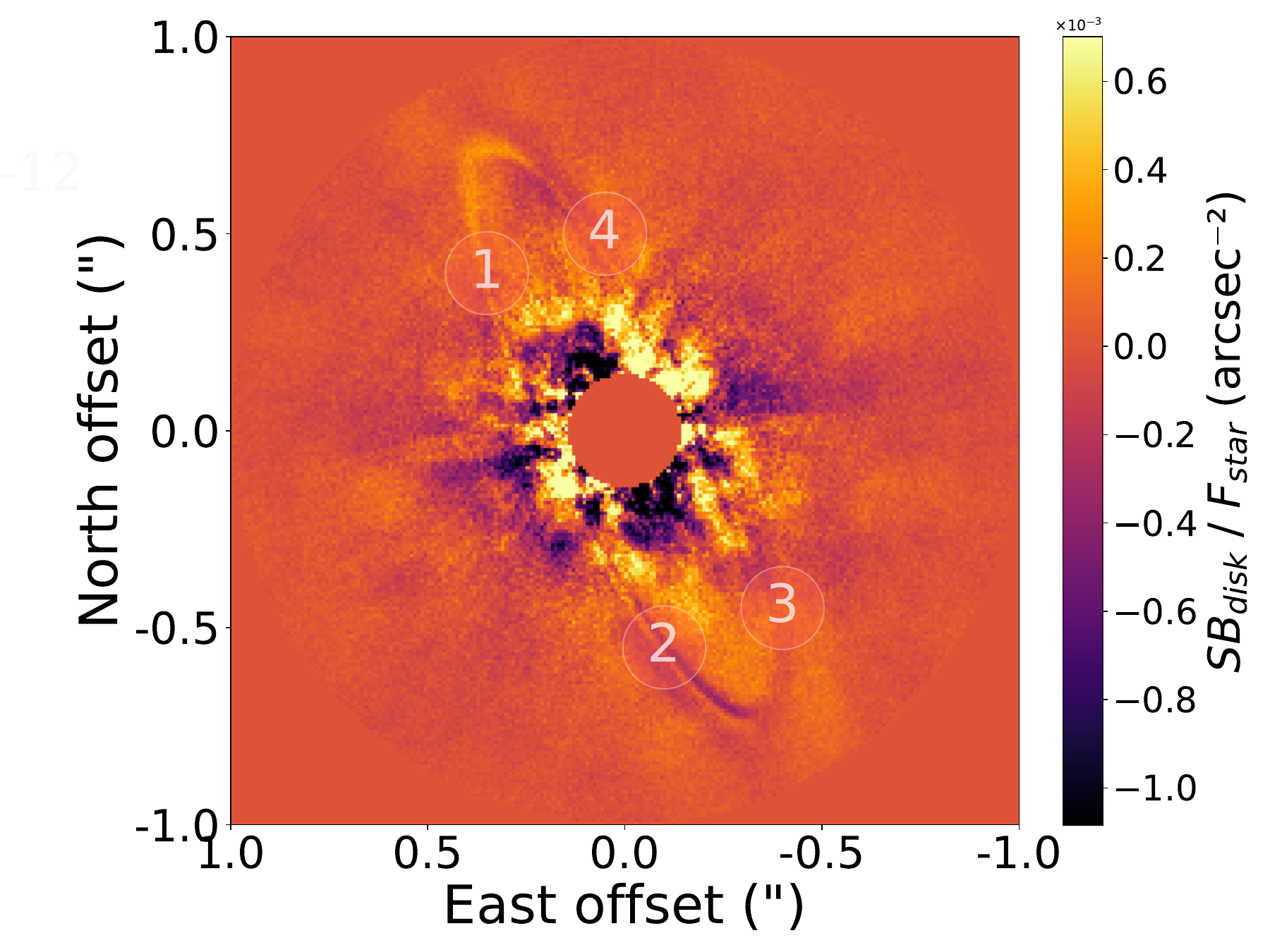} \\
    \includegraphics[width=0.5\linewidth]{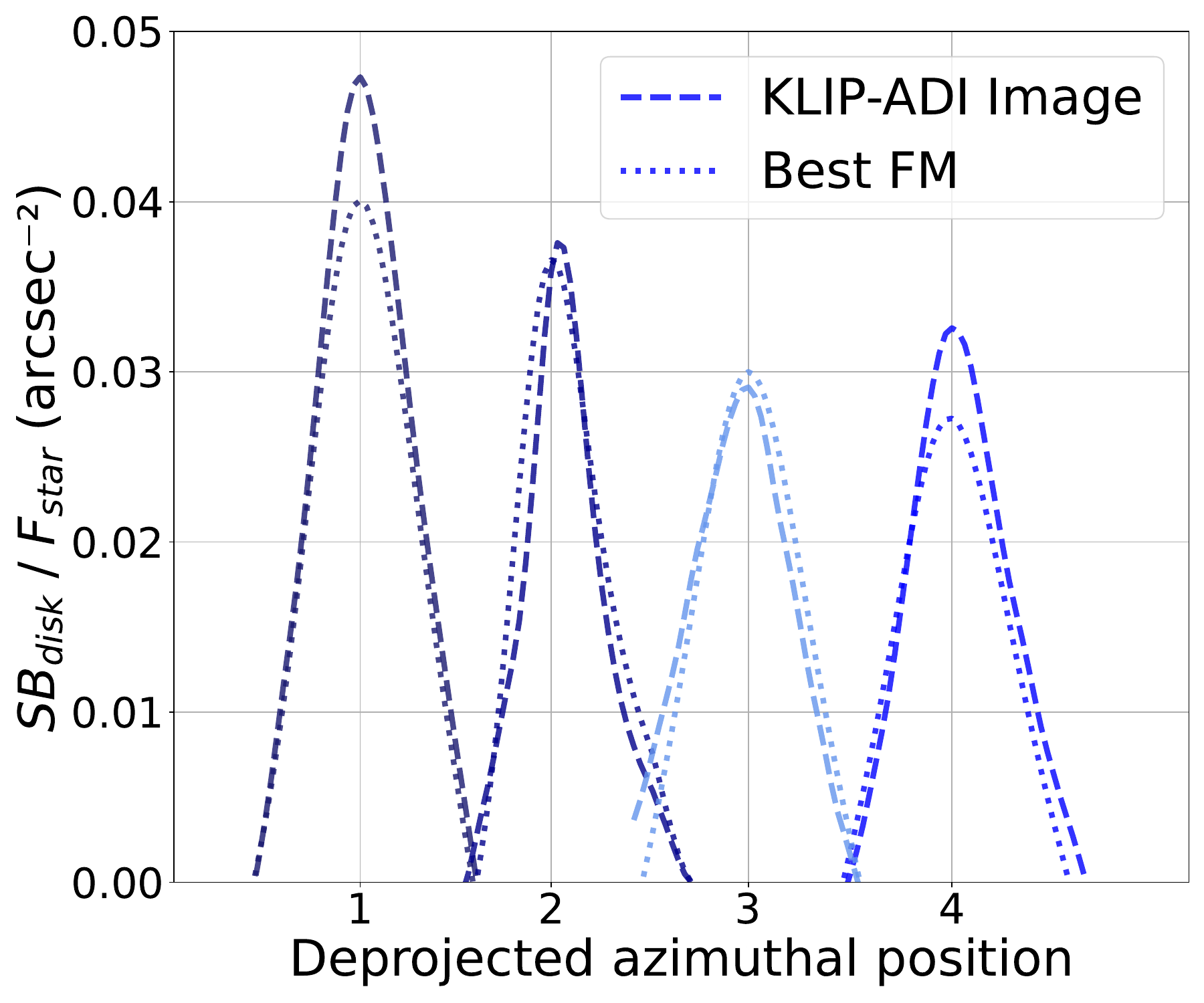}
    \caption{Residuals map between our best-fit forward models and our KLIP-reduced ADI images for \rband{} showing a consistent halo of underfit disk flux over the whole azimuthal range. We extracted the radial dust profiles at four distinct locations in the deprojected disk image and best-fit forward model as annotated in the top panel. Radial cuts 1 through 4 correspond to $45^{\circ}, 135^{\circ}, 225^{\circ}$, and $305^{\circ}$ measured counter-clockwise from the major axis of the disk, respectively.}
    \label{fig:residuals}
\end{figure}

From our physical disk modeling results, we recover a thin eccentric ring with a very steep inner edge ($\alpha_{in} \sim 45$) and stellar offsets in both the $x-$ and $y-$directions on the order of 1.5---2.5 au shifting the model toward the southeast. We fit for the opening angles $h_o$ for our KLIP-ADI images but held them fixed at $h_0 = 1.0\%$ for our KLIP-RDI models as we found that those images did not constrain the aspect ratio parameter well. Our values for the KLIP-ADI models at \iband{} and \zband{} tended toward $h_0 \approx 1\%$ indicating a thin disk in vertical direction. In the case of the \gband{} and \rband{} images, this parameter tended to something slightly larger ($h_0 \approx 1.6\%$). We also detect a significant brightness asymmetry between the north and south ansae (Table \ref{tab:photometry}) that cannot be fit with our model. This asymmetry has been reported in prior works (\citealt{schneider_stis_2009}; \citealt{milli_optical_2019}; \citealt{chen_multiband_2020}; \citealt{arriaga_multiband_2020}). To quantify the difference in brightness between the ansae, we computed the SW/NE flux ratios based on the procedure outlined in \cite{schneider_stis_2009}'s Section 4.5. Our values are within $3\sigma$ of, but consistently closer in brightness than, those reported in \cite{schneider_stis_2009} and \cite{milli_optical_2019}. We note that we deprojected our disk image first before extracting the flux from the ansae which might be the source of the slight discrepancy since it is unclear if this step was followed in the aforementioned studies. This brightness enhancement in the north ansa is most evident in the residuals from our KLIP-ADI images (Figure \ref{fig:diskfm_all} and Figure \ref{fig:residuals}).

Another notable feature that we recover in the dust density is a radial profile which is not well-described with our disk model. This faint halo of underfit disk flux can be seen most clearly in the residuals between the KLIP-ADI images at $g'$, $r'$, and \iband{} and our best-fit forward models (columns 4 and 5, Figure \ref{fig:diskfm_all}). We also include a larger \rband{} residuals image and show an analysis of select radial cuts in the residuals in Figure \ref{fig:residuals}. We chose this image because we detected the disk at the highest signal-to-noise at this passband. The fact that the residual halo is the faintest at \zband{} (almost undetectable by eye) is an interesting result given both the high signal-to-noise for most of the disk in this image (Figure \ref{fig:snr_adi}) and comparable integration time to that of our KLIP-ADI \iband{} image. A very extended and diffuse cloud of dust grains was detected (out to about $12\farcsec$ or 875 au) at visible wavelengths by \cite{schneider_hr_2018} using HST/STIS and is one explanation for this enigmatic feature (i.e., the data at \rband could be more sensitive to scattering by these small grains than \zband). N-body simulations of small, bound dust grains interacting with the local interstellar medium by \cite{olofsson_dust_2019} showed that if the dust around \hr{} is orbiting in a counter-clockwise direction, their simulation produced results consistent with what was observed by \cite{schneider_hr_2018}.

Our freeform disk models reveal slightly different dust distribution results. Namely, the outer radial power law for the dust distribution tends to a much steeper value of $\alpha_{out} \sim 22$. Additionally, the aspect ratio $h_0$ is more consistent between passbands and closer to about $1\%$ for all \adi{} images. The inner radial power law $\alpha_{in}$ and critical radius $R_c$ are consistent with the physical model DiskFM results.

\subsection{SPF models}
\label{sec:results-spfs}

\begin{figure}[htbp]
    \centering
    \includegraphics[width=0.75\linewidth]{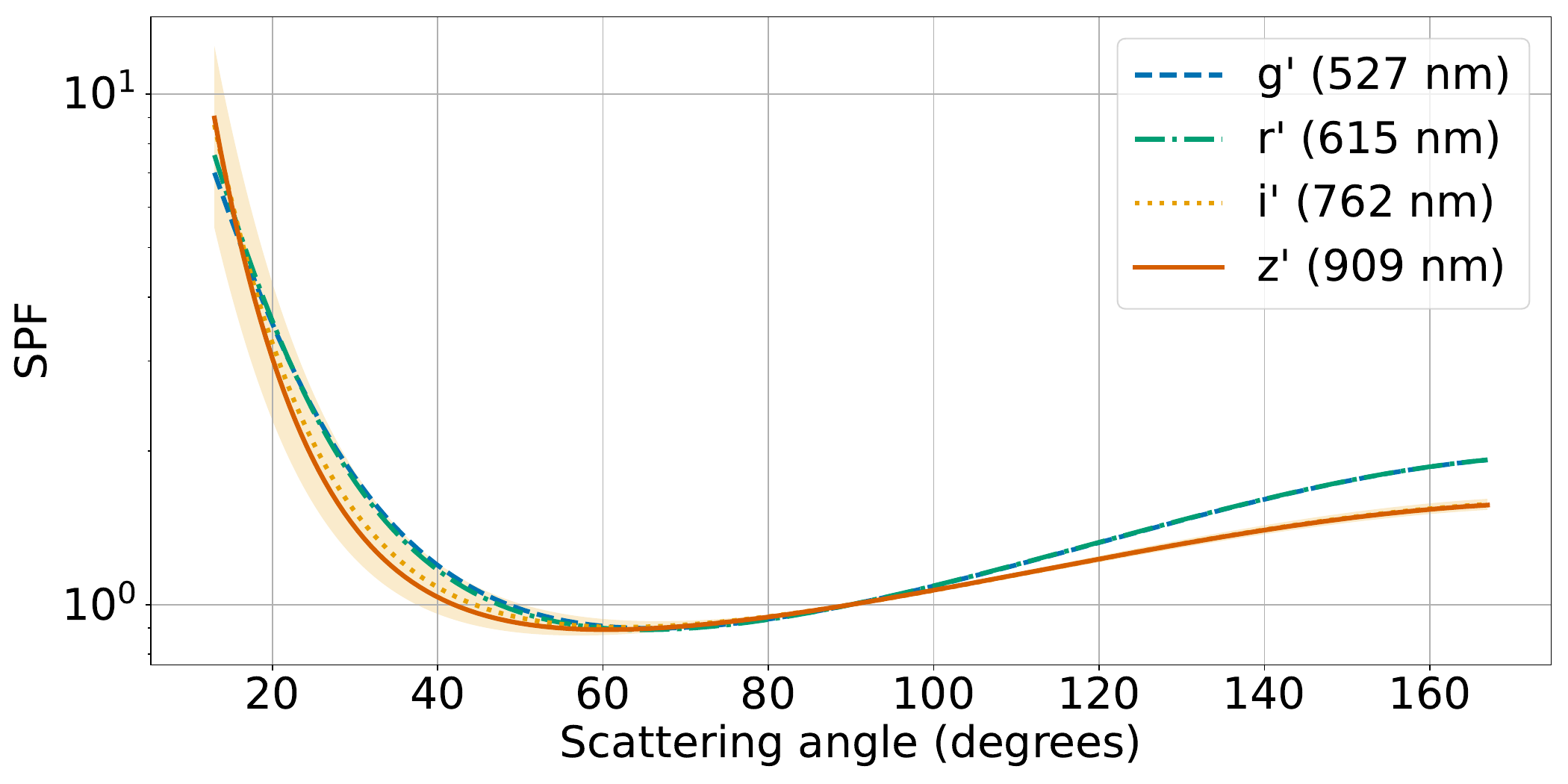}
    \caption{Comparison of the best-fit HG solutions from our forward modeling analysis on our ADI images at $g'$, $r'$, $i'$, and $z'$. The SPF is plotted on a log scale on the y-axis and the scattering angle in degrees is plotted linearly on the x-axis. We only include the $3\sigma$ uncertainty associated with the \iband{} filter results as a representative uncertainty estimation for clarity.}
    \label{fig:spf_hg}
\end{figure}

From our disk models that best fit our KLIP-ADI images, we recover very strongly forward-scattering SPFs with HG asymmetry parameters $g1$ ranging from $0.794$ to $0.967$ (Table \ref{tab:diskfm-results}). The $g2$ (back-scattering) asymmetry parameter ranges from $-0.145$ to $-0.203$ with the \rband{} and \gband{} results attaining the highest values. For \zband{}, the weighting factor $\alpha1$ is poorly-constrained especially in comparison to the other passbands. We show the HG solutions that best fit the images in Figure \ref{fig:spf_hg}. These SPFs feature a strong forward-scattering peak with a moderate amount of back-scattering, consistent with previous studies (\citealt{milli_near-infrared_2017}, \citealt{chen_multiband_2020}, \citealt{hom_uniform_2024}).

\begin{figure}[htbp]
    \centering
    \includegraphics[width=0.75\linewidth]{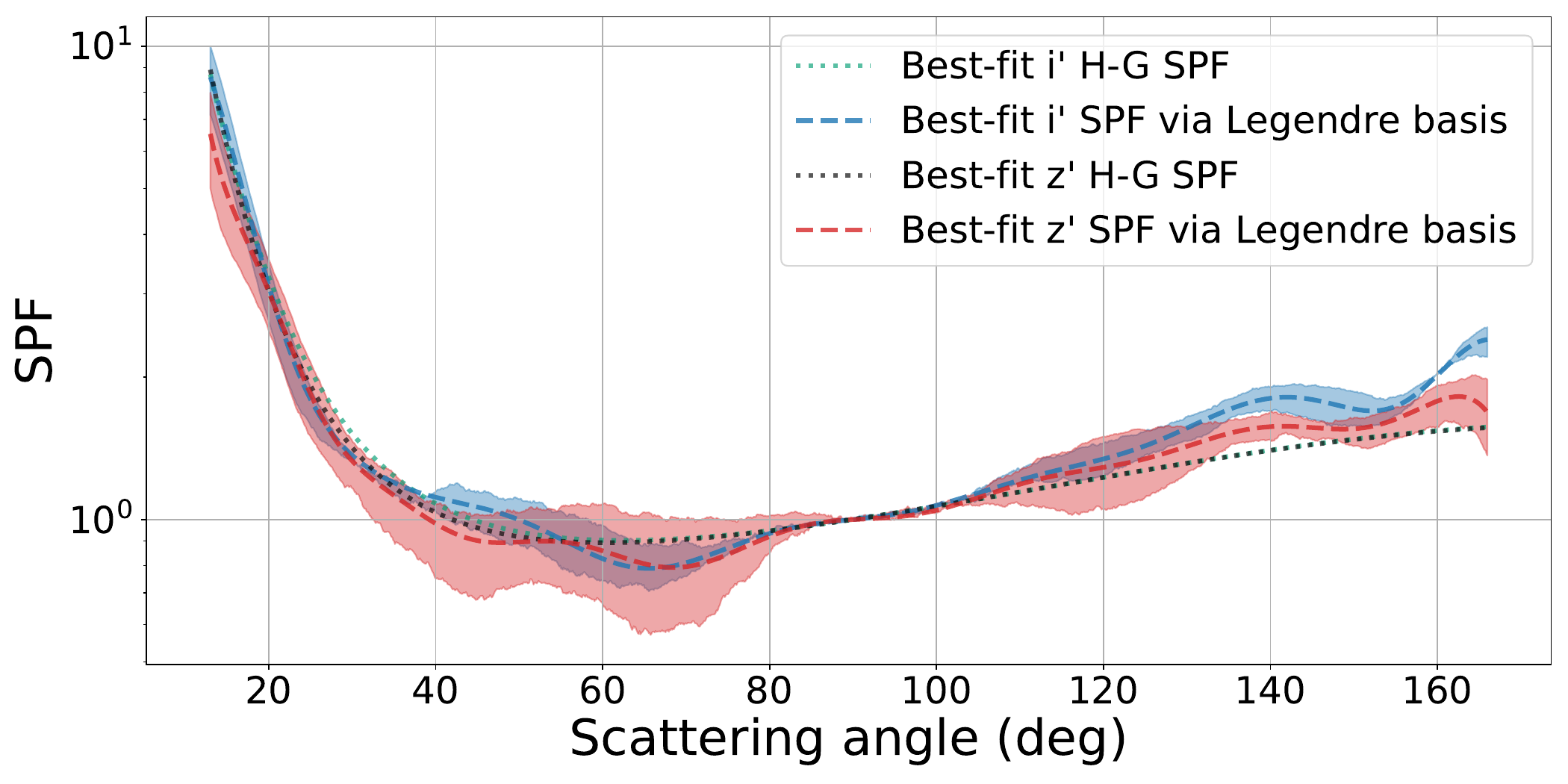}
    \caption{Comparison of our best-fit model SPFs to our \rdi{} and \adi{} data at $i'$ and $z'$. For the \rdi{} images, we fit a custom SPF made from a basis of Legendre polynomials shown as the blue and red dashed lines for \iband{} and \zband{}, respectively. For the \adi{} images, we fit HG SPFs which we plot as green and black dotted lines for \iband{} and \zband{}, respectively.  We plotted the SPF along the y-axis in log-scale and the scattering angles in degrees linearly along the x-axis. The uncertainty in the Legendre SPFs corresponds to the shaded regions. More details for the Legendre polynomial solutions and the uncertainties can be found in Appendix \ref{sec:app_mcmc}.}
    \label{fig:spf_custom}
\end{figure}

We achieved excellent fits to our KLIP-RDI data by instead using a custom SPF constructed from the first 16 Legendre polynomials for the $i'$-band image and the first 18 for the \zband{} image. Fitting the coefficient values for these bases from our MCMC calculations resulted in fitting for 23 and 27 free parameters respectively. This number of basis components allowed for accurate fitting of the azimuthal brightness distribution of the disk without significant overfitting. Our custom SPFs are overplotted with our best-fit HG solutions for both $i'$-band and $z'$-band KLIP-ADI images in Figure \ref{fig:spf_custom}. As with the HG solutions, we recover a strong forward-scattering peak but with a degree of back-scatter that the HG SPF cannot reproduce. The forward scattering peak for the custom SPF differs more in the \zband{} case compared to \iband{} for the HG solution, which might explain the aforementioned poor constraint on the \zband{} HG weighting factor parameter $\alpha1$. A higher-order feature that we recover in our custom SPFs is a dip centered around the $\sim70^{\circ}$ scattering angle (see Appendix B for uncertainties for these SPFs). This dip in brightness can also be seen in Figure \ref{fig:wedge_phot_adi} at approximately 12 and 4 o'clock in the radar charts.

We described in Section \ref{sec:freeform-describe-subsec} how extracting the SPF from our freeform disk models is a useful way to get scattering information from the dust for our \adi{} images since they suffer from self-subtraction at the disk's minor axis. We show our individual SPF extraction results in Figure \ref{fig:spf_freeform} and our averaged freeform SPF in Figure \ref{fig:freeform_averaged} (bottom panel). At all passbands, we observe general agreement, within the uncertainties, for the HG and Legendre polynomial-based SPFs. The averaged freeform SPF reveals a minimum at scattering angle $\sim60^\circ$ for both halves of the disk. This minimum is also seen in the same general area in the individual freeform SPFs, though the higher noise makes this dip less distinct in the red south-side SPF curves. However, the sharp features in the SPF at \textit{south side} \iband{} and \zband{} spanning $\sim140$ to 160 degrees are due to a WDH artifact that is not easily removed from the final image. Additionally, there is a prominent quasistatic speckle pinned at the very back part of the disk in the \adi{} \rband{} image (it can be seen in Figure \ref{fig:finalklip_adi}, top-right panel) that is causing the spikes at the highest scattering angles in those freeform SPF curves. The \gband{} \adi{} image exhibits the lowest signal-to-noise among all our reduced images which is reflected by the many sharp features and higher uncertainties in those freeform SPF curves.

\begin{figure*}[!ht]
    \centering
    \includegraphics[width=\linewidth]{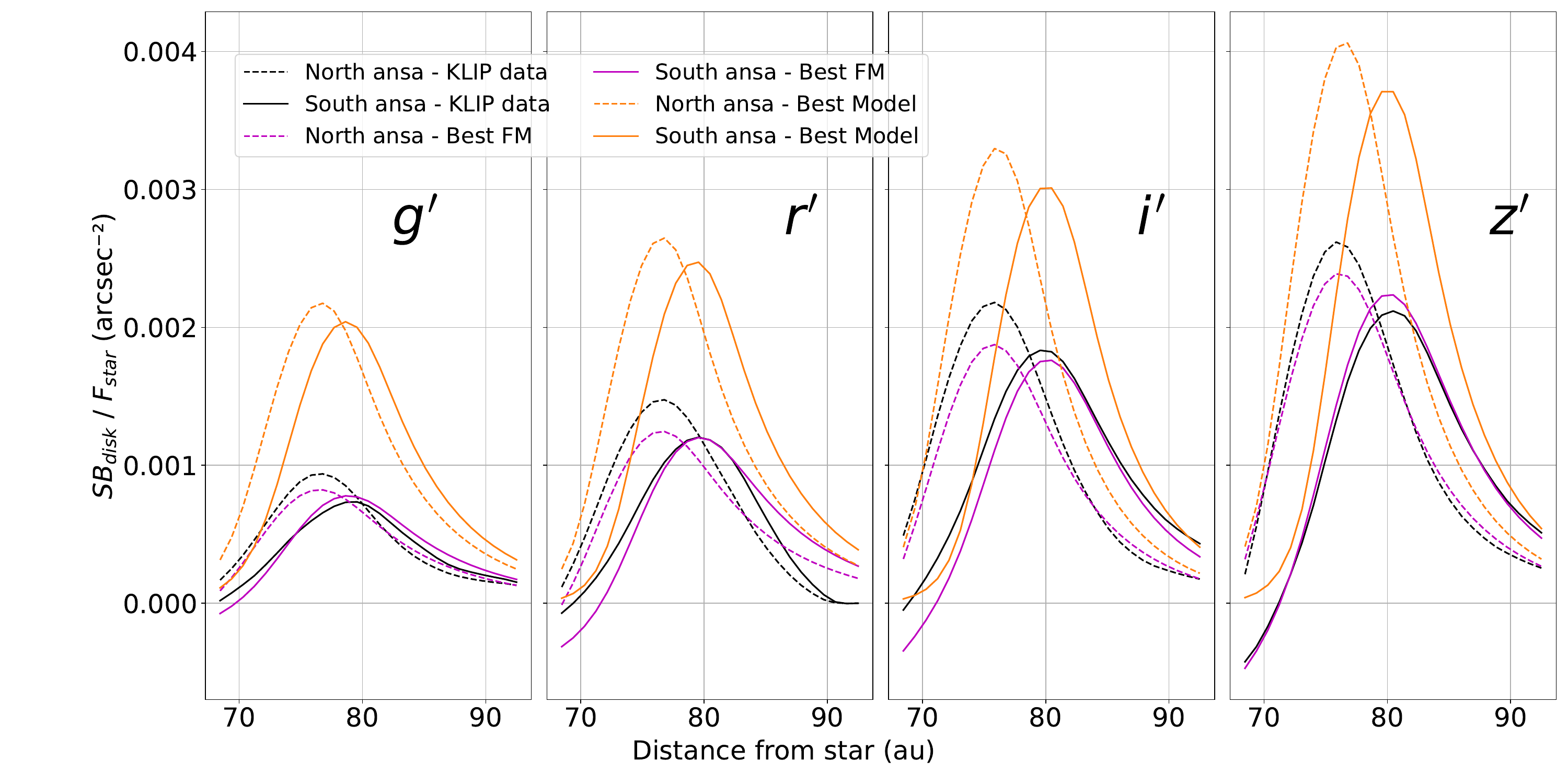}
    \caption{Radial brightness profiles extracted along the disk's major axis using our deprojected \adi{} images. We used circular apertures centered on every pixel along the major axis of radius 63 mas. We denote the surface brightness profiles of the north and south ansae as the dashed and solid lines respectively. Black and magenta lines representing profiles from the KLIP-reduced data and best-fit forward models complement the radial cuts in Figure \ref{fig:residuals} from 4 other distinctly-different locations in the disk.}
    \label{fig:sb_profiles}
\end{figure*}

\begin{figure*}[htbp]
    \centering
    \includegraphics[width=0.88\linewidth]{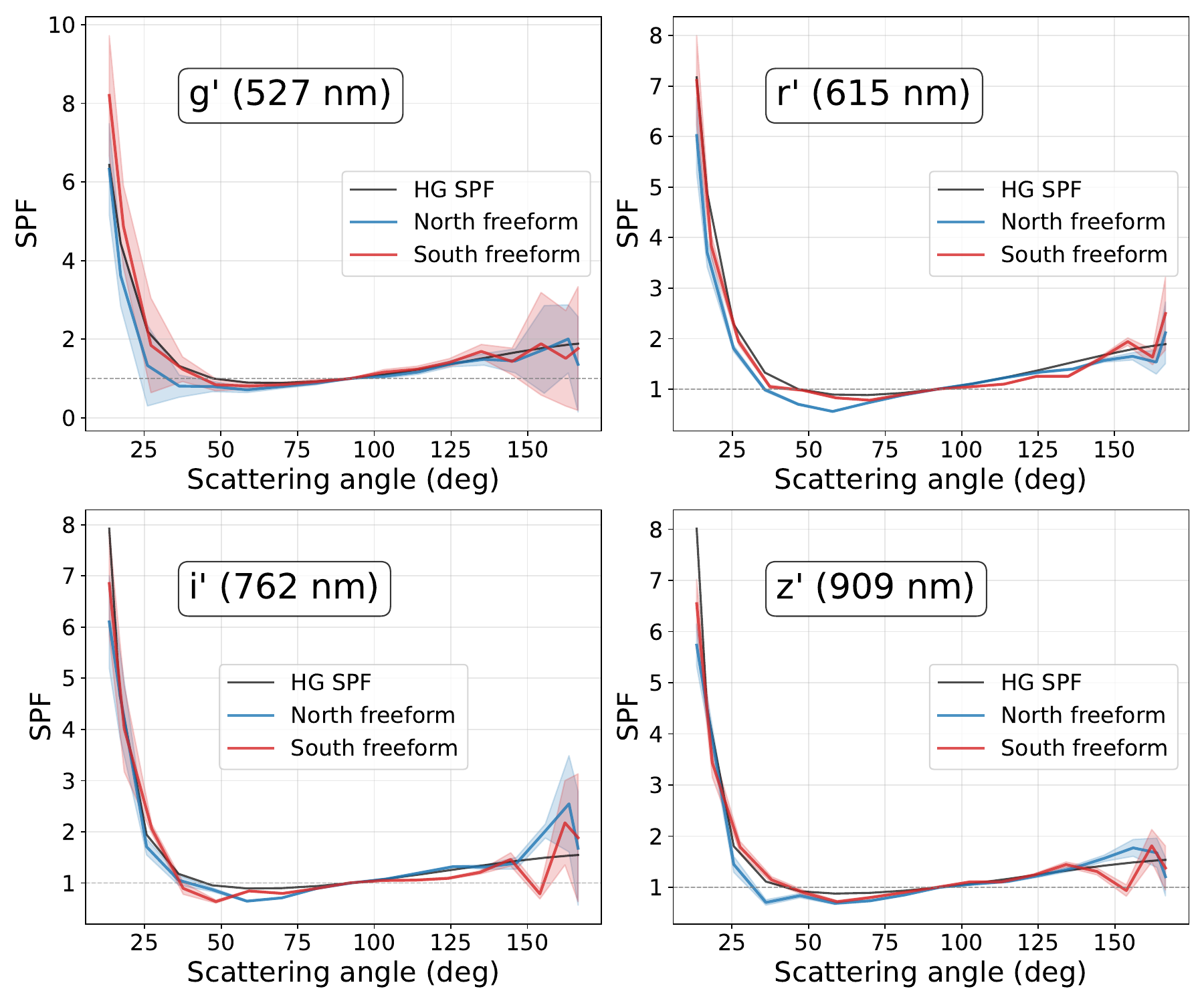} \\
    \includegraphics[width=0.88\linewidth]{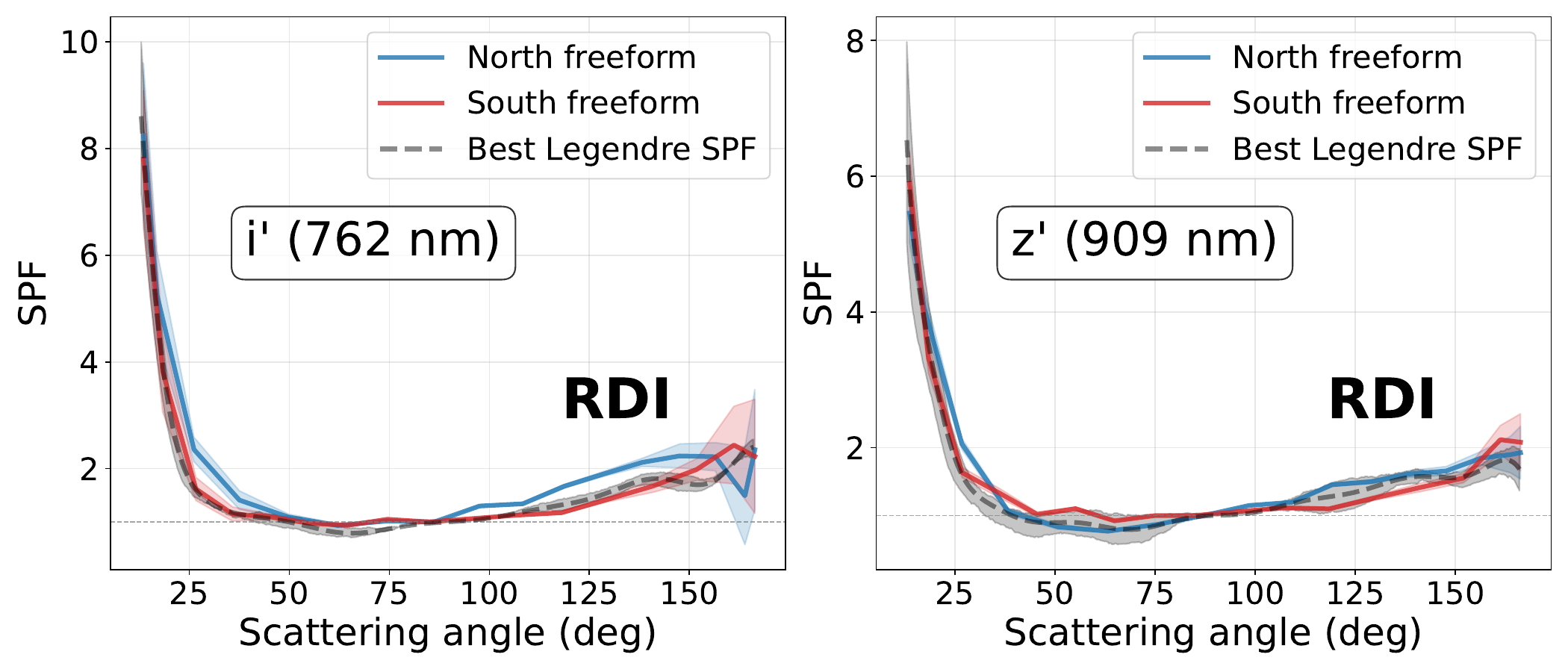}
    \caption{Comparison of the SPFs extracted from our freeform disk models to our data at $g'$, $r'$, $i'$, and $z'$. We extracted the SPF for the north and south halves of the disk separately for analyzing asymmetry. We illustrate each SPF in log scale along the y-axis and linearly along the x-axis. We illustrate uncertainties as shaded regions. Top and Middle rows: In the ADI data, the dip in the \textit{south side} SPF from scattering angles $\sim140$ to 160 degrees at $i'$ and $z'$ is due to a persistent WDH artifact. For comparison, we include the best-fit HG SPFs as solid gray lines. Bottom row: The \rdi{} freeform model SPFs compared to Legendre polynomial-based SPFs from Figure \ref{fig:spf_custom} as the dashed gray lines.}
    \label{fig:spf_freeform}
\end{figure*}

\begin{table}[htbp]
\centering
\caption{Extracted disk photometry for each dataset.}
\label{tab:photometry}
\begin{tabular}{lccc}
\hline
         & Ansae Avg.      & Whole Disk Avg. & Ansae Ratio                         \\
         & ($\frac{SB_{\text{disk}}}{F_{\text{star}}}$) $\text{(")}^{-2}$ & ($\frac{SB_{\text{disk}}}{F_{\text{star}}}$) $\text{(")}^{-2}$ & \\
         & $\times$1e-3    & $\times$1e-3    &     SW/NE                                \\ \hline
$g'$ ADI & $1.35 \pm 0.19$ & $1.73 \pm 0.18$ & \multicolumn{1}{c}{$0.79 \pm 0.06$} \\
$r'$ ADI & $1.59 \pm 0.12$ & $2.17 \pm 0.21$ & \multicolumn{1}{c}{$0.81 \pm 0.05$} \\
$i'$ ADI & $1.94 \pm 0.32$ & $2.71 \pm 0.32$ & $0.83 \pm 0.05$                     \\
$z'$ ADI & $2.4 \pm 0.37$  & $3.22 \pm 0.39$ & $0.85 \pm 0.05$                     \\
$i'$ RDI & $1.98 \pm 0.37$ & $2.98 \pm 0.33$ & $0.76 \pm 0.08$                     \\ 
$z'$ RDI & $2.73 \pm 0.44$ & $3.63 \pm 0.34$ & $0.75 \pm 0.03$                     \\ \hline
\end{tabular}
\end{table}

\subsubsection{Disk Photometry}
\label{sec:results_sb}

We extracted the disk surface brightness along the major axis in our best-fit models and KLIP-ADI reduced images using aperture photometry with circular aperture radii of $r = 63$ mas. We report the surface brightness profiles for the best-fit models (orange lines), forward models (magenta lines), and the reduced images (black lines) in Figure \ref{fig:sb_profiles}. We note that the degree of brightness asymmetry at the ansae is clearly portrayed in the traces extracted from the KLIP-ADI images which our forward models are unable to reproduce at all passbands even with stellocentric offsets. Additionally, the attenuation of the disk signal due to convolution and PSF-subtraction is portrayed by the difference in height between the orange and magenta curves.

\begin{figure}[!ht]
    \centering
    \includegraphics[width=0.75\linewidth]{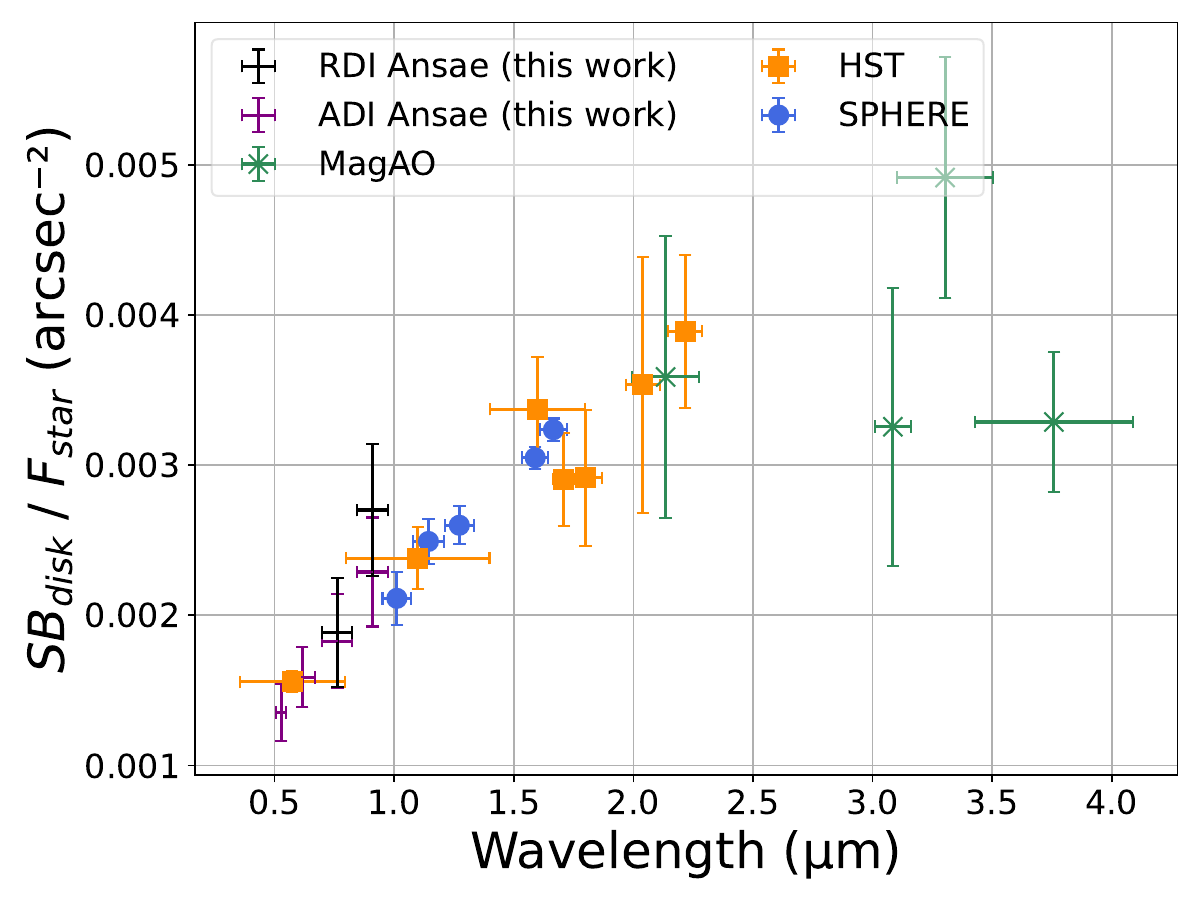}
    \caption{Comparison of our extracted photometry at the disk ansae (purple and black points) and values reported in the literature. We include Hubble Space Telescope and MagAO data from \cite{rodigas_morphology_2014}'s Figure 6 as the yellow square and green star points respectively. We also include data from VLT/SPHERE represented by the blue circle points. We recover a red spectral slope in our photometry and show consistency with the photometry from the aforementioned studies.}
    \label{fig:photometry}
\end{figure}

Figure \ref{fig:photometry} compares the photometry we extracted from the ansae of our best-fit disk models with other values reported in the literature. We include data points measured by HST/NICMOS \citep{debes_complex_2008} as yellow square points, Magellan/MagAO \citep{rodigas_morphology_2014} as green star points, and VLT/SPHERE \citep{milli_near-infrared_2017} and blue circle points and demonstrate that our data (purple and black points) agree with these values and support a red spectral slope for the scattered light. We also integrate the total flux density in each passband for our best-fit disk models and report them in the bottom of Table \ref{tab:diskfm-results} and with the disk ansae photometry over the entire disk in Table \ref{tab:photometry}. For the photometry averaged over the entire disk, we mirrored the procedure in \cite{milli_near-infrared_2017} and placed non-overlapping circular apertures of radius $r = 63$ mas along a fitted ellipse over the disk model and averaged over the ensemble of aperture sums. All of our reported photometry values have been normalized by the estimated flux densities of the star at their respective passbands to remove the star's color contribution.

\begin{table*}[ht!]
    \centering
    \small
    \caption{Best Disk Model Geometry Parameters (Disk Plane).}
    \label{tab:disk_plane_params}
    \begin{tabular}{rccccc}
        \hline
                               & $R_c$ (mas)        & $e$                & $i$ (deg)        & $\omega$ (deg)    & $\Omega$ (deg)   \\ \hline
        $g'$ KLIP-ADI (UT 2025-04-08) & $1060.03 \pm 1.17$ & $0.024 \pm 0.0008$ & $76.48 \pm 0.02$ & $-57.48 \pm 0.78$  & $26.26 \pm 0.12$ \\
        $r'$ KLIP-ADI (UT 2023-03-12) & $1061.89 \pm 1.05$ & $0.033 \pm 0.0006$ & $76.72 \pm 0.02$ & $-58.32 \pm 0.92$ & $26.90 \pm 0.12$ \\
        $i'$ KLIP-ADI (UT 2023-03-09) & $1063.69 \pm 1.44$ & $0.034 \pm 0.0007$ & $76.60 \pm 0.02$ & $-47.08 \pm 1.20$ & $26.68 \pm 0.12$ \\
        $z'$ KLIP-ADI (UT 2023-03-09) & $1065.47 \pm 4.4$  & $0.033 \pm 0.0023$  & $76.62 \pm 0.05$ & $-50.69 \pm 3.4$  & $26.72 \pm 0.13$ \\
        $i'$ KLIP-RDI (UT 2024-03-28) & $1061.73 \pm 0.97$ & $0.034 \pm 0.0006$ & $76.49 \pm 0.02$ & $-38.65 \pm 1.12$ & $26.33 \pm 0.15$ \\
        $z'$ KLIP-RDI (UT 2024-03-28) & $1062.27 \pm 1.05$ & $0.041 \pm 0.0007$ & $76.38 \pm 0.02$ & $-57.51 \pm 0.75$ & $26.14 \pm 0.15$ \\ \hline
        \end{tabular} \\
    \footnotesize
    \vspace{2mm}
    \raggedright \textbf{Notes.} The uncertainties for each parameter are the 16th, 50th, and 84th percentiles of that parameter's marginalized posterior distribution. We included the uncertainty in the direction to true north (\citealt{long_astrometric_2025} and in Appendix \ref{sec:app_astrometry} for 2025A) for longitude of the ascending node.
\end{table*}

\section{Dust Modeling Discussion}
\label{sec:discussion}

\subsection{Best-fit geometric parameters}
\label{sec:discuss-offsets}

Our modeling results show preference for disk models with a $dx \sim 1-2$ au offset in the negative x-direction and about the same amount in the positive y-direction. Since these offsets are applied to the disk in the stellocentric frame before the position angle and inclination transformations, this gives the disk models an effective movement in the southeast direction in the detector frame, putting pericenter northwest of the star as found by prior works \citep{milli_near-infrared_2017, milli_optical_2019, olofsson_dust_2019, chen_multiband_2020, arriaga_multiband_2020}. We point out that a greater spread in best-fit values for the $dx$ parameter might be expected due to the \adi{} self-subtraction at the minor axis. Additionally, because of our $2\times2$ binning and the steep inclination of the disk, the model disk can move about 4 au in $dx$ before it shifts the width of a pixel in our final (projected) forward model images. Coupling this with the many opportunities for sub-pixel shifts in the forward modeling pipeline (convolution, image rotation, etc.), the $dx$ parameter will be less constrained. In any case, we report stellocentric offsets consistent with the results presented in \cite{chen_multiband_2020}.

We include the results for the best-fit disk model geometry parameters with respect to the disk plane in Table \ref{tab:disk_plane_params}. Comparing our results to those reported in \cite{chen_multiband_2020}, we note good agreement between our values and their Table 3 with the exception of pericenter. Though our values for pericenter do not fall within the uncertainties, we echo the point made by \cite{chen_multiband_2020} that the near circular nature of the disk makes this value very sensitive to small offsets. Our values for the semi-major axis of the disk in the second column ($R_c$ parameter) are with respect to a distinctly different part of the disk and are better compared to \cite{chen_multiband_2020}'s results from a geometric ellipse-fitting as shown in their Table 1. As with \cite{chen_multiband_2020}, our fitted disk eccentricities are lower than the values reported in \cite{milli_near-infrared_2017, milli_optical_2019} and \cite{olofsson_dust_2019}. The exact sources for these discrepancies are not known, but we postulate that our modeling methods might have a significant effect on the best-fit eccentricities of the disk models. For example, we model an offset circle in the disk plane, consistent with \citet{chen_multiband_2020}, rather than a true ellipse as was done in \cite{olofsson_dust_2019}. Additionally, the analyses conducted by \cite{milli_near-infrared_2017, milli_optical_2019} and \cite{olofsson_dust_2019} did not perform forward modeling to estimate the effects of PSF subtraction on the disk image.

\subsection{Radial Profile}
\label{sec:discuss-radprof}

Our best-fit inner and outer power laws indicate that the disk assumes the morphology of a thin ring with a very steep inner edge as found by prior works \citep[e.g.,][]{schneider_stis_2009}. The inner power law $\alpha_{in}$ assumes a very steep value of 40---50 across all datasets. If we calculate the vertical scale height from the best-fit \adi values for the aspect ratio we get $0\farcsec018$, $0\farcsec016$, $0\farcsec011$, and $0\farcsec011$ for \textit{g'r'i'z'}, respectively. However, comparing these values with the expected FWHM (i.e., $\lambda/D$) of the PSFs at \textit{g'r'i'z} (0\farcsec017, 0\farcsec020, 0\farcsec024, and 0.029 respectively) we see that the vertical scale height of the disk is likely unresolved especially given that we binned our images 2x2 (pixel scale $0\farcsec012$). As part of our preliminary exploration of the disk modeling, we tried the disk model as described in \cite{millar-blanchaer__2015} which fits an infinitely-steep inner edge followed by a dust density fall-off described by a single power law. However, even though the scale height of the disk is likely unresolved, we observed worse fits with the single power law disk model compared to the model from \cite{ren_exo-kuiper_2019} that fits the dust density profile with a broken power law.

One theory for the steep inner edge of the ring is an unseen massive planet shepherding the dust. This idea, coupled with the measurable eccentricity of the dust grains (see above, Section \ref{sec:discuss-offsets}) lends compelling evidence for an unseen planetary companion influencing the dynamics of the dust which is a concept that has been discussed in detail (\citealt{milli_near-infrared_2017}; \citealt{chen_multiband_2020}; \citealt{rodigas_morphology_2014}; \citealt{arriaga_multiband_2020}). Deriving planetary detection limits for these data is one way to advise these exciting ideas, but is something we encourage for a future study as it is beyond the scope of this manuscript.

The residuals maps in Figure \ref{fig:diskfm_all} reveal a diffuse halo of disk flux that cannot be described by our SCL disk model. This halo is brightest in the residuals map at \rband{}, but is also seen to a lesser extent in all other passbands in our KLIP-ADI image residuals. This halo is very difficult to make out by eye in the residuals made from our KLIP-RDI images, but we point out that these data were the result of about 45 minutes of integration time (Table \ref{tab:obslog}). This integration time is substantially less than the $\sim2$ to 2.5 hours of integration time contained in our \adi{} images which is probably the reason for the general discrepancies in the residuals between the data from the two observing modes. A detailed look at the \rband{} residuals and the widths of radial slices are shown in Figure \ref{fig:residuals}. The colorbar for this figure was given a harder stretch versus the one in the main DiskFM results figure (Figure \ref{fig:diskfm_all}) for a better look at the faint halo of underfit disk flux. To illustrate the goodness-of-fit for the forward model to our reduced image across several points in azimuth, we deprojected both the \rband{} KLIP-ADI disk image and the best-fit forward model and extracted profiles at the four annotated locations in the top panel of the figure. These locations correspond to $45^{\circ}, 180^{\circ}, 225^{\circ}$, and $305^{\circ}$ counter-clockwise from the major axis of the disk. The poor fit between the forward model and the reduced image data from Position 1 in this plot stems from the brightness asymmetry between the north and south ansae as previously mentioned (see Section \ref{sec:results_dust}). We also see a similar discrepancy in Position 4 which is also in the north ansa. Position 3 represents the region of the disk with the best agreement between our data and the forward model and Position 2 shows a disk flux overestimation resulting in a distinct over-subtraction artifact. One possible explanation for these residuals is the extended halo of dust detected by \cite{schneider_hr_2018} using HST. From their images, this halo appears to be brightest toward the northeast. \cite{olofsson_dust_2019} attempted to model these results (see their Figure 7) and were able to reproduce the general shape of the extended halo through N-body simulations of dust grains interacting with the local interstellar medium. If our faint halo of residuals is due to this halo of small grains, it would make sense that we are only detecting them close to the bright ring (i.e., the birth ring) where their density would still be relatively high. Looking at the right panel of \cite{olofsson_dust_2019}'s Figure 7, we can begin to qualitatively understand the origin of the northeast to southwest ansa asymmetry which is readily seen in Figure \ref{fig:residuals}.

\subsection{Potential dust clump}
\label{sec:subsec_clump}

Our optimized freeform models are one way to infer spatial structure in the dust finer than a resolution element since we optimize the model before convolution with the instrument PSF and forward modeling through KLIP. As described in Section \ref{sec:results-freeform}, we took our ensemble of fitted freeform models and averaged them to better study the distribution of small-scale features in the dust (Figure \ref{fig:freeform_averaged} and Figure \ref{fig:deconvolution}). To perform the averaging operation, we first divided each model by its maximum pixel value to remove the effects of the color of the dust in the averaged freeform model. Importantly, while we still see remnants of strong speckle noise near the minor axis of the disk (marked by white arrows), the clump-like feature is far enough away from the star that any stray speckles are very likely to average out between the 6 total datasets. Furthermore, artifacts stemming from the edge of the AO control region are also unlikely as the radial distance of the edge of the control region changes with wavelength (this clump feature is well-outside of the control region at all MagAO-X bands bluer than \zband{}). We also note that this feature is seemingly at the same location across all four passbands so effects from diffraction are also unlikely to be at play.

If we consider the other planetary signposts in the disk such as the nonzero eccentricity of the dust, the very sharp inner edge, and the brightness asymmetry between the ansae, one conclusion is that the existence of this dust clump stems from an unseen companion. The same conclusion was also presented in \cite{booth_clumpy_2023} for the clumpy disk around $\epsilon$ Eridani. If the clumps around $\epsilon$ Eridani and \hr{} are real, a likely explanation for these disturbances is resonances induced by a massive companion somewhere in the system. Indeed, we have observed direct evidence of this in our own solar system through the resonances induced by Neptune in the Kuiper Belt \citep{clement_stability_2021}. An additional example is the outer Adams ring around Neptune, which shows a train of bright ``arcs" signaling discrete areas in a portion of the ring where the dust is piling up, likely due, in part, to moon \textit{Galatea} (e.g., \citealt{madeira_numerical_2022}; \citealt{giuliatti_winter_neptunes_2020}). Although only one highly-localized clump is clearly seen in our freeform models, it could be the case that other unresolved clumps exist in other parts of the disk that present unfavorable viewing geometries (e.g., they might be on the back side). An alternative explanation for this feature is that we are detecting the aftermath of a catastrophic collisional or breakup event involving one or more large bodies. This was postulated for the HD~141569A inner disk due to the highly asymmetric brightness distribution \citep{singh_revealing_2021}. However, since the clump is very compact, this collisional event would have to have happened very recently meaning sufficient time has not passed to allow the dust to spread out along the ring. In any case, we encourage dynamical simulation work to help cement the origin of this enigmatic feature.

\subsection{Color}
\label{sec:discuss-color}

Photometric points representing the average across the whole disk in Table \ref{tab:photometry} show higher values than those extracted from the ansae only because the whole-disk photometry includes the very bright forward-scattering at the minor axis. Because of this, we refrained from plotting these points in Figure \ref{fig:photometry} as they might be initially misleading. Furthermore, our RDI point at \zband{} is slightly higher than the point from the KLIP-ADI image, but it is still within error. When taking the unsaturated calibration PSF images for this dataset there was an error in the camera parameters such that the counts from the star (and thus the calibration speckles, too; see Section \ref{sec:data_reduction}) were very low. Thus, we attribute these low quality calibration PSF images as a probable source of this discrepancy.

Our extracted disk photometry at the ansae show excellent agreement with the data from prior works (\citealt{debes_complex_2008}; \citealt{schneider_stis_2009}; \citealt{rodigas_morphology_2014}; \citealt{milli_near-infrared_2017}) and we recover a red spectral slope for the dust. These prior analyses have noted the red spectral slope for the dust, and \cite{debes_complex_2008} attribute the color to organic materials such as tholins which, interestingly, are seen on many bodies in our outer solar system (e.g., \citealt{stern_initial_2019}; \citealt{emery_tale_2024}). In contrast, \cite{kohler_complex_2008} found that scattering from porous grains rather than the composition could cause the red slope. Findings by \citet{mulders_why_2013} and \citet{tazaki_effect_2019} suggest that compact aggregate grains of typical compositions with a large size parameter (i.e., the ratio between the grain size to the wavelength) can explain a red-colored debris disk, supporting the aforementioned idea. These degenerate solutions underscore the need for testing advanced grain models on this system. We discuss our SPF results and the implications for the properties of the dust grains in the next section.

\section{Scattering phase function}
\label{sec:spf-discussion-sec}

\subsection{Modeling results}
\label{sec:spf_discuss-results_subsec}

\subsubsection{Henyey-Greenstein solutions}
\label{sec:discuss-spf-hgfits}

Our HG SPF model solutions for our \adi{} images recover a very strongly forward-scattering SPF with a moderate amount of back-scattering (Figure \ref{fig:spf_hg}). We emphasize that the HG SPF is a mathematical function widely used to model SPFs because of the small number of free parameters needed, but it often disagrees in shape significantly with empirically-measured SPFs from other debris disks and dust populations in the solar system \citep{hughes_debris_2018}. In any case, one interesting trend that we observe with our best-fit HG solutions is a muting of the height of the peak of the SPF with bluer bandpasses. This effect is also seen in the polarized SPF curves reported in \cite{arriaga_multiband_2020} and \cite{milli_optical_2019}.

\subsubsection{Custom SPF solutions} 
\label{sec:discuss-spf-customfits}

\begin{table}[htbp]
\centering
\caption{Best-fit Legendre polynomial coefficients to construct the SPF for \iband{} and \zband. The uncertainties for each coefficient are shown the in corner plots in Appendix \ref{sec:app_mcmc}}
\label{tab:coefficients}
\begin{tabular}{lcc}
\hline
$a_k$  & \iband{} & \zband{} \\ \hline
$a_0$  & 1.620  & 1.615  \\
$a_1$  & -0.721 & -0.965 \\
$a_2$  & 1.732  & 1.088  \\
$a_3$  & -2.451 & -2.605 \\
$a_4$  & 1.071  & 0.757  \\
$a_5$  & -2.053  & -2.032 \\
$a_6$  & 0.558  & 0.705  \\
$a_7$  & -1.422 & 1.116  \\
$a_8$  & 0.092  & 0.784  \\
$a_9$  & -0.671 & -0.293 \\
$a_{10}$ & -0.386  & 0.938  \\
$a_{11}$ & 0.114 & 0.120  \\
$a_{12}$ & -0.151  & 1.004  \\
$a_{13}$ & 0.289 & 0.032  \\
$a_{14}$ & 0.217  & 0.674  \\
$a_{15}$ & 0.162  & -0.087 \\
$a_{16}$ & -0.262 & 0.340  \\
$a_{17}$ & ---    & -0.173 \\
$a_{18}$ & ---    & 0.229  \\ \hline
\end{tabular}
\end{table}

Our custom-fit SPFs made from a basis of the Legendre polynomials (see Section \ref{sec:spf_description} for details) reveal a taller forward scattering peak at $i'$-band compared to $z'$-band at scattering angle $\theta \sim 13^{\circ}$ (Figure \ref{fig:spf_custom}). However, this is likely the effect of the Legendre-based model fitting to under-subtracted starlight at the minor axis of the disk. Further evidence of this notion is given by the disconnect in freeform SPF solutions at \iband{} between the \adi{} and \rdi{} datasets, and the lower amount of forward scattering in the freeform SPFs in general. One could attempt to probe for chromatic effects by generating SPF curves from simulated agglomerated dust particles (see Sec. \ref{sec:spf-shape-discuss_subsubsec}; \citealt{arnold_stumbling_2022}) for a variety of grain sizes and grain size distributions at our 4 bandpasses of interest, but this is beyond the scope of this study. Additional details for our Legendre SPF modeling results can be found in Appendix \ref{sec:app_mcmc}.

\subsubsection{SPFs extracted from the freeform disk models}
\label{sec:discuss-spf-freeform}

Recall, that through the efficiency of \textsc{Jax}, we were able to make use of datasets twice the size as those used for the physical disk modeling (see Section \ref{sec:freeform-describe-subsec}). Despite the higher quality for these images, there were lingering noise artifacts affecting the backside (faintest) part of the disk. These noise artifacts are likely quasistatic speckles which can persist for hours during an observation (see, e.g., \citealt{martinez_speckle_2013} and \citealt{males_mysterious_2021} for details on this type of speckle noise) and the WDH. As an example, the sharp dip at scattering angles $\sim140$ to 160 degrees for the \textit{south side} SPF at \iband{} and \zband{} in Figure \ref{fig:spf_freeform} is likely due to the WDH. We note that these images were recorded simultaneously, so a noise feature in the disk due to the WDH is expected to be coincident. Additionally, we suspect the WDH because we see a region of oversubtraction consistent with the lobes of the WDH in a roughly North-South orientation in these images (see Figure \ref{fig:finalklip_adi}). While the double-lobed WDH also crosses the north half of the disk, it does so on the forward side which is not as faint as the back-side of the disk that is affected in the south half. The images at \gband{} also were subject to strong WDH artifacts which is part of the reason for the greater uncertainties in that freeform SPF. The current state of the freeform forward modeling pipeline is such that there is a tradeoff between regularization toward physical features in the disk and noise artifacts (Kueny et al. 2025, submitted). Work is ongoing to better characterize speckles, the WDH, and background artifacts to improve inference of the intensity distribution due to the disk. In any case, we recover some interesting features in the SPFs extracted from the freeform models.

When comparing the extracted freeform SPFs with our HG solutions, the freeform SPFs suggest that the HG models overestimate the brightness at scattering angles $\lesssim40^\circ$. This is likely because the HG SPF is not ideal for fitting the observed brightness of \hr{} and other model parameters, such as the stellar offsets, are drifting to account for this. Additionally, the \adi{} freeform SPFs tend to exhibit a lower amount of forward scattering compared to the SPFs associated with the \rdi{} data. It is not known whether this is due to biases stemming from the \adi{} self-subtraction or residual stellar halo pinned to the bright forward part of the disk in our \rdi{} images.

We note that our freeform models represent the deconvolved image of the disk; one could theoretically extract the disks's azimuthal brightness distribution using thin slits with widths on the order of a pixel. We attempted this using thin slits distributed along the deprojected freeform disk models. The slit dimensions were $0\farcsec012$ in width and $0\farcsec1$ in length, positioned such that the long axis of the slit was oriented in line with the image center. However, we did not observe any significant deviations from the overall shapes of the lower-resolution freeform SPFs shown in Figures \ref{fig:freeform_averaged} and \ref{fig:spf_freeform}, so we do not include them in this manuscript. However, we mention these ideas since they could be used in future grain modeling studies where higher-resolution SPFs can be useful.

\subsection{Implications for the Composition of the Dust}
\label{sec:composition_subsec}

\subsubsection{Recent efforts to model the dust in the HR 4796A disk}
\label{sec:spf-pastmodeling-discuss_subsubsec}

The VLT/SPHERE $H$-band study of the \hr{} disk by \citet{milli_near-infrared_2017} empirically measured the SPF of the disk from $13.6^{\circ}$ to $166.4^{\circ}$ for the first time. To fit their impressive degree of forward-scattering of the dust, they concluded that the dust grains would have to be quite large (i.e., on the order of tens of microns). However, it is important to note that their models used Mie theory (solid spheres) and a Distribution of Hollow Spheres \citep[DHS;][]{min_modeling_2005,min_multiwavelength_2016} to approximate the scatterers, which are too simple to accurately portray the behavior of realistic extrasolar dust grains, as pointed out in their conclusions. A study by \cite{brunngraber_influence_2017} found that using compact spheres to model dust grains that are, in actuality, fluffy or aggregate in nature leads to overestimation of the grain size power law distribution and minimum grain size. This implies that aggregate grains with significantly smaller radii compared to Mie or DHS grains offer an explanation for the results of \cite{milli_near-infrared_2017}. The conclusion of this study was that more advanced grain modeling is needed to fit both the SPF and the spectrum.

Building upon the work of \cite{milli_near-infrared_2017}, \cite{chen_multiband_2020} studied the \hr{} disk using an approach similar to ours in that they implemented a \textsc{DiskFM} routine to put constraints on the physical properties of the disk. They used a simple SCL disk model with a 2-parameter HG SPF in their model fitting. \citet{chen_multiband_2020} converged to HG SPF solutions similar to ours (Figure \ref{fig:spf_hg}) featuring a strong forward peak for scattering angles $\theta \leq 45^\circ$ and a moderate degree of back-scatter beyond $\theta \geq 120^\circ$. In addition to their \textsc{DiskFM}-based modeling, \cite{chen_multiband_2020} performed a reanalysis of the $H$-band total intensity image from SPHERE using methods consistent with those reported in \citet{milli_near-infrared_2017}. Namely, \citet{chen_multiband_2020} modeled their HG SPF solution using a DHS model with \texttt{MCFOST} \citep{pinte_monte_2006} and Bayesian inference to assign limits to the chemical composition of the dust. The specific parameters probed by \citet{chen_multiband_2020} were minimum grain size ($s_{min}$), size distribution power law $\nu$, porosity $P$, maximum void fraction $V_{max}$, amorphous silicate ($f_{Si}$), amorphous carbon ($f_{c}$), and metallic iron ($f_{Fe}$). They found that their SPF led to a best-fit chemical composition of $f_{Si}\approx42\%$, $f_c \approx 17\%$, and $f_{Fe} \approx 37\%$ with a best-fit minimum grain size of $s_{min} \approx 25\mu$m, maximum void fraction $V_{max} \approx 76\%$, porosity $P \approx 15\%$, and power law size distribution of $\nu \approx -3.74$.

\citet{rodigas_morphology_2014} showed that spherical grain models do not fit both the scattered light and the emission from the disk. These findings were reinforced through the analysis of ALMA images of \hr{} by \citet{kennedy_alma_2018}. Mie spheres show spectral and SPF resonances that are unphysical and trends not observed in laboratory experiments or in astrophysical data; hence the need for a more realistic grain model.

\subsubsection{Inferred SPF shape}
\label{sec:spf-shape-discuss_subsubsec}

\begin{figure}
    \centering
    \includegraphics[width=0.75\linewidth]{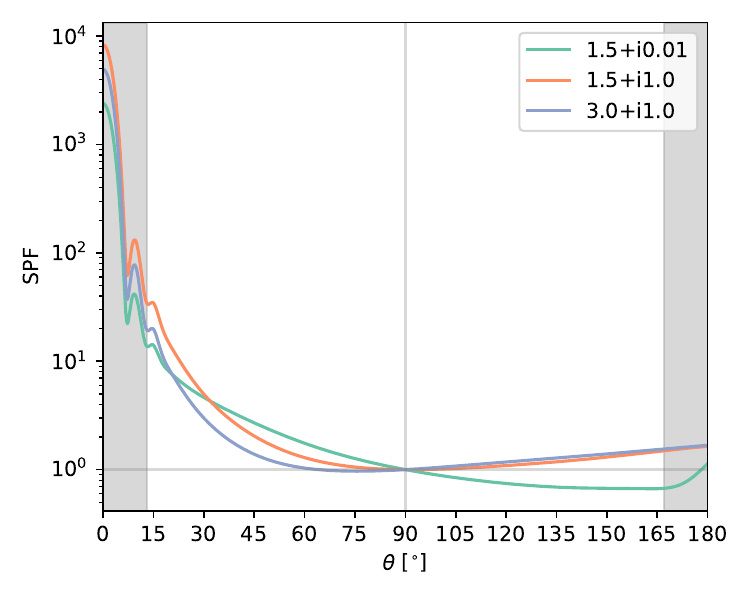}
    \caption{SPFs for irregular dust particles from \texttt{glitterin} adopting a single size parameter $x = (2\pi r)/\lambda = 35.0$ where we have varied the complex index of refraction as labeled in the plot legend. The shaded regions are $13^{\circ}$ away from $\theta=0^{\circ}$ and $180^{\circ}$ which are not observable based on the inclination of the disk. The SPF calculated using the particles with the highest amount of absorption (blue line) most-closely align with both the forward-scattering peak and location of the minimum SPF value that we observe in Figure \ref{fig:spf_freeform}. The back-scattering half of the blue and orange lines (high imaginary refractive index) are also close in shape to our measured SPFs. These simulations suggest the scattering in the \hr{} disk is dominated by highly-absorptive particles that are at least a few microns in size.}
    \label{fig:daniel}
\end{figure}

Our custom Legendre-based and freeform SPFs display a distinct angular behavior that differs from previously reported phase functions for \hr{} and debris disks more broadly (e.g., \citealt{hughes_debris_2018}; \citealt{engler_hd_2020}; \citealt{desgrange_dust_2025}). We recover a minimum at $\sim 60^\circ$ to $65^\circ$ and a steady, monotonic rise from that minimum through the largest accessible angles at $g'r'i'z'$. This behavior appears in both the parametric Legendre basis fits and the non-parametric freeform reconstructions. The Discrete Dipole Approximation (DDA; \citealt{purcell_scattering_1973}; \citealt{draine_discrete-dipole_1988}) lends a promising solution to realistic modeling of the SPFs. For an arbitrarily-shaped scatterer, the DDA facilitates construction of a grain from an array of polarizable dipoles which can better mimic the scattering properties of realistic dust grains \citep{zubko_validity_2010}. One caveat is that the computational time needed to calculate the scattering properties of these models grows exponentially as the size parameter of the grain increases. However, due to the recent advances and availability of computational resources, modeling preferences are beginning to shift from approximations like solid or hollow spheres to those based on the DDA \citep[e.g.,][]{pawellek_debris_2024} as calculations involving grains with larger size parameters are beginning to come to fruition \citep{lin_glitterin_2025}. We note that our new freeform pipeline forward models the disk without assuming anything about the grains, so these results can be leveraged for simulating grains without the \textsc{DiskFM} operations.

We utilize \texttt{glitterin}, which is a neural network trained on scattering of irregular grains using the Discrete Dipole Approximation \citep{yurkin_discrete-dipole-approximation_2011, lin_glitterin_2025}, to identify the types of particles that could reproduce our measurements. We find that the general features of the \hr{} SPF are naturally reproduced by irregular particles with radii of at least several microns that are also highly absorptive (i.e., have large imaginary refractive index, $k$; Figure \ref{fig:daniel}). For these simulations, we fixed the size parameter of the particle (i.e., the ratio of the radius of the particle to wavelength of observation) to $x=35$, which corresponds to $\sim 3.9$~$\mu$m at $\lambda=700$~nm, and vary the complex refractive index $m$. When $m=1.5+0.01i$, just as an example of moderate real ($n$) but low imaginary ($k$) index (green line), the SPF is a monotonic decrease from $\theta=15^{\circ}$ to $165^{\circ}$, which is significantly different from what is observed (Figure \ref{fig:spf_freeform}). As a result, we further explore $m=1.5+i$ and $3+i$ to demonstrate the trends when increasing the refractive index. 
The $m=1.5+i$ case demonstrates that the gradually increasing back-scattering region ($\theta>90^{\circ}$) requires an increase in the imaginary part. When increasing the real part to $m=3+i$, the minimum of the SPF shifts to smaller $\theta \sim 65^{\circ}$, while maintaining the monotonic increase to $\theta=180^{\circ}$. The exploration here suggests that the observed SPF requires a fairly high $k$ compared to silicates and a modest increase in $n$. 

These results are broadly consistent with prior studies conducted primarily in the near-infrared (\citealt{chen_multiband_2020}; \citealt{arriaga_multiband_2020}) that also require relatively high $m$ by incorporating higher volume fractions of carbon and iron. Since our analysis involves images of the disk at visible wavelengths, these results provide complementary constraints on the size and composition of the grains. The rapid oscillations at $\theta<15^{\circ}$ are due to diffraction that would be averaged out when considering a size distribution. We highly encourage future efforts towards a more quantitative analysis to place stricter constraints on the sizes and composition of the dust utilizing SPFs from this and previous works.

Subtle differences in slope between the SPFs extracted for the north- and south-halves of the disk may indicate real azimuthal variations in grain species. Note that this type of extraction is one advantage associated with our freeform models as it does not assume a fixed parametric form for the disk structure or SPF. A simple assumption would be that the two sides of the disk have homogeneous grain populations as the orbital timescale at 77 AU (several hundreds of years) is much less than the age of the star ($\sim10$ Myr). In this case, the difference between the SPFs from the separate halves would have to be attributed to residual systematic error. However, all our \rdi{} and \adi{} data are quite consistent in displaying a shallower slope for the southern half. This could be interpreted as a difference in the size or composition of the grains. This is perhaps consistent with the clump observed in the southwest, which if real, would indicate that there are large individual collisions happening frequently and not yet smoothed out. Alternatively, these variations may reflect residual systematics arising from quasistatic speckles, WDH artifacts, or spatially varying throughput in the PSF-subtraction process. Distinguishing between astrophysical and instrumental origins will require both improved calibration of spatially correlated noise and future observations through slower turbulence.

In sum, the distinct SPF morphology we recover suggests a dust population dominated by large, absorptive, aggregate grains whose scattering behavior departs strongly from classical Mie or DHS. These results provide valuable empirical targets for future efforts in grain scattering simulations. As modeling capabilities grow, the SPF shape we report here will serve as a useful constraint on the physical nature of the dust around \hr{} and extrasolar dust in general.

\section{Summary}
\label{sec:summary}

We observed \hr{} during three epochs in 2023A, 2024A, and 2025A using MagAO-X at \gband{} (527 nm), \rband{} (615 nm), \iband{} (762 nm), and \zband{} (908 nm). We detected the disk at high signal-to-noise in all passbands, offering a glimpse of the complex dust dynamics occurring in this system. For the 2024A epoch, we observed the disk using ``star hopping" \citep{wahhaj_search_2021} and subtracted the PSF using the KLIP-RDI algorithm to detect the entirety of the disk including the highly-coveted forward-scattering peak; these are the bluest images of the whole disk at the time of writing. We passed our reduced images through many variations of a forward modeling pipeline based on the DiskFM \citep{mazoyer_diskfm_2020} open-source Python tool. Most of our modeling efforts entailed fitting SCL disk models to our data using a MCMC framework to tightly constrain the geometrical parameters of the disk. Since our KLIP-RDI images offered a clean detection of the disk's minor axis, we also implemented SPF modeling by tying our SCL disk forward modeling routine to a custom SPF constructed with a basis of the Legendre polynomials. The final variation of our modeling efforts resulted in the implementation of a \textsc{Jax}-based freeform disk modeling routine to fit a distribution of pixel intensities to our data, permitting SPF extraction for our \adi{} images despite heavy PSF subtraction artifacts. Our open-source freeform modeling pipeline \texttt{ffortissimo} (Kueny et al. 2025; submitted) outputs the deconvolved image of the disk and is able to fit complex brightness distributions.

Our modeling efforts revealed:

\begin{itemize}
    \item A thin ring with a sharp inner edge and eccentric nature, which potentially points to an undetected massive planet sculpting the ring.
    \item Confirmation of a complex azimuthal and radial dust distribution that is not well-described by our parametric SCL disk models.
    \item Confirmation of a red spectral slope for the dust at visible wavelengths.
    \item Empirically-measured SPFs at $g'r'i'z'$ from scattering angles $13.6^{\circ}$ to $166.4^{\circ}$ that show a high degree of forward-scattering and are distinct from the best-fitting HG SPF solutions.
    \item SPF shapes that suggest the scattering is dominated by highly-absorptive, large irregular grains that are several microns in size.
    \item A clump-like artifact in the southwest portion of the disk visible in all filters and at all epochs (2023---2025) that could be additional indirect evidence for an unseen planet or the result of a very recent collisional breakup event.
\end{itemize}

Our results have unveiled both the advantages and the present limitations associated with direct imaging of debris disks in scattered light. The complex scattering behavior that we detected is hard to model using classical or simplified techniques. Additionally, our measured SPFs reaffirm that real extrasolar grains scatter unlike any theoretical models and that accurate inference will depend increasingly on data-driven techniques and/or highly-complex grain models. In the near future, the next generation of extremely large ground-based telescopes (e.g., the GMT) will offer the sensitivity and angular resolution needed to measure SPFs at even smaller scattering angles and to detect subtle structures in the dust, enabling a deeper physical understanding of debris disks and their composition.

\section*{Acknowledgements}

J.K.K., J.R.M., Y.D.L., and A.J.W. acknowledge support from the NSF, grant no. AST-2307613. J.D.L. was supported by the Flatiron Software Research Fellowship at the Flatiron Institute, a division of the Simons Foundation.

We thank Prof. Ewan Douglas and the University of Arizona Space Research Lab for access to computing resources during critical phases in our modeling. These computing resources were supported by funding by generous anonymous philanthropic donations to the Steward Observatory of the College of Science at the University of Arizona.

This work used Jetstream2 at Indiana University through allocation PHY250222 from the Advanced Cyberinfrastructure Coordination Ecosystem: Services \& Support (ACCESS) program, which is supported by National Science Foundation grant nos. 2138259, 2138286, 2138307, 2137603, and 2138296.

This material is based upon High Performance Computing (HPC) resources supported by the University of Arizona TRIF, UITS, and Research, Innovation, and Impact (RII) and maintained by the UArizona Research Technologies department.

MagAO-X was developed with support from the NSF MRI Award No. 1625441. The Phase II upgrade program is made possible by the generous support of the Heising-Simons Foundation.

\ This work has made use of data from the European Space Agency (ESA) mission
{\it Gaia} (\url{https://www.cosmos.esa.int/gaia}), processed by the {\it Gaia}
Data Processing and Analysis Consortium (DPAC,
\url{https://www.cosmos.esa.int/web/gaia/dpac/consortium}). Funding for the DPAC
has been provided by national institutions, in particular the institutions
participating in the {\it Gaia} Multilateral Agreement.

\ This material is based on work supported by the National Science Foundation Graduate Research Fellowship Program under grant No. 2020303693. Any opinions, findings, and conclusions or recommendations expressed in this material are those of the author(s) and do not necessarily reflect the views of the National Science Foundation.

\ This research has made use of SAO Image DS9, developed by Smithsonian Astrophysical Observatory \citep{joye_new_2003}.

\ This work made use of the \textit{astropy} \citep{astropy_collaboration_astropy_2013} and SciPy \citep{scipy_10_contributors_scipy_2020} software packages.

\ This publication makes use of data products from the Two Micron All Sky Survey, which is a joint project of the University of Massachusetts and the Infrared Processing and Analysis Center/California Institute of Technology, funded by the National Aeronautics and Space Administration and the National Science Foundation.

\vspace{5mm}

\appendix

\section{MagAO-X 2025A Astrometric Solution}
\label{sec:app_astrometry}

\renewcommand\thefigure{\thesection\arabic{figure}}
\setcounter{figure}{0}
\renewcommand\thetable{\thesection.\arabic{table}}
\setcounter{table}{0}

\begin{figure}
    \centering
    \includegraphics[width=0.5\linewidth]{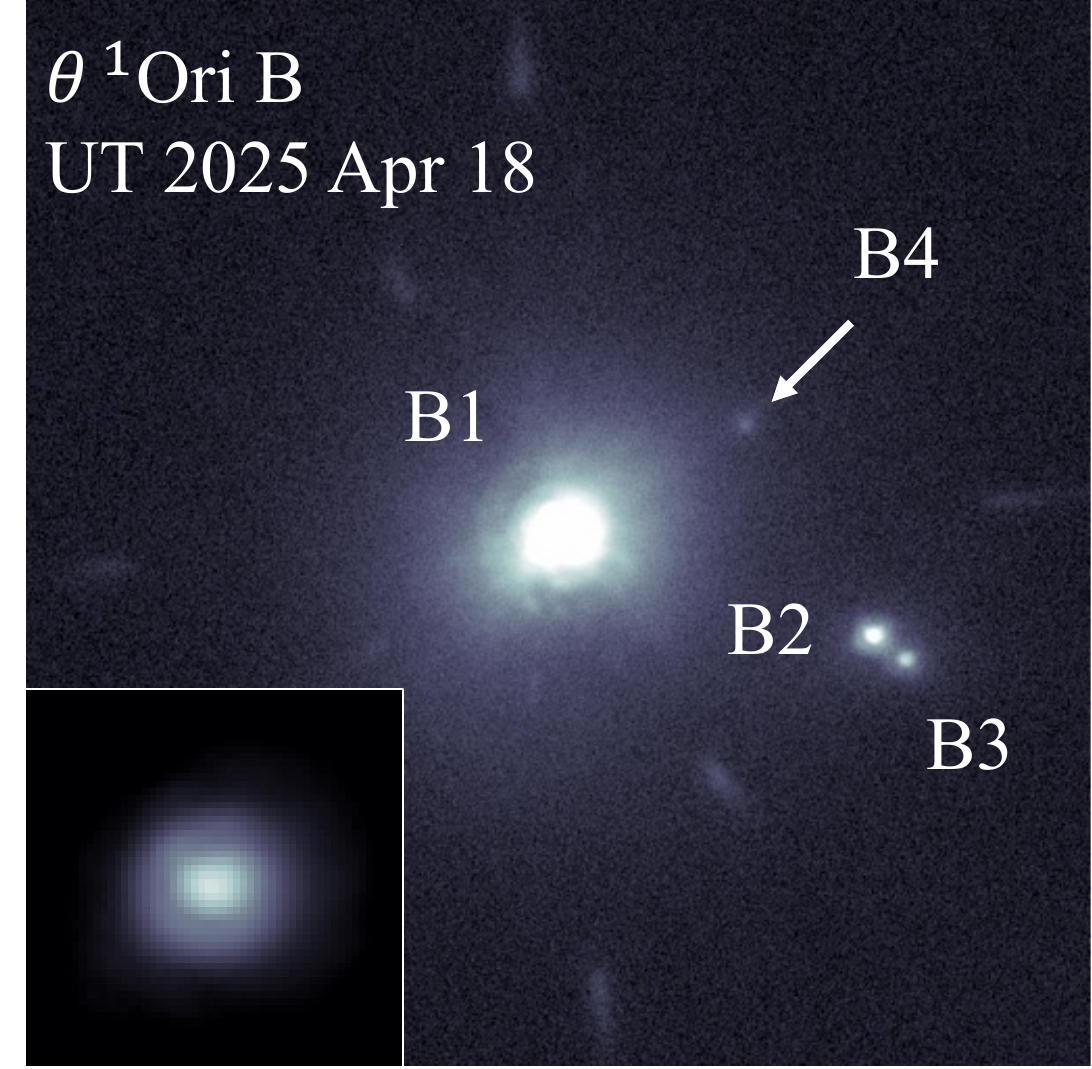}
    \caption{Final reduced image of $\theta^1$ Ori B shown in a log stretch to reveal all members of the cluster (B1 through B4). The radially-elongated sources scattered throughout the image are artifical speckles caused by MagAO-X's DMs. We display the PSF core of B1 with a softer log stretch in the inset image in the bottom left.}
    \label{fig:tetOriB_appendix}
\end{figure}

We observed $\theta^1$ Ori B at \zband{} on the night of UT 2025 Apr 18 from 23:55:00 to 23:55:40 in variable seeing conditions (median seeing of $0\farcsec6$). Here, we describe how we use this image to measure the position angle of $\theta^1$ Ori B2 (B2 hereafter) with respect to $\theta^1$ Ori B1 (B1 hereafter) to determine the angular offset $\Delta \theta$ needed to correct our MagAO-X images taken during 2025A to true north. We collected observations with MagAO-X's camsci1 (see \citealt{long_astrometric_2025}'s Figure 1 for a diagram of MagAO-X's science cameras) with the exposure time set to 4 Hz yielding a total raw image count of 174. To reduce these data, we median-stacked all 174 frames and then determined the pixel coordinates of the brightest star B1 using this stacked image. We then shifted each individual raw frame to establish the measured centroid coordinates of $\theta^1$ Ori B1 as the new image center. Then, we derotated the shifted images by the header value for the parallactic angle. Finally, we stacked all shifted and derotated images by averaging and then cut out a sub-image of size $256\times256$ pixels centered on B1. We resolved all four members of the cluster in our final image cutout which we show in Figure \ref{fig:tetOriB_appendix}.

\begin{figure}
    \centering
    \includegraphics[width=0.45\linewidth]{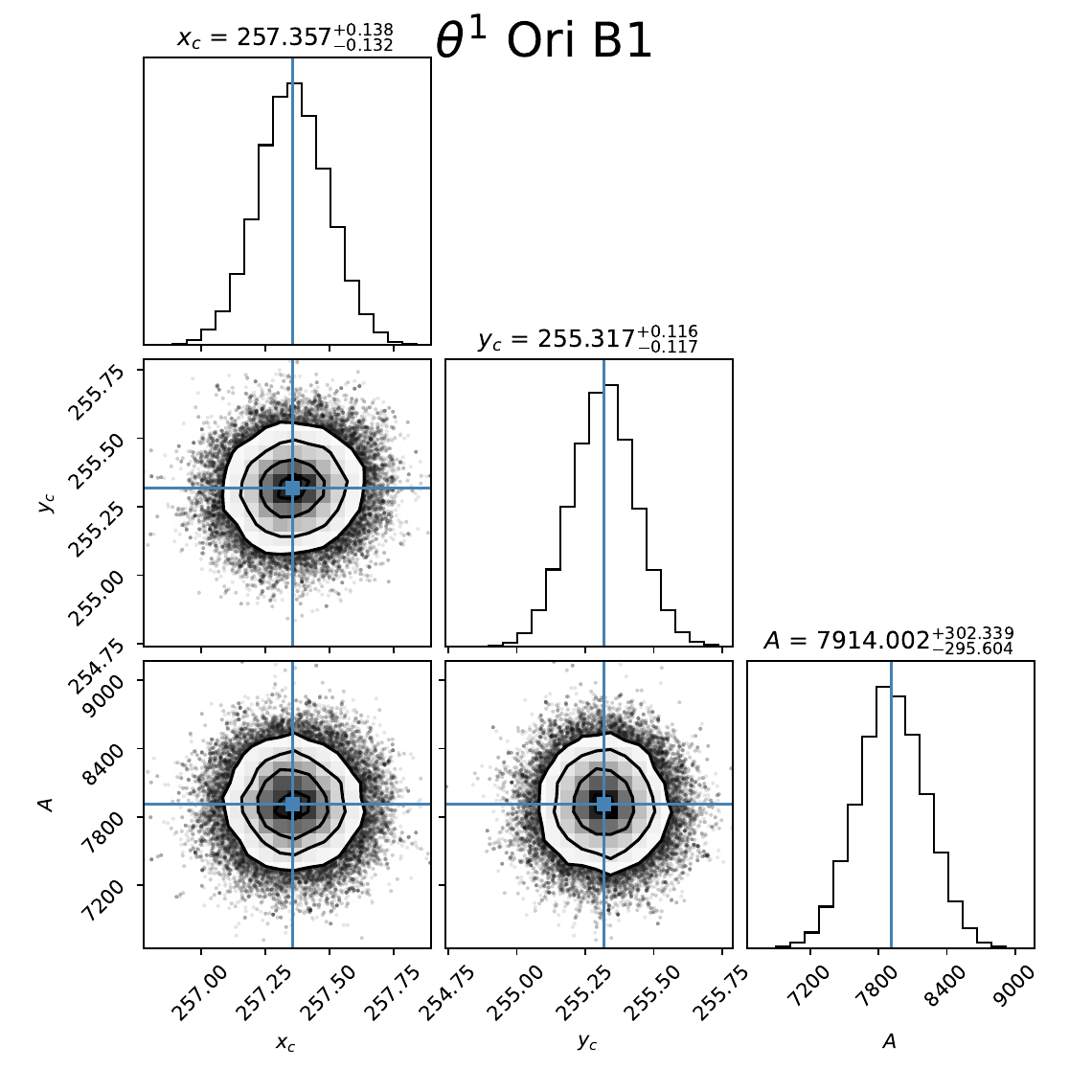}
    \includegraphics[width=0.45\linewidth]{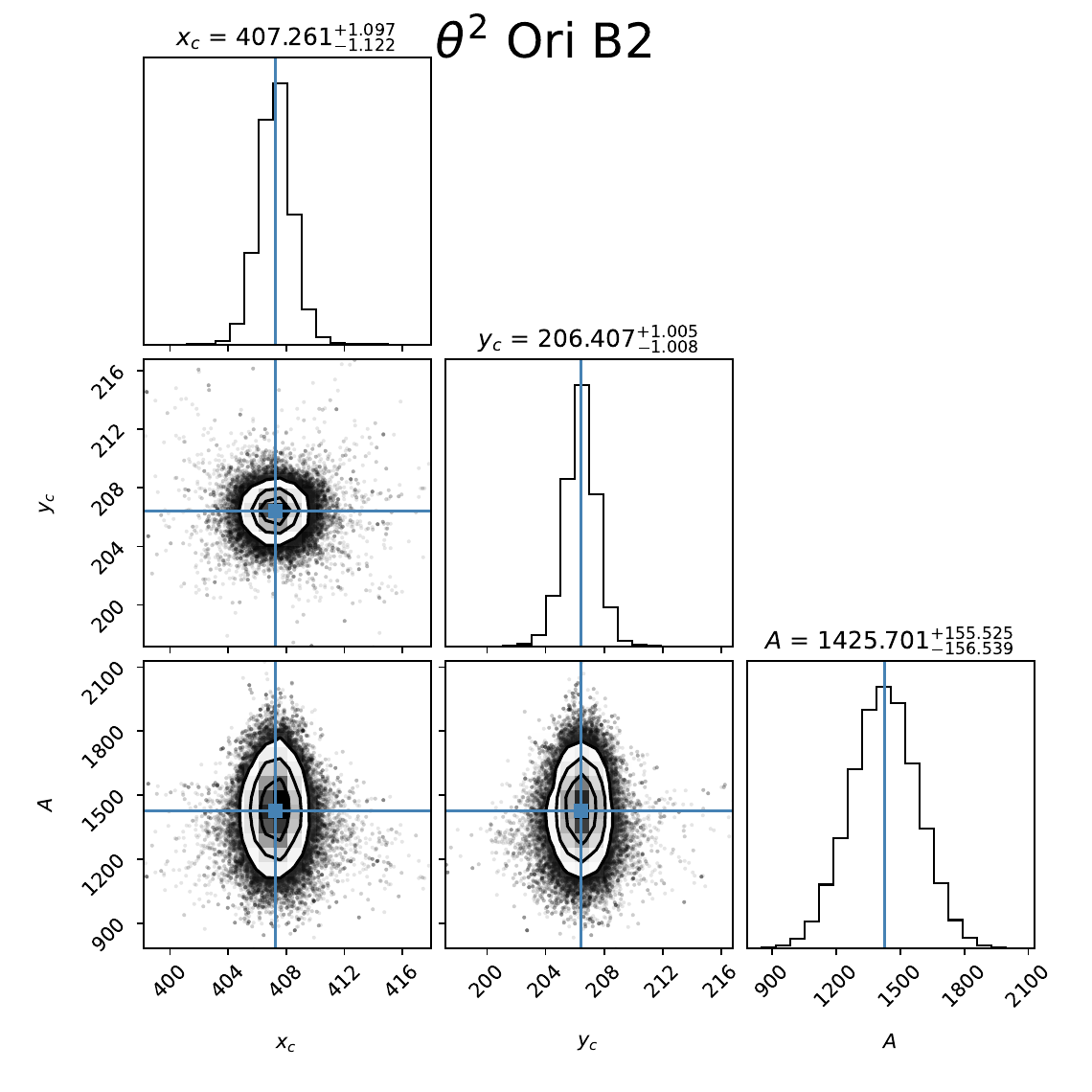}
    \caption{Corner plot detailing our MCMC results for PSF centroid fitting to B1 (left) and B2 (right). Histograms on the main diagonal show the posterior probability distributions for each fitted parameter marginalized over all other fitted parameters where the dashed vertical lines show the 16th, 50th, and 84th percentiles. The off-diagonal scatter plots show the joint probability distributions overlaid with contours also representative of the aforementioned percentiles.}
    \label{fig:corner_B1B2_appendix}
\end{figure}

To determine the uncertainties associated with the pixel coordinates of the PSF centroids for each star, we fitted 2D Gaussian models to each star separately in an MCMC wrapper facilitated by \texttt{emcee}. We chose a 2D Gaussian as our PSF model due to the poor PSF quality in our image brought by the variable seeing conditions. We defined our likelihood function as:

\begin{equation} \label{chisquare}
    \ln \mathcal{L}(\boldsymbol{\Theta|D}) = 
    - \frac{1}{2} \sum_{ij} \left( \frac{D_{ij} - G_{ij}(\boldsymbol{\Theta})}{\sigma} \right)^2
\end{equation}

\noindent where $D$ represents our reduced data, $G$ is the 2D model Gaussian, $\sigma$ is the noise, $\boldsymbol{\Theta}$ is the vector of fitted parameters, and $ij$ are the pixel indices in our fitting ROI. $\boldsymbol{\Theta} = (x_c, y_c, A)$ are our 3 fitted parameters for each star representing the x- and y-coordinates for the peak of the Gaussian $(x_c, y_c)$ and its amplitude $A$. We defined the region of interest where the likelihood was maximized using a circular region with a radius of 18 pixels (about $3\times$FWHM of the PSF) centered on the initial guess for the centroid of the stellar PSF. We estimated the noise $\sigma$ by measuring the noise contributed from the detector in annuli centered on the center of the image and by the measured photon noise associated with each star. For the detector noise estimation for B1, we computed the standard deviation in an annulus with an inner and outer radius of 200 and 240 pixels (40 and 48 $\lambda/D$), respectively. For B2, we used an annulus centered on the radial distance of B2 from B1 because B2 was clearly within the stellar halo of B1. This resulted in using an annulus with inner and outer radii 140 and 180 pixels (28 and 36 $\lambda/D$) from the center of the image, respectively. Both of these background noise measurement annuli coincided with MagAO-X's DM speckle artifacts and other stars in the cluster, so we created a bespoke software mask to exclude them from our standard deviation calculation. We also measured the photon noise for each star by summing the counts in an 18 pixel radius aperture, converting the counts to photons by dividing by the camera gain (2.32 e$^-/$ pixel, from the MagAO-X instrument handbook\footnote{\url{https://magao-x.org/docs/handbook/observers/cameras.html}}), and then computing the square root of this value. We then added the measured background and photon noise for each star in quadrature to use as the noise $\sigma$ in Equation \ref{eq:chisquare} during the MCMC sampling procedure. We show the posterior distributions in the corner plots in Figure \ref{fig:corner_B1B2_appendix}. 

\subsection{Rotation Angle Offset}
\label{sec:app_rotation_offset_subsec}

To calculate the position angle $\theta_{B1-B2}$ of B2 relative to B1, we used: 

\begin{equation} \label{eq:pa_app}
    \theta_{B1-B2} = f(m_{B1}, m_{B2}) = \text{atan2}(\Delta x, \Delta y)
\end{equation}

\noindent using the centroids of the PSFs of $m_{B1} = (257.36, 255.32)$ and $m_{B2} = (407.26, 206.41)$. So $\Delta x = 149.90$ and $\Delta y = -48.91$.

Using Equation \ref{eq:pa_app}, we obtain $\theta_{PA} = \text{atan2}(149.94, -48.93) = 251.93^{\circ}$. Propagating our uncertainties entails addition in quadrature:

\begin{equation} \label{eq:prop}
    \sigma_V^2 = \left(\frac{\partial f}{\partial x}\right)^2  \sigma^2_{B1} + \left(\frac{\partial f}{\partial y}\right)^2  \sigma^2_{B2}
\end{equation}


\noindent where 
\begin{equation}
  \frac{\partial f}{\partial x} = \frac{\partial}{\partial x} \arctan{\frac{\Delta x}{\Delta y}} = \frac{\Delta y}{\Delta x^2+ \Delta y^2}
\end{equation}

\noindent and

\begin{equation}
    \frac{\partial f}{\partial y} = \frac{\partial}{\partial y} \arctan{\frac{\Delta x}{\Delta y}} = -\frac{\Delta x}{\Delta x^2+ \Delta y^2}.
\end{equation}
 
Plugging in our numbers and propagating our uncertainties using Equation \ref{eq:prop} yields $\sigma_V = 0.378^\circ$. This gives our final measured position angle:

\begin{equation}
    \theta_{B1-B2} = 251.93^\circ \pm 0.38^\circ.
\end{equation}

We can compare our $\theta_{B1-B2}$ to the PA values in \cite{close_diffraction-limited_2013} for stars B1 and B2. If we take the last two PA values for stars B1-B2 in \citet{close_diffraction-limited_2013}'s Table 1 and average them, we get $254.60^\circ \pm 0.42^\circ$. We chose to consider these two values specifically due to their lower uncertainties. Comparing this result with our value for $\theta_{B1-B2}$ by taking the difference between them gives:

\begin{equation}
    \Delta\theta_{B1-B2} = 2.58^\circ \pm 0.57^\circ.
\end{equation}

We can compare $\Delta\theta_{B1-B2}$ with the difference in $\theta_{PA}$ values from our \gband{} results that used data collected in 2025A with those presented in \citet{chen_multiband_2020}. \citet{chen_multiband_2020} also performed a comprehensive forward modeling study on \hr{} using \textsc{DiskFM} using total intensity images from GPI/Gemini and SPHERE/VLT. Looking at their values for the best-fit disk PA for the SPHERE IRDIS $H$ image and the GPI $J$ image we see $26.94^\circ \pm 0.04^\circ$ and $26.83^\circ \pm 0.11^\circ$, respectively. We chose to compare these two specific PA values to ours due to the low uncertainty associated with the $H2$-band result and given that $J$-band was collected the most recently in that study. The SPHERE IRDIS $H$ and GPI $J$ data were collected on UT 2014 May 19 and UT 2016 Mar 23 respectively. Taking the difference between these values and our \gband{} result:

\begin{equation}
    \Delta\theta_{g'-H} = 2.67^\circ \pm 0.37^\circ
\end{equation}

\noindent and 

\begin{equation}
    \Delta\theta_{g'-J} = 2.56^\circ \pm 0.39^\circ.
\end{equation}

\subsection{Pixel Scale}

The as-designed pixel scale for MagAO-X is 6 mas pixel$^{-1}$. Since we have estimates of the pixel coordinates of the PSF centroids for B1 and B2, we can estimate the distance between them: $d_{\text{pix}} = \sqrt{\Delta x^2 + \Delta y^2} = 157.680 \pm 5.20e-3$ pixels. We can again make use of the results presented in \citet{close_diffraction-limited_2013} who found a separation of $\theta_{\text{sep}} = 0.941 \pm 1.30e-3$ arcseconds (taking the last two entries in their Table 1 and averaging them as in Section \ref{sec:app_rotation_offset_subsec}). Given these values, we can estimate the 2025A pixel scale:

\begin{equation}
    S = 5.970 \pm 0.195 \text{ mas pixel}^{-1}.
\end{equation}

\section{MCMC Modeling Corner Plots}
\label{sec:app_mcmc}

\renewcommand\thefigure{\thesection\arabic{figure}}
\setcounter{figure}{0}
\renewcommand\thetable{\thesection.\arabic{table}}
\setcounter{table}{0}

\begin{figure}
    \centering
    \includegraphics[width=\linewidth]{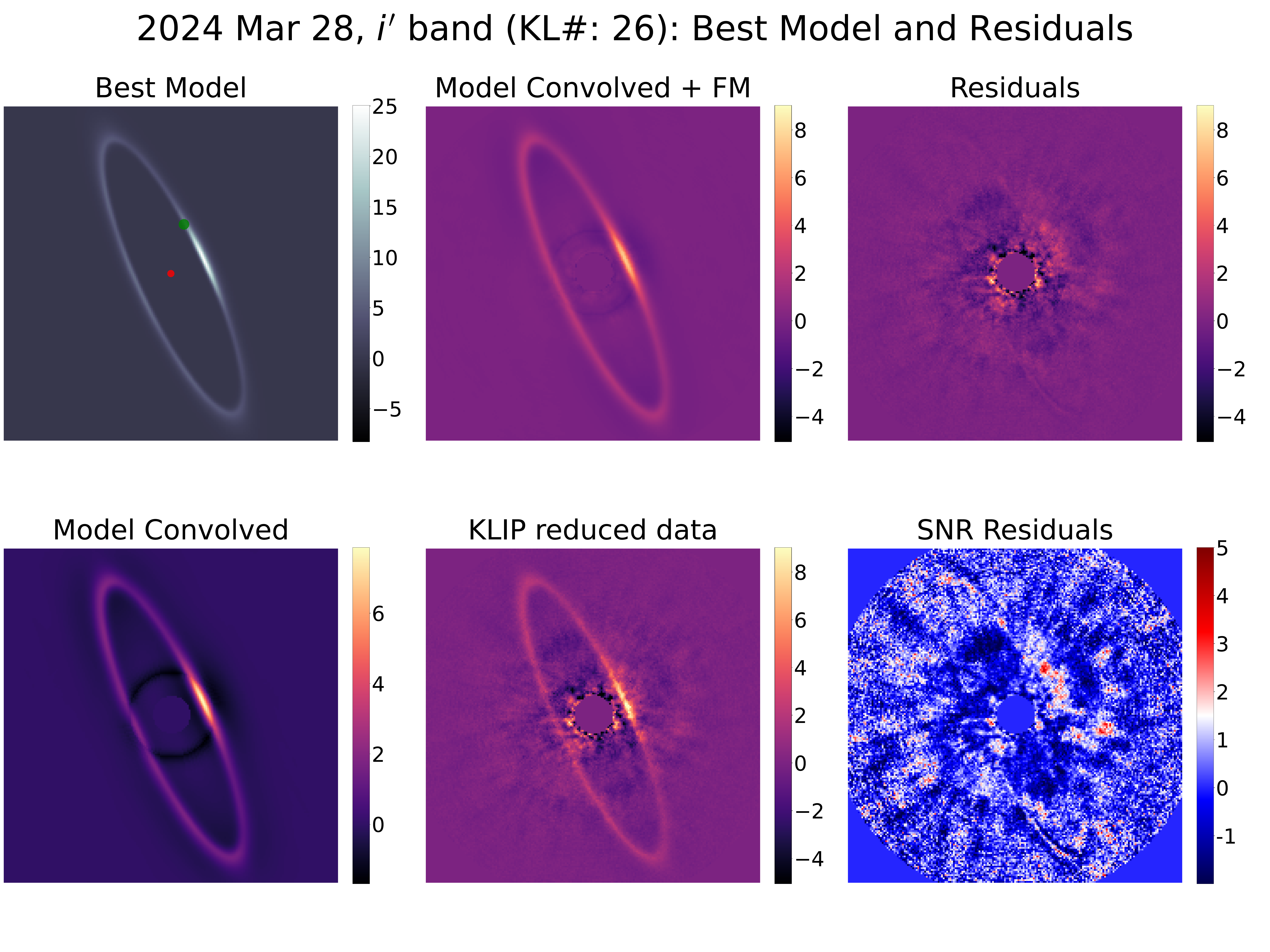}
    \caption{DiskFM results for the \iband{} \rdi{} dataset from UT 2024 Mar. 28. \textbf{First column:} the top panel shows the best-fit parametric disk model after our MCMC run while the bottom panel shows this model convolved with the instrument PSF. The red and green circles correspond to the location of the star and argument of pericenter, respectively. \textbf{Second column:} the top panel shows the best-fit forward model while the bottom panel shows the \rdi{} reduced image used in this MCMC DiskFM analysis. \textbf{Third column:} the top panel illustrates the residuals between the \rdi{} image and the best-fit forward model while the bottom panel shows this residuals map divided by the empirically-measured noise map to show the significance of the residual structure.}
    \label{fig:appendix_diskfm_irdi}
\end{figure}

\begin{figure}
    \centering
    \includegraphics[width=\linewidth]{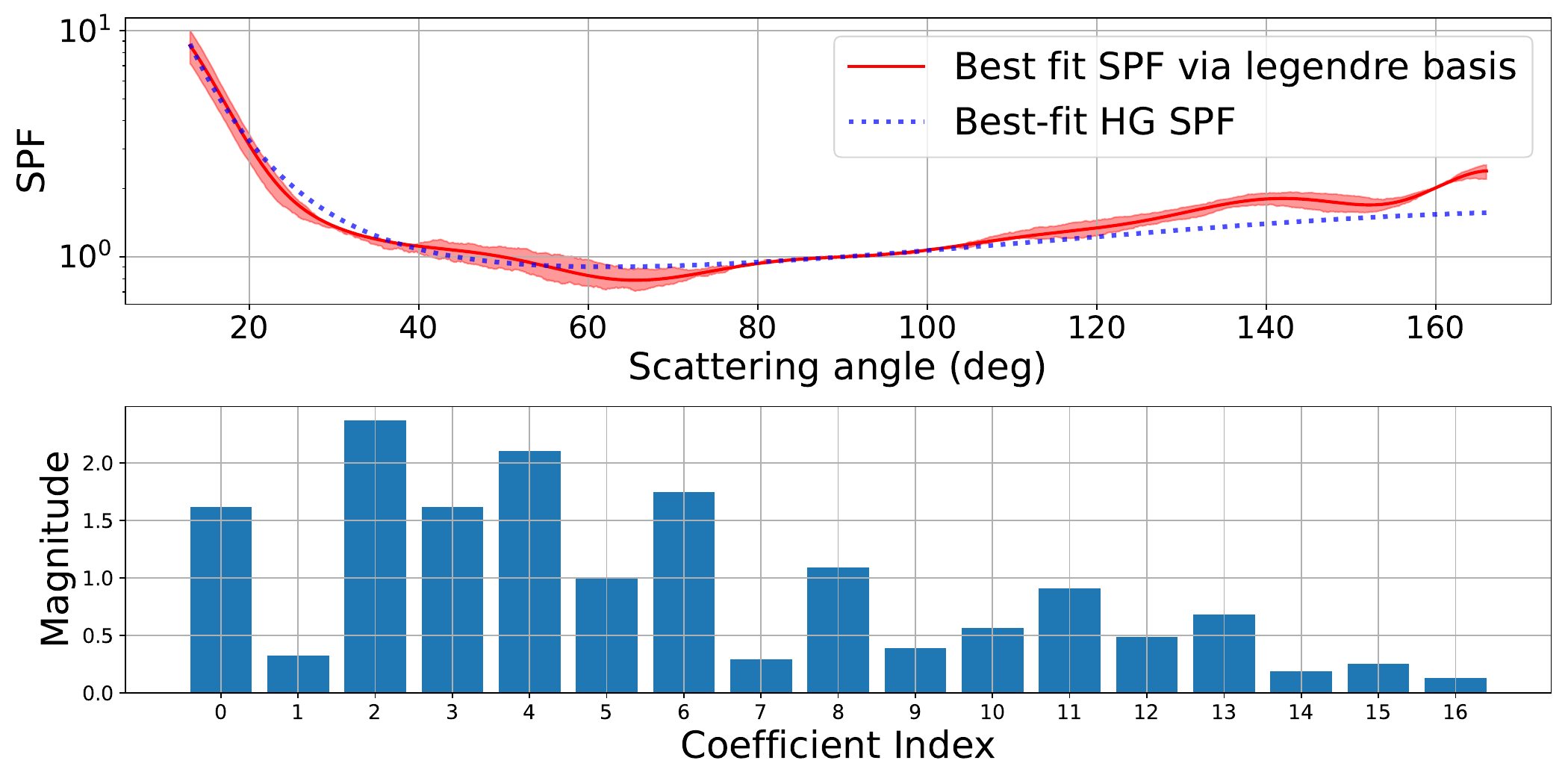}
    \caption{Comparison between the measured SPF using our Legendre-based methods and the best-fit HG SPF for the \rdi{} dataset from UT 2024 Mar. 28. The top plot shows the Legendre polynomial-based SPF as the solid red line and the HG SPF as the dotted blue line. The bottom histogram shows the magnitudes of the Legendre coefficients that makes up the Legendre-based SPF, also shown in Table \ref{tab:coefficients}.}
    \label{fig:appendix_legendre_spf_irdi}
\end{figure}

\begin{figure}
    \centering
    \includegraphics[width=0.75\linewidth]{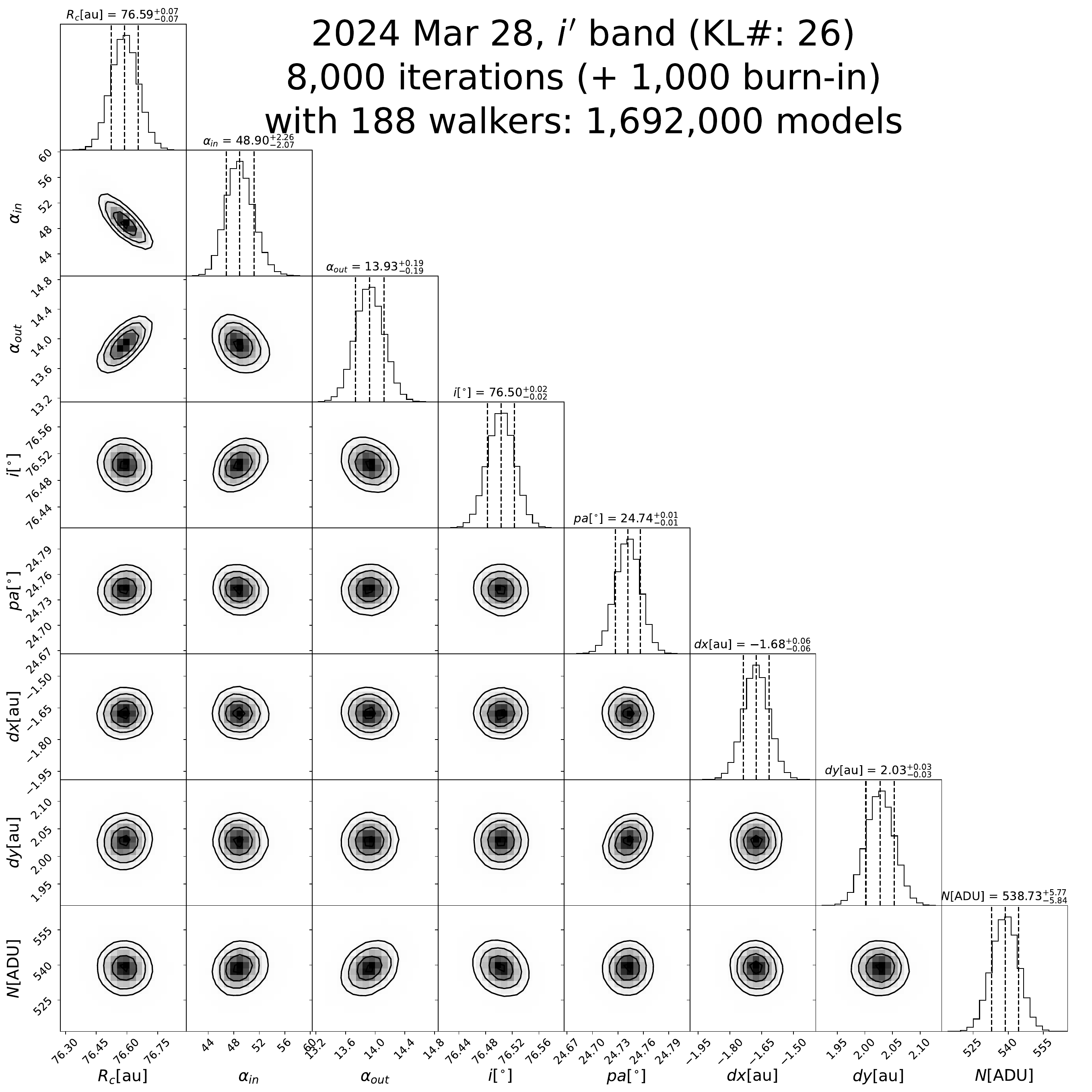}
    \caption{Posterior distributions for the fitted geometrical disk parameters to the \rdi{} image at \iband{}. Histograms on the main diagonal show the posterior probability distributions for each fitted parameter marginalized over all other fitted parameters where the dashed vertical lines show the 16th, 50th, and 84th percentiles. The off-diagonal scatter plots show the joint probability distributions overlaid with contours also representative of the aforementioned percentiles.}
    \label{fig:appendix_corner_irdi-geo}
\end{figure}

\begin{figure}
    \centering
    \includegraphics[width=\linewidth]{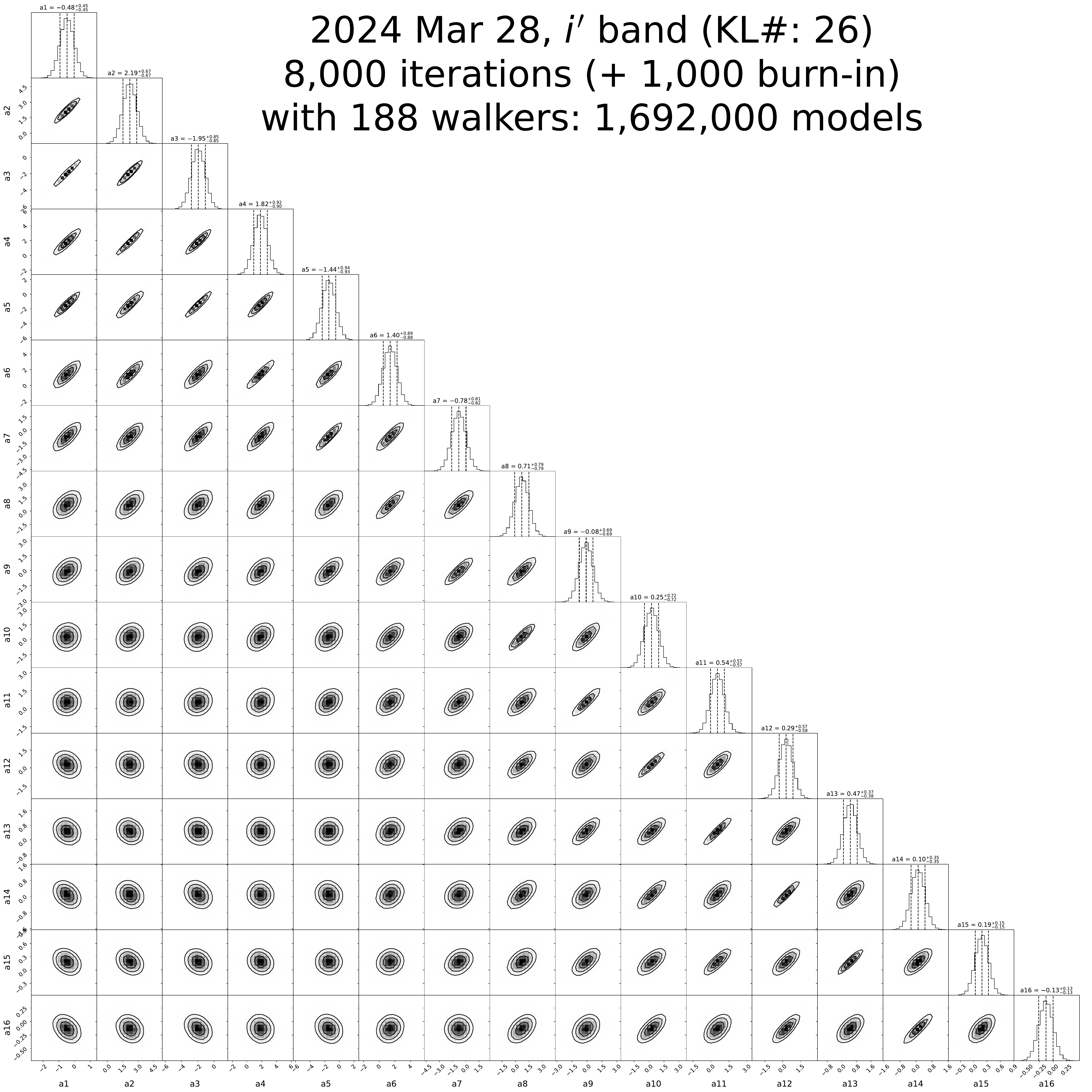}
    \caption{Posterior distributions for the fitted coefficient values for Legendre polynomials 1 through 16 (we fixed the zeroth Legendre polynomial).}
    \label{fig:appendix_corner_irdi-coeffs}
\end{figure}

\begin{figure}
    \centering
    \includegraphics[width=\linewidth]{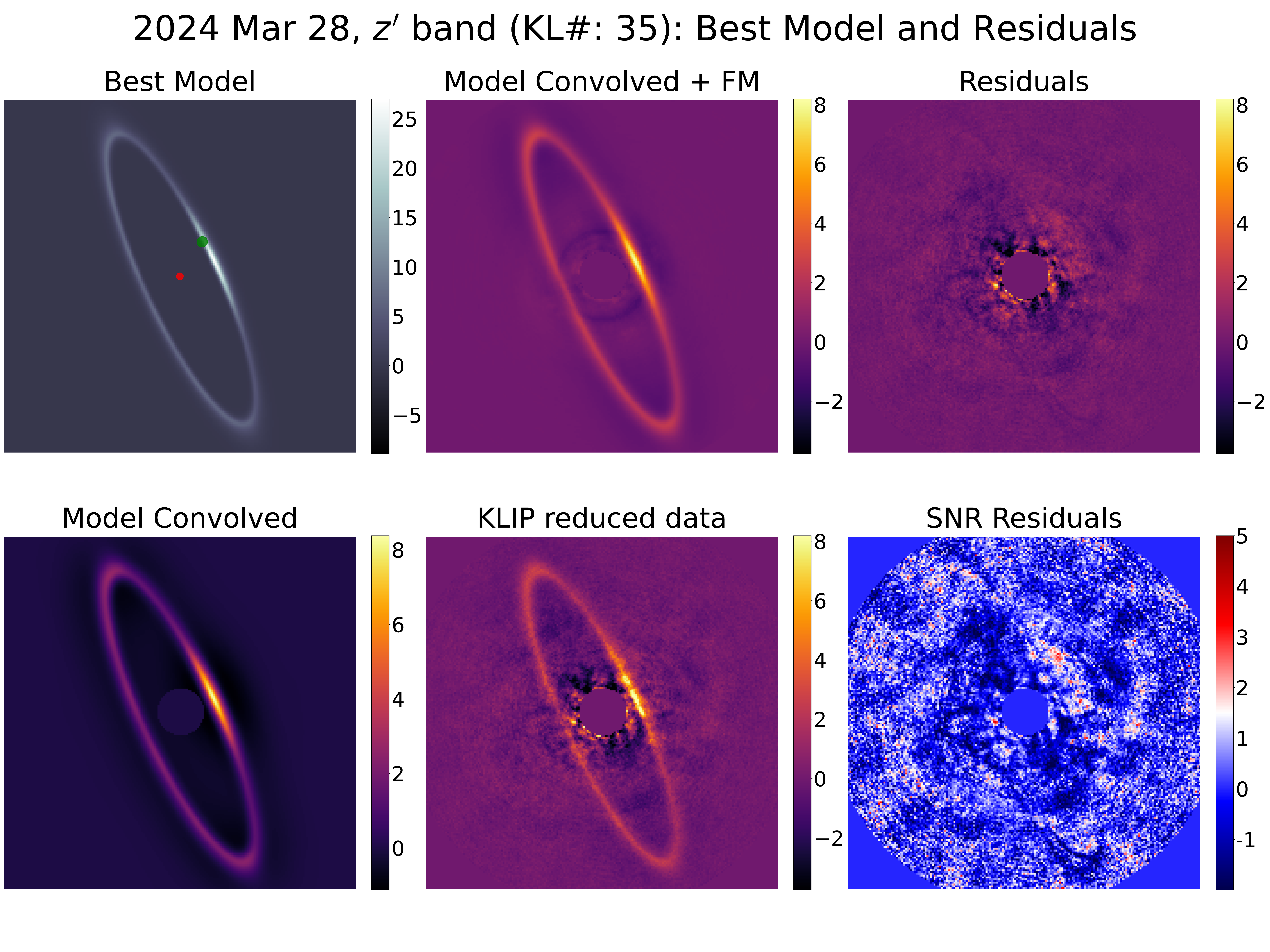}
    \caption{The same as Figure \ref{fig:appendix_diskfm_irdi}, but for the \rdi{} \zband{} image.}
    \label{fig:appendix_diskfm_zrdi}
\end{figure}

\begin{figure}
    \centering
    \includegraphics[width=\linewidth]{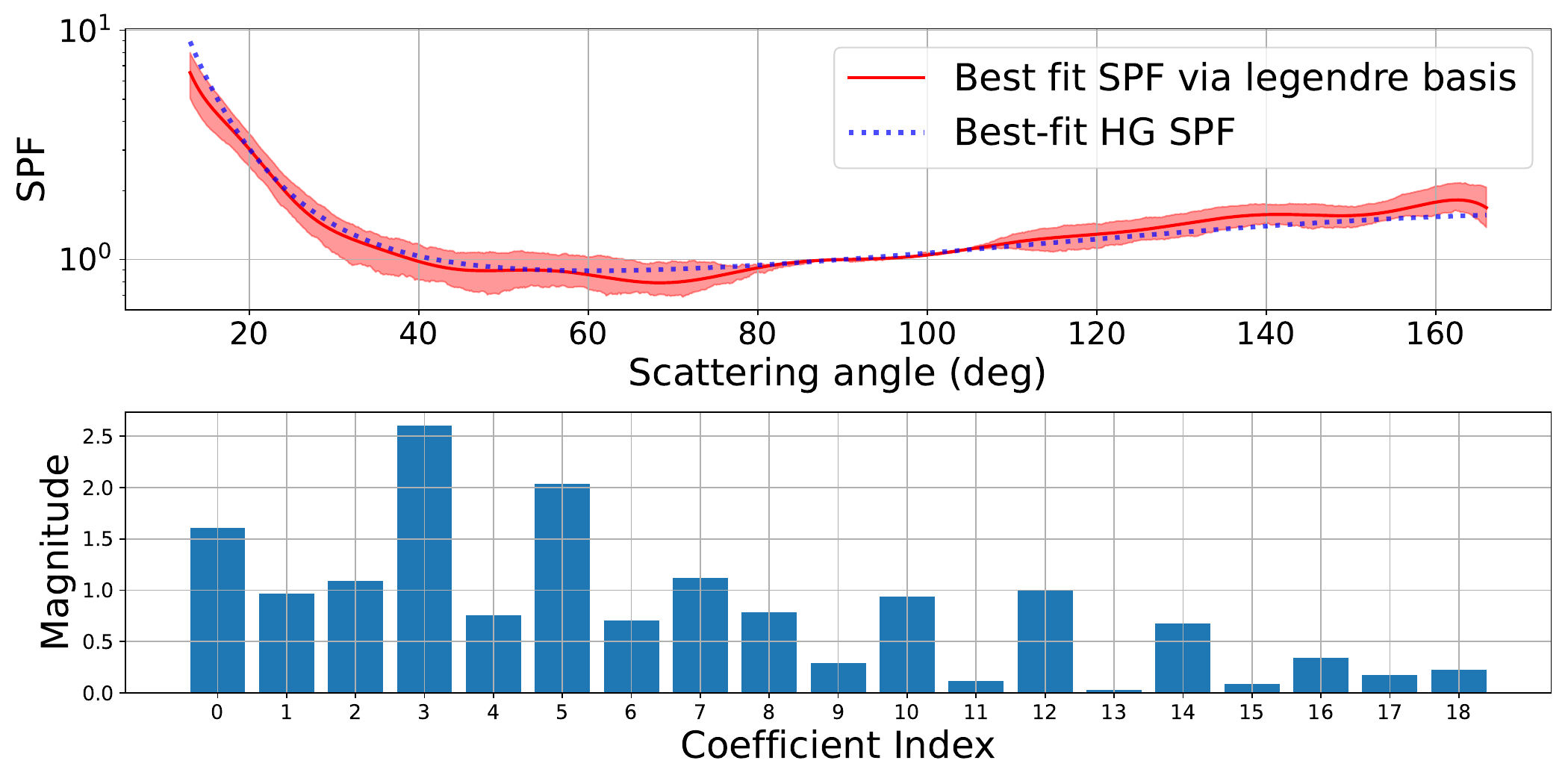}
    \caption{The same as Figure \ref{fig:appendix_legendre_spf_irdi}, but for the \rdi{} \zband{} image.}
    \label{fig:appendix_legendre_spf_zrdi}
\end{figure}

\begin{figure}
    \centering
    \includegraphics[width=0.75\linewidth]{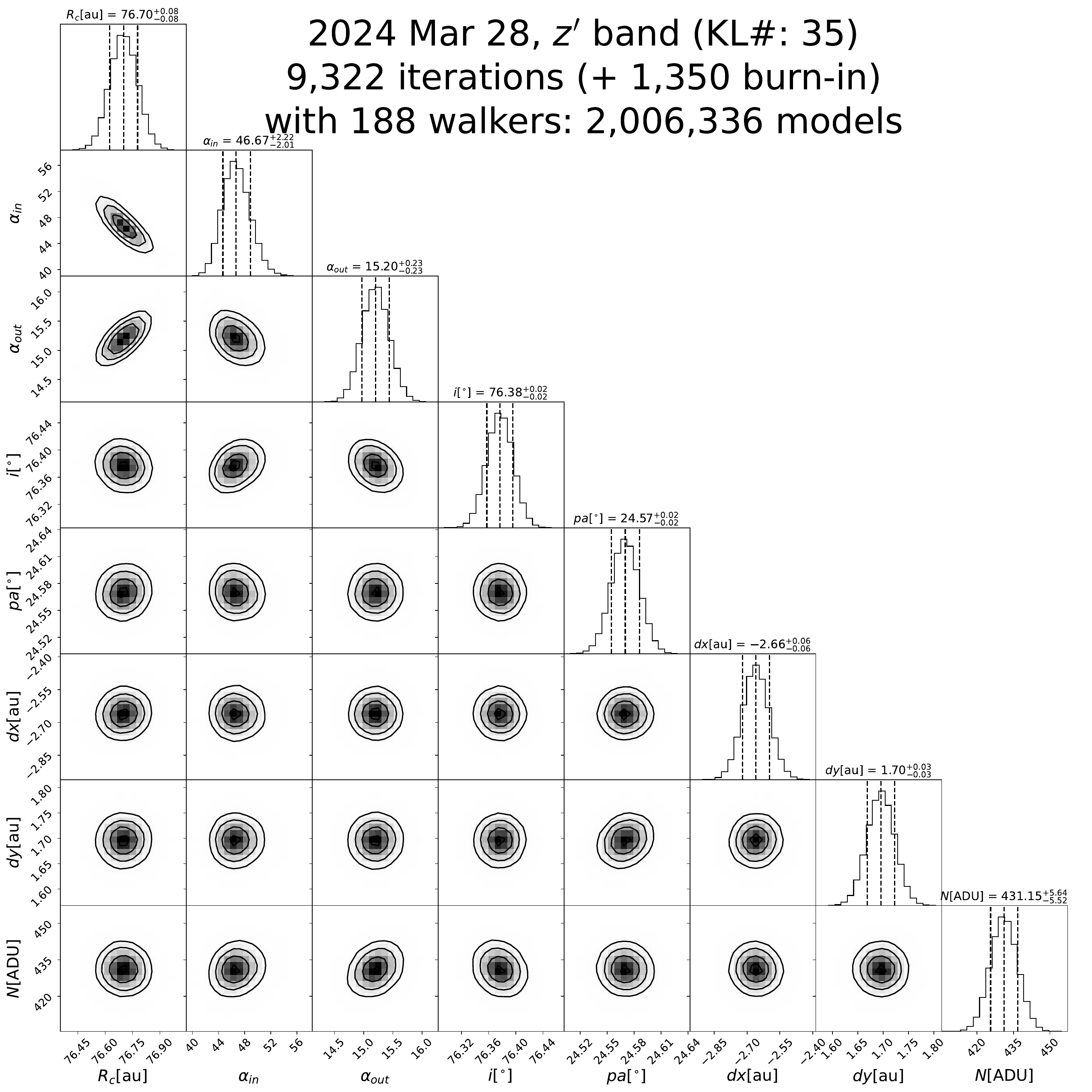}
    \caption{Same as Figure \ref{fig:appendix_corner_irdi-geo}, but for the \rdi{} \zband{} image.}
    \label{fig:appendix_corner_zrdi-geo}
\end{figure}

\begin{figure}
    \centering
    \includegraphics[width=\linewidth]{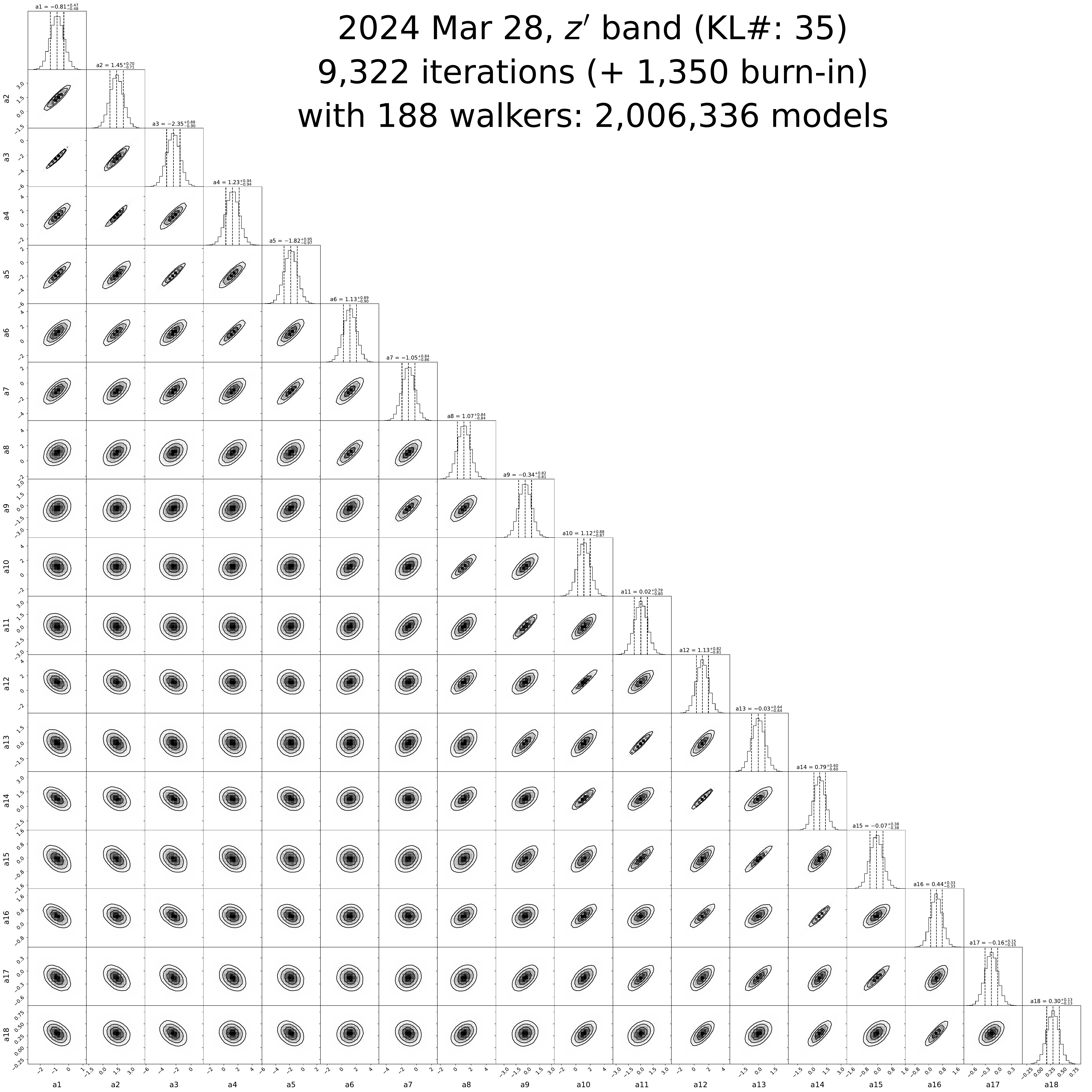}
    \caption{The same as Figure \ref{fig:appendix_corner_irdi-coeffs}, but for the \rdi{} \zband{} image. We used the coefficients for Legendre polynomials 1 through 18.}
    \label{fig:appendix_corner_zrdi-coeffs}
\end{figure}

We display the \textsc{DiskFM} MCMC results pertaining to our \rdi{} data in this appendix. Figures \ref{fig:appendix_diskfm_irdi} - \ref{fig:appendix_corner_zrdi-coeffs} illustrate the best SCL model, Legendre SPF, and associated corner plots for \iband{} and \zband{}. We ran all MCMC runs until the number of iterations exceeded 50 times the maximum autocorrelation time of the walker chains (computed using the included functionality in the \pkg{emcee} package; \citealt{foreman-mackey_emcee_2013}) and discarded the number of initial steps equal twice the autocorrelation time as recommended by the \pkg{emcee} documentation \footnote{\url{https://emcee.readthedocs.io/en/stable/user/autocorr/}}. As a sanity check, visual inspection of the traces of the walker paths showed that the walkers had stabilized well outside of both of these criteria.

\clearpage

\clearpage

\centering
\clearpage

\centering
\clearpage

\bibliography{HR4796}{}
\bibliographystyle{aasjournal}

\end{document}